\documentclass[a4paper,11pt]{article}
\usepackage{jheppub} 
\usepackage{cite}
\usepackage{tikz}
\usetikzlibrary{calc,decorations.pathmorphing}
\usepackage{amsmath, amssymb}
\usepackage{tikz-cd,feynmp-auto}
\usepackage{ytableau,youngtab}
\usepackage{pifont,subfig}
\usepackage{mdframed}

\tikzstyle{Orange Dot}=[fill={rgb,255: red,241; green,143; blue,31}, draw=black, shape=circle]
\tikzstyle{Green Dot}=[fill={rgb,255: red,120; green,151; blue,95}, draw=black, shape=circle]
\tikzstyle{Cyan Dot}=[fill={rgb,255: red,0; green,159; blue,223}, draw=black, shape=circle]
\tikzstyle{White Dot}=[draw=black, shape=circle]

\tikzstyle{Dashed}=[-, dashed]
\tikzstyle{Average}=[-,dashed, draw = {rgb,255: red,120; green,151; blue,95},ultra thick]
\tikzstyle{Orange}=[-, draw = {rgb,255: red,241; green,143; blue,31},ultra thick]
\tikzstyle{Cyan}=[-, draw = {rgb,255: red,0; green,159; blue,223},ultra thick]
\tikzstyle{Arrowright}=[->]

\title{\boldmath Five points for the Polyakov Bootstrap }

\author[a,b]{Ant\'onio Antunes,}
\author[b]{Sebastian Harris,}
\author[c,b]{Apratim Kaviraj}
\affiliation[a]{Laboratoire de Physique, \'Ecole Normale Sup\'erieure, Universit\'e  PSL, CNRS, Sorbonne Universit\'e, Universit\'e  Paris Cit\'e, 24 rue Lhomond, F-75005 Paris, France}
\affiliation[b]{Deutsches Elektronen-Synchrotron DESY, Notkestr. 85, 22607 Hamburg, Germany}
\affiliation[c]{Department of Physics, Indian Institute of Technology - Kanpur, Kanpur 208016, India}

\emailAdd{antonio.antunes@phys.ens.fr, 
sebastian.harris@desy.de,
akaviraj@iitk.ac.in}

\abstract{
    Higher-point correlation functions encode the data of infinitely many 4-point correlators in conformal field theory (CFT).
    In this paper, we develop new tools to efficiently extract this data from  multi-point crossing equations.
    Concretely, we generalize the functionals constituting the so-called Polyakov bootstrap of 4-point correlators to the case of 5-point functions in one-dimensional CFTs. 
    We first construct the crossing symmetric Polyakov blocks, and then derive sum-rules by requiring consistency with the operator product expansion (OPE). 
    This procedure leads to two classes of functionals controlling OPE coefficients of double- and triple-twist families. After extensively checking the validity of the associated sum-rules, we apply our functionals to the truncated 5-point bootstrap where we find several advantages with respect to more standard derivative functionals.
}

\begin{document}
\maketitle
\flushbottom

\section{Introduction}
\label{sec:Introduction}

    In an idealized version of the 4-point bootstrap approach to CFT, one would determine the local CFT data, i.e.~scaling dimensions and OPE coefficients of local operators by solving crossing for \textit{all} 4-point functions in the theory. 
    In particular, this includes operators in arbitrary representations of the rotation/Lorentz group as well as in any representation of the global symmetry. 
    Consistency of this infinite set of 4-point functions implies consistency of all higher-point correlation functions. In this paper, following \cite{Bercini:2020msp,Antunes:2021kmm,Kaviraj:2022wbw,Harris:2024nmr}, we turn this logic around: a single higher-point correlation function of light operators contains information about an infinite number of 4-point correlators, including arbitrarily heavy external states.
    This suggests that studying higher-point correlation functions could be a realistic alternative to an idealized bootstrap involving all 4-point constraints. 
    \smallskip
    
    The first step in this program is to consider 5-point functions. Such correlators are not reflection positive, and therefore are not amenable to standard positive semi-definite programming methods.\footnote{Instead, the six-point function does have this property in one of its topologies and this was explored in \cite{Antunes:2023kyz,Harris:2025cxx}.} Instead, we can consider a truncated version of the bootstrap equation \cite{Gliozzi:2013ysa,Gliozzi:2015qsa,Gliozzi:2016cmg,Li:2023tic}, and obtain estimates for the scaling dimensions and OPE coefficients in the 5-point function. In fact, with a refinement of this simple idea, estimates for $TT\epsilon$ OPE coefficient in the 3d Ising CFT, as well as other spinning-scalar-spinning three-point functions, have been obtained in this way \cite{Poland:2023vpn,Poland:2023bny,Poland:2025ide}. Crucially, when truncating the crossing equations, a choice of functionals to act on this equation is made, and, as we would like to argue in this paper, it is 
    of key importance to select functionals that capture information as efficiently as possible.

    For 4-point correlators, the theory of bootstrap functionals is well-developed. While most numerical approaches simply Taylor expand the crossing equation, many analytical studies introduce functionals that satisfy properties which lead to a more detailed and concise extraction of CFT data.\footnote{See however \cite{Paulos:2019fkw,Paulos:2021jxx,Ghosh:2023onl,Ghosh:2025sic} for numerical applications of such analytic functionals.} 
    In particular, \cite{Mazac:2016qev, Mazac:2018mdx, Mazac:2018ycv} introduced a set of `analytic functionals’ which provide bootstrap sum-rules that are trivially solved by a special OPE spectrum. This spectrum corresponds to a generalized free theory i.e. massive free theories in AdS$_2$ which are analytically solvable. 
    The analytic functionals are integrals with special kernels that account for all the singularities and asymptotic properties of conformal blocks and correlation functions in the entire complex plane of cross-ratios. 
    As a result the bounds and solutions derived from a single analytic  functional cannot be realistically obtained with derivative functionals as it would require an infinite number of the latter. 
    Beyond the framework of 1d CFT identical scalar correlators  analytic functionals have been constructed for global symmetries \cite{Ghosh:2021ruh}, higher dimensions \cite{Paulos:2019gtx, Mazac:2019shk}, boundary CFTs \cite{Kaviraj:2018tfd, Mazac:2018biw} and CFTs in real projective space \cite{Giombi:2020xah}. 

    The integral kernels of analytic functionals are notoriously difficult to work with and to generalize beyond the case of 4-point identical scalar correlators. 
    Fortunately this complication can be bypassed with an alternative method, the Polyakov bootstrap. 
    The idea here is to express a CFT correlator as an expansion in  Polyakov blocks, which are crossing symmetric combinations of tree-level Witten diagrams. 
    The equality of this description with the usual conformal block expansion furnishes a set of conditions on the OPE data that is equivalent to analytic functional sum-rules. 
    Tree-level Witten diagrams are well-studied objects and deriving the sum-rules from them can be achieved using standard techniques. 
    The Polyakov block is also a natural replacement of a conformal block in Mellin space, where it perfectly mimics all the analytic and crossing properties of a CFT correlation function. 
    In fact, the Polyakov block expansion can be rigorously established from crossing symmetric dispersion relations for CFTs in general spacetime dimensions \cite{Gopakumar:2021dvg, Bhat:2025zex}. See \cite{Penedones:2019tng,Carmi:2020ekr, Kaviraj:2021cvq, Carmi:2024tmp} for related works.

    In this paper, we derive analytic functionals for the 5-point correlator in the one-dimensional context, making use of the Polyakov bootstrap formalism. 
    We begin by assuming the existence of an expansion of the 5-point correlator $\mathcal{G}$ of identical operators in terms of crossing-symmetric Polyakov blocks $P$ which must be compatible with the usual OPE, i.e.~with the expansion in terms of ordinary conformal blocks $G$. 
    Concretely, defining the five point correlator $\mathcal{G}$ as
    \begin{equation}
         \langle \phi(x_1)\phi(x_2)\phi(x_3)\phi(x_4)\phi(x_5)\rangle (\mathcal{L}(x_i))^{-1} = \mathcal{G}(\chi_1, \chi_2)\,,
    \end{equation}
    where $\mathcal{L}$ is a scaling covariant prefactor and $\chi_{1,2}$ are two invariant cross-ratios of five points, we must have
    \begin{equation}
      \mathcal{G}(\chi_1, \chi_2)=\sum_{\Delta_1,\Delta_2} a_{\Delta_1,\Delta_2} G_{\Delta_1,\Delta_2}(\chi_1,\chi_2) = \sum_{\Delta_1,\Delta_2} a_{\Delta_1,\Delta_2} P_{\Delta_1,\Delta_2}(\chi_1,\chi_2)\,,  
    \end{equation}
    where $a_{\Delta_1,\Delta_2}$ is the appropriate product of OPE coefficients. Our first main result is an explicit construction of the Polyakov block. As in the 4-point case, this object can be written as a linear combination of tree-level exchange Witten diagrams, with some subtractions determined by contact interactions. We show that the Polyakov block can be written as
    \begin{equation}
    P_{\Delta_1, \Delta_2} =
    \left(
    \tikz[baseline=0ex]{\draw[thick] (0,0) circle(1.);
    
    \foreach \angle in {200, 160, 90, 20, -20} {
        \fill[black] (\angle:1.) circle(0.03);
    }
    
    \coordinate (L) at ($(0,0) + (-0.5,0)$);
    \coordinate (R) at ($(0,0) + (0.5,0)$);
    \coordinate (C) at ($(L)!0.5!(R)$);
    
    \draw[very thick, blue] (L) -- (C) ;\node[below] {$\Delta_1,\Delta_2$};
    \draw[very thick, red] (C) -- (R);
    ;
    
    \draw[thick] (160:1.) -- (L);
    \draw[thick] (200:1.) -- (L);
    
    \draw[thick] (20:1.) -- (R);
    \draw[thick] (-20:1.) -- (R);
    
    \draw[thick] (90:1.) -- (C);
    }
    + \dots
    \right)
    + c_1 \left(
    \tikz[baseline=0ex]{\draw[thick] (0,0) circle(1.);
    
    \foreach \angle in {200, 160, 30, 0, -30} {
        \fill[black] (\angle:1.) circle(0.03);
    }
    
    \coordinate (L) at ($(0,0) + (-0.5,0)$);
    \coordinate (R) at ($(0,0) + (0.5,0)$);

    \draw[very thick, blue] (L) -- (R);\node[below] {$\Delta_1$};

    \draw[thick] (160:1.) -- (L);
    \draw[thick] (200:1.) -- (L);
    
    \draw[thick] (30:1.) -- (R);
    \draw[thick] (-30:1.) -- (R);
    \draw[thick] (0:1.) -- (R);
    }
    + \dots
    \right)
    + c_2\,
    \tikz[baseline=0ex]{
      \draw[thick] (0,0) circle (1);
      \coordinate (C) at ($(0,0)$);
      \foreach \angle in {30,90,150,-30,210} {
        \fill[black] (\angle:1.);
        circle(0.03); }
      \draw[thick] (30:1) -- (C);
      \draw[thick] (90:1) -- (C);
      \draw[thick] (150:1) -- (C);
      \draw[thick] (-30:1) -- (C);
      \draw[thick] (210:1) -- (C);
    }\quad,
    \end{equation}
    where the $c_i$ coefficients depend on $\Delta_i$ and on the external dimension $\Delta_\phi$, the middle term is understood to be symmetrized under $(\Delta_1 \leftrightarrow \Delta_2)$ and the ellipsis denote all permutations of the external legs of the diagrams. Clearly, along with the exchanges of the form $\{\Delta_1,\Delta_2\}$, the Polyakov blocks contain several other families of exchanges in their OPE, including $\{\Delta_1,2\Delta_\phi+n\}\,,\{\Delta_1,\Delta_1+\Delta_\phi+n\}\,,\{2\Delta_\phi+n,2\Delta_\phi+m\}$ and $\{2\Delta_\phi+n,3\Delta_\phi+m\}$. Consistency with the OPE leads to the requirement that the exchanges of these spurious families of operators cancel out when summing over all Polyakov blocks. We then find several families of sum-rules for the 5-point function, with the most important ones being
    \begin{equation}
        \sum_{\Delta_1,\Delta_2} a_{\Delta_1,\Delta_2} \alpha^{(2\Delta_\phi,2\Delta_\phi)}_{n,m}(\Delta_1,\Delta_2)=0\,, \qquad  \sum_{\Delta_1,\Delta_2} a_{\Delta_1,\Delta_2} \alpha^{(2\Delta_\phi,3\Delta_\phi)}_{n,m}(\Delta_1,\Delta_2)=0\,,
    \end{equation}
    where we introduced the crossing functionals $\alpha$ labeled by the different families of spurious exchanges in the obvious way. These functionals depend on the two variables $\Delta_1,\Delta_2$ and capture information about OPE coefficients of the form $C_{\phi,[\phi \phi]_n, [\phi \phi]_m}$ and $C_{\phi,[\phi \phi]_n, [\phi \phi \phi]_m}$; they are therefore a vast generalization of the usual 4-point functionals. With these functionals at hand, it is then straightforward to use them in the truncated 5-point bootstrap as a replacement (or complement) to the standard derivative functionals. We find that our functionals have multiple advantages over the familiar derivative basis.

    \medskip

    The rest of this paper is structured as follows. In Section \ref{sec:recap} we review the Polyakov bootstrap approach to the 4-point crossing equation, emphasizing the extraction of double-trace coefficients in Witten diagrams through a one-dimensional Mellin formalism. In Section \ref{sec:5pt}, which is the core of the paper, we first recall basic facts about conformal 5-point functions, and then write down an ansatz for the crossing symmetric Polyakov Block as a sum of tree-level Witten diagrams. We subsequently derive several families of two-variable sum-rules by requiring consistency with the OPE. We also test these sum-rules in several non-trivial correlators confirming the validity of our results. In Section \ref{sec:Appli} we give a simple application of our functionals, using them to bootstrap the OPE coefficients of a non-trivial solution to crossing through the truncated bootstrap approach. We conclude in Section \ref{sec:Conclusions} where we point out  directions for future developments. The paper is complemented by several appendices.  Appendix \ref{app:Mellin} develops in detail the Mellin techniques used throughout the main text. Appendix \ref{app:spectral} reviews the spectral representation which makes the structure of Witten diagrams more manifest. Appendix \ref{app:integratedvertex} reviews the so-called integrated vertex identities which can be used as a cross-check of the Mellin computations and streamline certain computations. Finally, Appendix \ref{app:recrel} establishes Casimir recursion relations for the 5-point blocks and Witten diagrams which serve as a further cross-check to the other techniques used in the paper, but could also be of independent interest. 

\section{Polyakov Bootstrap recap}
\label{sec:recap}
    In this section, we review the necessary ingredients to derive Polyakov Bootstrap sum-rules for 4-point functions, emphasizing the techniques that will be needed in the 5-point case. We first review the basic kinematics of 1d CFTs in Section \ref{ssec:4pt} and then the principles of Polyakov bootstrap for the simple case of 4-point correlation function of identical scalars in Section \ref{sec:Polyakov4pt}, followed by the necessary 1d Mellin amplitude methods in Section \ref{ssec:mellin}.
    
\subsection{Four-point kinematics}
\label{ssec:4pt}
 In general dimensions the 4-point correlator is given by 
\begin{align}
&\langle \phi(x_1)\phi(x_2)\phi(x_3)\phi(x_4)\rangle=\frac{\mathcal{G}(u,v)}{x_{12}^{2\Delta_\phi}x_{34}^{2\Delta_\phi}}\,,\\ \text{ \  where \  } &u=\chi\bar{\chi}=\frac{x_{12}^{2}x_{34}^{2}}{x_{13}^{2}x_{24}^{2}}\,, \ \ v=(1-\chi)(1-\bar{\chi})=\frac{x_{14}^{2}x_{23}^{2}}{x_{13}^{2}x_{24}^{2}}\nonumber\,.
\end{align}
In 1d the two crossratios are related as $\chi=\bar{\chi}$, so we may write $\mathcal{G}(u,v)=\mathcal{G}(\chi)$. \\ \\
If the $\phi(x_i)\phi(x_j)$ OPE is composed of operators with dimensions denoted by $\Delta$ with respective OPE coefficients $C_\Delta$, then the 
$s$-channel conformal block expansion of the function $\mathcal{G}(\chi)$ is given by
\begin{equation}\label{blockexp4pt}
\mathcal{G}(\chi)=\sum_{\Delta}C_{\Delta}^2G_{\Delta}(\chi)\,.
\end{equation}
where $G_{\Delta}(\chi)=\chi^{\Delta}{}_2F_1(\Delta,\Delta,2\Delta,\chi)$ denotes SL$(2,\mathbb{R})$ conformal blocks. \\ \\
For bosonic operators, the correlation function is crossing symmetric in the exchanges $x_2\leftrightarrow x_4$ as well as $x_2\leftrightarrow x_4$. This implies the conditions:
\begin{equation}\label{cross4pt}
    \mathcal{G}(\chi)=\left[\frac{\chi}{1-\chi}\right]^{2\Delta_\phi}\mathcal{G}(1-\chi)=\chi^{2\Delta_\phi}\mathcal{G}(1/(1-\chi))
\end{equation}
The r.h.s. of the first and second equalities give the $t$ and $u$ channel conformal block expansions. The bootstrap equation for the $s$-$t$ channel crossing equation is given by
\begin{equation}\label{bootseqn}
    \sum_{\Delta}C_\Delta^2\,F_{\Delta}(\chi)=0
\end{equation}
where $F_\Delta(\chi)=\chi^{-2\Delta_\phi}G_\Delta(\chi)-(1-\chi)^{-2\Delta_\phi}G_\Delta(1-\chi)$  is the bootstrap vector. \\ \\
We will focus on correlation functions that diverge at most with the rate $\chi^{2 \Delta_\phi}$ in the limit of large $\chi$, i.e.~correlators that satisfy
\begin{equation}\label{Regge4pt}
    \chi^{-2\Delta_\phi-\epsilon}\, \mathcal{G}(\chi)\stackrel{\chi\to \infty }{=\joinrel = }0 \, \text{ \  for any \  } \epsilon >0 \,.
\end{equation}
This is called boundedness in the $u$-channel Regge limit and is a consequence of unitarity \cite{Mazac:2018ycv}. In conventional bootstrap methods  \eqref{Regge4pt} is not assumed for bounding OPE data, but it is expected to  hold for a complete solution to crossing symmetry. 

\paragraph{Generalized free theory: } An important solution to crossing, that we will keep coming back to, is the 4-point function of a generalized free theory (GFF). It is given by the correlation function 
\begin{equation}\label{gff4pt}
\mathcal{G}(\chi)=1+\chi^{2\Delta_\phi}+\left[\frac{\chi}{1-\chi}\right]^{2\Delta_\phi}  \,.
\end{equation}
This corresponds to the CFT at the boundary of a bulk free massive bosonic theory in AdS$_2$. The mass is related to scaling dimension by $m^2=\Delta_\phi(\Delta_\phi-1)$. Since the theory is free its correlators are given by Wick contractions. From the conformal block decomposition one can see the operator dimensions $\Delta=\Delta_n:= 2\Delta_\phi+2n$ in the OPE with OPE coefficients 
\begin{equation}
    C_{\Delta_n}^2=\frac{2\Gamma(4\Delta_\phi+2n-1)((2\Delta_\phi)_{2n})^2}{\Gamma(2n+1)\Gamma(4\Delta_\phi+4n-1)}\label{gffopecoeff}\,.
\end{equation}
Clearly \eqref{gff4pt} satisfied both conditions \eqref{cross4pt} and \eqref{Regge4pt}. In a bootstrap problem the GFF corresponds to a spectrum with the maximum OPE coefficient of the first non-identity operator whose dimension is fixed at $2\Delta_\phi$. 

\subsection{Polyakov bootstrap}
\label{sec:Polyakov4pt}
The main idea in Polyakov bootstrap is to replace the conformal blocks in the expansion \eqref{blockexp4pt} with a new object that is manifestly crossing symmetric and Regge bounded. We write this demand as follows
\begin{equation}\label{Polyblockexp4pt}
\mathcal{G}(\chi)=\sum_{\Delta}C_{\Delta}^2\, P_{\Delta}(\chi)\,,
\end{equation}
where $P_{\Delta}(\chi)$ is called a Polyakov block  that satisfies
\begin{equation}\label{Polycross4pt}
    P_{\Delta}(\chi)=\left(\chi/(1-\chi)\right)^{2\Delta_\phi}P_{\Delta}(1-\chi)=\chi^{2\Delta_\phi}P_{\Delta}(1/(1-\chi))
\end{equation}
and also $\chi^{-2\Delta_\phi}\, P_\Delta(\chi)=0$ is finite as $\chi\to \infty$. These requirements indicate that a Polyakov block is itself a consistent solution of the crossing equation, but it does not have a unique form. A possible definition is in terms of tree level Witten diagrams as follows
\begin{equation}\label{Polydef}
    P_{\Delta}(\chi)=W^{(s)}_{\Delta}(\chi)+W^{(t)}_{\Delta}(\chi)+W^{(u)}_{\Delta}(\chi)+c_{\Delta}W^{\text{con}}(\chi)
\end{equation}
Here $W^{(i)}_{\Delta}(\chi)$ denotes a tree level Witten diagram exchanging a bulk operator corresponding to dimension $\Delta$ in the $i=s,t,u$ channel, and $W^{\text{con}}(\chi)$ is a 4-point contact diagram. The term $c_\Delta$ can be chosen in multiple ways. One possible choice is given below \eqref{Polyblockdeco} and is discussed further at the end of this subsection. \\
We discuss the structure of the Witten diagrams in more detail in the next subsection. For now let us note two important properties. Firstly, all the diagrams in \eqref{Polydef} are Regge bounded. 
Also each of them can be written as a sum over conformal blocks and their derivatives as follows:
\begin{align}
    W^{(i)}_{\Delta}(\chi)=&\delta_{i,s}G_{\Delta}(\chi)+\sum_{n}\Big[a^{(i)}_{n}(\Delta,\Delta_\phi)G_{\Delta_n}(\chi)+b^{(i)}_{n}(\Delta,\Delta_\phi)\partial_{\Delta}G_{\Delta_n}(\chi) \nonumber\\
    & \hspace{2cm} + \tilde{a}^{(i)}_{n}(\Delta,\Delta_\phi)G_{\Delta_n+1}(\chi)+\tilde{b}^{(i)}_{n}(\Delta,\Delta_\phi)\partial_{\Delta}G_{\Delta_n+1}(\chi)\Big]\,,
    \label{Wsdeco}\\
     W^{\text{con}}(\chi)=&\sum_{n}\Big[a^{(c)}_{n}(\Delta_\phi)G_{\Delta_n}(\chi)+b^{(c)}_{n}(\Delta_\phi)\partial_{\Delta}G_{\Delta_n}(\chi) \label{Wcondeco}\Big]\,.
\end{align}
We additionally have $\tilde{a}_n^{(s)}=\tilde{b}_n^{(s)}=0$ and $\tilde{a}_n^{(t)}=-\tilde{a}_n^{(u)}, \tilde{b}_n^{(t)}=-\tilde{b}_n^{(u)}$\,. Putting these together in \eqref{Polydef} gives the conformal block decomposition of a Polyakov block
\begin{equation}\label{Polyblockdeco}
    P_{\Delta}(\chi)=G_{\Delta}(\chi)+\sum_{n}\Big[\alpha_{n}(\Delta,\Delta_\phi)G_{\Delta_n}(\chi)+\beta_{n}(\Delta,\Delta_\phi)\partial_{\Delta}G_{\Delta_n}(\chi)\Big]\,,
\end{equation}
where $\alpha_n:=\sum_i a_n^{(i)}+c_\Delta \, a_n^{(c)}$ and $\beta_n:=\sum_i b_n^{(i)}+c_\Delta \, b_n^{(c)}$\,. 
The blocks and derivatives of blocks corresponding to $\Delta_n$ values are sometimes called `spurious blocks'. Finally let us point out that we choose $c_\Delta$ so that $\beta_0(\Delta,\Delta_\phi)=0$.
\\ \\
The statement of Polyakov bootstrap is that the Polyakov block expansion is the same as conformal block expansion:
\begin{equation}\label{Polyboot4pt}
    \sum_{\Delta}C_{\Delta}^2\, P_{\Delta}(\chi)=\sum_{\Delta}C_{\Delta}^2\, G_{\Delta}(\chi)\,.
\end{equation}
This means if we choose the right dimensions in the OPE in a $P_\Delta(\chi)$ expansion, the spurious blocks cancel among themselves leaving only the sum over physical block as in r.h.s. of \eqref{Polyboot4pt}. So the Polyakov bootstrap conditions can be restated as follows:
\begin{align}
\sum_{\Delta}\, C_{\Delta}^2 \, \alpha_n(\Delta,\Delta_\phi)=\sum_{\Delta}\, C_{\Delta}^2 \, \beta_n(\Delta,\Delta_\phi)=0\label{Polysumrule}
\end{align}
Let us mention here, in words, that there is an alternative way to derive \eqref{Polysumrule} from the crossing symmetry equation. This is by using analytic functionals which are a set of dual functionals (call $\alpha_n$ and $\beta_n$) to a basis of the crossing vectors $F_{\Delta_n}(\chi)$ and $\partial_\Delta F_{\Delta_n}(\chi)$ respectively. The fact that any $F_\Delta(\chi)$ can be expanded in such a basis is nothing but the consequence of a Polyakov block $P_{\Delta}(\chi)$ being crossing symmetric on its own and its decomposition satisfies eqn. \eqref{bootseqn}. The exact forms of the functionals will not be important in this paper , but their actions on $F_{\Delta}(\chi)$ are precisely $\alpha_n(\Delta,\Delta_\phi)$ and $\beta_n(\Delta,\Delta_\phi)$ \cite{Mazac:2016qev,Mazac:2018mdx,Mazac:2018ycv}, and acting on the bootstrap equation \eqref{bootseqn} they give the sum-rules \eqref{Polysumrule}.\\ \\
An important feature of the Polyakov block coefficients (or analytic functional actions) is the orthogonality property
\begin{align}
\label{4ptortho}
    &\alpha_n(\Delta_m,\Delta_\phi)=\delta_{mn}\hspace{1cm} \partial_\Delta\alpha_n(\Delta_m,\Delta_\phi)=c_n\delta_{m0}\nonumber\\
    &\beta_n(\Delta_m,\Delta_\phi)=0\hspace{1.5cm} \partial_\Delta\beta_n(\Delta_m,\Delta_\phi)=\delta_{mn}+d_n\delta_{m0}\,.
\end{align}
These are consequences of the dualities with the basis vectors $F_{\Delta_n}, \partial_\Delta F_{\Delta_n}$ mentioned in the   paragraph above. All $\beta$ sum-rules also satisfy $\beta_n(\Delta=0)=0$ so they vanish identically on the GFF spectrum. The $\alpha$'s also trivialize, but $\alpha_n(\Delta=0)=-C_{\Delta_n}$ so that each $\alpha_n$ sum-rule corresponds to the solution of the $n$-th GFF OPE coefficient. \\ \\ %
Furthermore since the sum-rules have mostly double zeroes, they let us solve any small deformation around GFF also very easily. To be concrete, if we consider a deformation 
\begin{equation}
    \Delta_{(n)}=\Delta_n+\gamma_n\,, \ \ C^2_{\Delta_{(n)}}=C^2_{\Delta_n}+\delta_n.
\end{equation}
and assume $\gamma_0$ to be a fixed parameter. Then all other $\gamma_n$ and $\delta_n$'s could be obtained as 
\begin{equation}
    \delta_n=-\gamma_0\, c_n\,, \ \ \gamma_n=-\gamma_0d_n\,.
\end{equation}
The fact that we chose $\gamma_0$ as the deformation parameter goes back to our choice of  the Polyakov block.  We chose $c_\Delta$ in \eqref{Polydef} such that there is no $\beta_0$ sum-rule, so we cannot solve $\gamma_0$ from the sum-rules. Alternatively we could have chosen to eliminate any other sum-rule. The justification of this is that given a Regge-bounded crossing solution $\mathcal{G}_0(\chi)$ we can define, with an arbitrary parameter $\lambda$, another correlator $\mathcal{G}_{\lambda}(\chi)=\mathcal{G}_0(\chi)+\lambda\,W^{\text{con}}(\chi)$ which is still Regge-bounded and crossing symmetric.
\subsection{Mellin approach in one dimension}
\label{ssec:mellin}
In this subsection we will review Mellin transforms  of CFT correlators, especially Witten diagrams in CFT Mellin space. We will show how the Mellin space is a natural framework to set up the Polyakov bootstrap. \\
The inverse Mellin transform of a correlation function of four identical scalars is defined by
\begin{equation}\label{Mellin-def}
    \mathcal{G}(u,v)=\int_{-i\infty}^{i\infty}\, \frac{u^{\Delta_\phi-s_{12}}v^{-s_{14}}}{(2\pi i)^2}\, M(s_{12},s_{14})\,\Gamma(s_{12})^2\Gamma(s_{14})^2\Gamma(\Delta_\phi-s_{12}-s_{14})^2\,.
\end{equation}
In 1d we simply set $u=\chi^2$ and $v=(1-\chi)^2$ and continue to work with two Mellin variables.\footnote{There is no ambiguity since we always compute the Mellin integrals in a way that is compatible with the OPE. This formalism has the advantage of simplifying the form of the Mellin amplitudes, unlike an intrinsically one-dimensional Mellin formalism \cite{Bianchi:2021piu}.} Here $M(s_{12},s_{14})$ is the Mellin transform or Mellin amplitude of $\mathcal{G}(\chi)$ which we will denote as $\mathcal{M}[\,\mathcal{G}\,]\,(s_{12},s_{14})=M(s_{12},s_{14})$\,.\\ \\
Now let us discuss  Witten diagrams in AdS$_2$. To define them let us consider a free bosonic theory with cubic and quartic interactions as follows
\begin{equation}
    S_{\text{AdS}_2}=\int d^2x \sqrt{g}\Big[(\partial \Phi)^2+m_{\Delta_\phi}^2\Phi^2+(\partial \mathcal{O})^2+m_{\Delta}^2\mathcal{O}^2+g_1\Phi^2\,\mathcal{O}+g_2\Phi^4\Big]
. 
\end{equation}
In this subsection we only consider Witten diagrams with four external $\Phi$'s on the boundary. We get an exchange diagrams using the cubic vertex. It exchanges the scalar bulk operator $\mathcal{O}$ with mass $m^2=\Delta(\Delta-1)$, and is proportional to $g_1^2$. The contact diagram is obtained from the quartic vertex and is proportional to $g_2$.  \\ \\
We will ignore the vertices $g_1$ and $g_2$ below and review only the kinematical part. An $s$-channel exchange diagram has the Mellin transform $\mathcal{M}[W_{\Delta}^{(s)}]=M^{(s)}_{\Delta}$ given by \cite{Penedones:2010ue,Fitzpatrick:2011ia,Paulos:2011ie}
\begin{equation}\label{Ms}
    M^{(s)}_{\Delta}(s_{12},s_{14})=\sum_{m=0}^{\infty}\frac{\mathcal{Q}_m^{(\Delta_\phi)}}{s_{12}+\frac{\Delta}{2}-\Delta_\phi+m}\,.
\end{equation}
The $\mathcal{Q}_m^{(\Delta_\phi)}$ is defined in App. \ref{app:Mellin1}. 
Notice that it is independent of $s_{14}$ which is a property of the spin 0 exchange diagram in $s$ channel. The Mellin transforms of $t$ and $u$ channel exchange diagrams are readily obtained from $M^{(s)}_{\Delta}$ by switching $s_{12}\leftrightarrow s_{14}$ and $s_{12}\to \Delta_\phi-s_{12}-s_{14}$ respectively:
\begin{align}
    M^{(t)}_{\Delta}(s_{12},s_{14})&=\sum_{m=0}^{\infty}\frac{\mathcal{Q}_m^{(\Delta_\phi)}}{s_{14}+\frac{\Delta}{2}-\Delta_\phi+m}\,, \\
        M^{(u)}_{\Delta}(s_{12},s_{14})&=\sum_{m=0}^{\infty}\frac{\mathcal{Q}_m^{(\Delta_\phi)}}{\frac{\Delta}{2}+m-s_{12}-s_{14}}\,.
\end{align}
The contact diagram Mellin transform $\mathcal{M}[W^{\text{con}}]=M^{\text{con}}$ is simply a constant. We may normalize it to be unity
\begin{equation}
    M^{\text{con}}(s_{12},s_{14})=1\,.
\end{equation}
In 1d, the poles in $M^{(s)}_{\Delta}$ under the inverse Mellin integral \eqref{Mellin-def} gives the block $G_{\Delta}(\chi)$ from the decomposition of s-channel diagram (see \eqref{Wsdeco}). 
Taking the poles in $s_{12}$ in the negative direction is the correct thing to do when we want a small $\chi$ expansion, as one has to close the integral contour at large negative $s_{12}$. This naturally corresponds to the $s$ channel conformal block expansion. \\ \\
The spurious blocks in that decomposition come from the $s_{12}$ poles of the $\Gamma$-functions in the integral measure of inverse Mellin transformation. This way the crossed channel and contact diagram Mellin amplitudes also follow the decompositions of \eqref{Wsdeco} and \eqref{Wcondeco} respectively.
We come back to them in a moment. \\ \\
We can now write  down the Mellin transform of a Polyakov block. For 4-point identical scalar correlator the Polyakov block in Mellin space is as follows 
\begin{align}\label{PolydefMellin}
     P_{\Delta}(s_{12},s_{14})=\sum_{m=0}^{\infty}\bigg[&\frac{\mathcal{Q}_m^{(\Delta_\phi)}}{s_{12}+\frac{\Delta}{2}-\Delta_\phi+m} +\frac{\mathcal{Q}_m^{(\Delta_\phi)}}{s_{14}+\frac{\Delta}{2}-\Delta_\phi+m}+\frac{\mathcal{Q}_m^{(\Delta_\phi)}}{\frac{\Delta}{2}+m-s_{12}-s_{14}}\bigg]
     +c_{\Delta}\,.
\end{align}
Let us now discuss why $P_{\Delta}(s_{12},s_{14})$ is the most natural analogue of a conformal block in Mellin space.
First let us point out that it is not possible to write a Mellin amplitude whose poles give only $G_{\Delta}(\chi)$ since it is not possible to close $s_{12}$ contour with those poles alone. So $s$ channel diagram $M^{(s)}_{\Delta}$ is the closest and simplest thing to a conformal block in Mellin space although it generates the spurious blocks in addition. The addition of $M^{(t)}_{\Delta}$ and $M^{(u)}_{\Delta}$ to it makes it crossing symmetric without adding any new spurious blocks.  \\ \\  
The addition of a constant $c_\Delta$ in the Polyakov block $P_{\Delta}(s_{12},s_{14})$ preserves the crossing symmetry and the spurious block content. It turns out that \eqref{PolydefMellin} is the most general expression we can write that does not grow at large $s_{12}$. This is the Mellin version of Regge boundedness condition. The general large $\chi$ behavior in Mellin space becomes \cite{Mazac:2018ycv}
\begin{equation}
    \mathcal{G}(\chi)\stackrel{\chi\to \infty}{\sim} \chi^{2j-1}\, \ \implies \  \mathcal{M}[\,\mathcal{G}\, ]=M(s_{12}) \stackrel{s_{12}\to \infty}{\sim}s_{12}^{2j}\,.
\end{equation}
A bounded correlator in Regge limit requires $j\le \frac 12$ or Mellin amplitude growing at most linearly in $s_{12}$. For the identical external scalar case the only completely crossing symmetric polynomial (contact diagram) with this property is a constant. The exchange diagrams also share the same property as can be checked easily. \\ \\
As discussed in Section \ref{sec:Polyakov4pt} the constant $c_{\Delta}$ is fixed by eliminating the coefficient of a spurious block. we may of course relax the Regge boundedness condition and require a softer $M\sim s_{12}^{2j}$, $j>\frac 12$ behavior for large $s_{12}$. We then have to add more crossing symmetric polynomials. So if there are $n$ independent polynomials we have to fix them by eliminating $n$ sum-rules. This represents the freedom of deforming a crossing solution by $n$ independent contact diagrams. 
\paragraph{Polyakov block rules:} Let us summarize the rules for writing a 4-point Polyakov block $P_{\Delta}(s_{12},s_{14})$ for a correlator $\mathcal{G}(\chi)$ in Mellin space are as follows:
\begin{itemize}
    \item $P_{\Delta}(s_{12},s_{14})$ is completely crossing symmetric in the Mellin variables $s_{12}, s_{14}$. 
    \item $P_{\Delta}(s_{12},s_{14})$ has simple poles in $s_{12}$ that reproduce $G_{\Delta}(\chi)$ and regular terms (polynomials). 
    
    \item The regular terms should respect the Regge boundedness of $\mathcal{G}(\chi)$. 
    \item The number $n$ of independent regular terms has to fixed by eliminating  $n$ sum-rules. 
\end{itemize}
These rules can be established from the consistency with the analytic functional method discussed in the previous subsection.  
But a more direct way to derive $P_{\Delta}(s_{12},s_{14})$ is from the crossing symmetric dispersion relation (CSDR) shown in  \cite{Gopakumar:2021dvg, Bhat:2025zex}. The CSDR is a dispersion relation that manifests the crossing symmetry of a function (Mellin amplitude in our case), with a special dispersive kernel, at the cost of locality or Regge boundedness. One has to further tune it to restore the missing properties - then CSDR reduces to a sum of the general spin of a Polyakov block. The spin 0 sector of this sum, relevant in 1d, corresponds to our $P_{\Delta}(s_{12},s_{14})$. 
\paragraph{Computation of sum-rules: } Finally let us discuss the computation of the coefficients $\alpha(\Delta,\Delta_\phi)$ and $\beta(\Delta,\Delta_\phi)$ from the decomposition \eqref{Polyblockdeco} of a Polyakov block in Mellin space. \\ \\
Let us start from the inverse Mellin transform \eqref{Mellin-def}. If we sum the residues of $s_{12}=-n$ we get 
\begin{align}
  \sum_n\, \int \frac{ds_{14}}{2\pi i}& \,  \chi^{2 \Delta \phi +2 n} (1-\chi)^{2 s_{14}}  \, \frac{\Gamma \left(s_{14}\right){}^2 \Gamma \left(n+\Delta_\phi -s_{14}\right)^2}{(n!)^2}\big(2 \log (\chi) \, M(-n,s_{14})  \nonumber\\
  & 2M(-n,s_{14})\big(\psi(n+\Delta_\phi+s_{14})-\psi(n+1)\big)-\partial_{s}M(s=-n,s_{14})\big)\,.\label{Mellindemolog}
\end{align}
For simplicity let us focus on the coefficient of $\chi^{2\Delta_\phi}\log (\chi)$ which gives $b_0(\Delta,\Delta_\phi)$ i.e. the coefficient of $\partial_{\Delta}G_{\Delta=2\Delta_\phi}(\chi)$ in the notation from \eqref{Wsdeco}, \eqref{Wcondeco}. The  easiest case is that of the contact diagram because the Mellin amplitude is unity. This gives an $s_{14}$ integral which can be carried out using Barnes' first lemma: 
\begin{equation}
    \chi^{2 \Delta \phi} \log(\chi)\int_{-i\infty}^{i\infty} \frac{ds_{14}}{2\pi i}\,    \, 2\, \Gamma \left(s_{14}\right){}^2 \Gamma \left(\Delta_\phi -s_{14}\right){}^2\, = \, \chi^{2 \Delta \phi} \log(\chi) \,\frac{2\Gamma(\Delta_\phi)^4}{\Gamma(2\Delta_\phi)}\, .\label{contactlog}
\end{equation} 
For the  $s$-channel diagram the $s_{14}$ integral is the same since $M^{(s)}$ is independent of $s_{14}$. So we only have the sum over $m$ in \eqref{Ms} which is straightforward to carry out.
For the $t$-channel it is is harder  since the $M^{(t)}$ has an $s_{14}$ in the denominator and so the $s_{14}$ integral cannot be evaluated using Barnes lemma. 
We show both of these cases in detail in App. \ref{app:Mellin1}.

\section{Five-point Polyakov Bootstrap}
\label{sec:5pt}
In this section we derive the main results of our paper, the 5-point Polyakov bootstrap sum-rules. We begin by reviewing 5-point function kinematics in Section \ref{ssec:setup} and move on to construct the Polyakov blocks in Section  \ref{ssec:Pblocks}. We systematically extract sum-rules from these blocks in Section
 \ref{ssec:sumrules} and describe general properties of the associated functionals, as well as provide some consistency checks in Section \ref{ssec:funcs}.
\subsection{Setup and conventions}
\label{ssec:setup}
To streamline notation, we will be working with 5-point functions of identical operators
\begin{equation}
    \langle \phi(x_1)\phi(x_2)\phi(x_3)\phi(x_4)\phi(x_5)\rangle  =  \mathcal{L}(x_i)\,\mathcal{G}(\chi_1, \chi_2)\,,
\end{equation}
with the generalization to external operators with different dimensions being straightforward. Here we defined the conformal covariant prefactor
\begin{equation}
    \mathcal{L}(x_i) = \left(\frac{x_{24}}{x_{12}^2  x_{23} x_{34}x_{45}^2} \right)^{\Delta}\,,
\end{equation}
as well as the two independent conformal cross-ratios
\begin{equation}
    \chi_1= \frac{x_{12} x_{34}}{x_{13}x_{24}}\,, \qquad   \chi_2= \frac{x_{23} x_{45}}{x_{24}x_{35}}\,,
\end{equation}
where we use the notation $x_{ij}\equiv |x_i-x_j|$. These
5-point conformal correlators can be decomposed into 5-point conformal blocks (note that we sum over \textit{pairs} of operators)
\begin{equation}
\label{OPE5pt}
    \mathcal{G}(\chi_1, \chi_2)= \sum_{(\Delta_1,\Delta_2)} C_{\phi\phi \Delta_1} C_{\Delta_1 \phi \Delta_2} C_{\phi \phi \Delta_2} G_{\Delta_1,\Delta_2}(\chi_1,\chi_2)\equiv \sum_{(\Delta_1,\Delta_2)} a_{\Delta_1,\Delta_2} G_{\Delta_1,\Delta_2}(\chi_1,\chi_2)\,,
\end{equation}
which are known explicitly in terms of a double power series \cite{Rosenhaus:2018zqn}\footnote{See also \cite{Goncalves:2019znr,Fortin:2019zkm,Buric:2020dyz,Buric:2021ywo} for relevant work on multi-point conformal blocks.}
\begin{equation}
    G_{\Delta_1,\Delta_2}(\chi_1,\chi_2)= \chi_1^{\Delta_1} \chi_2^{\Delta_2} \sum_{n_1,n_2=1}^\infty \frac{(\Delta_1)_{n_1}(\Delta_1+\Delta_2-\Delta_\phi)_{n_1+n_2} (\Delta_2)_{n_2}}{(2\Delta_1)_{n_1}(2\Delta_2)_{n_2}} \frac{\chi_1^{n_1}}{n_1!} \frac{\chi_2^{n_2}}{n_2!}\,.
\end{equation}
These correlators satisfy several crossing relations which can be generated by a single cyclic permutation $x_i\to x_{i+1}$ where $x_6\equiv x_1$. In terms of cross-ratios the crossing equation reads
\begin{equation}\label{eq:crossing}
    \mathcal{G}(\chi_1, \chi_2) = \left(\frac{\chi_1^2 \chi_2}{(\chi_1+\chi_2-1)^2}\right)^{\Delta_\phi}  \mathcal{G}\left(\chi_2, 1+\frac{\chi_2}{\chi_1-1}\right) \,.
\end{equation}

\paragraph{Disconnected correlator:} To illustrate the points above let us consider the simplest 5-point correlator, the disconnected correlator, which is a sum of products of two- and three- point functions $\langle\phi\phi\rangle \langle\phi \phi\phi\rangle$. Such a correlator naturally appears for singlet operators in $\mathbb{Z}_2$ preserving perturbative theories or for charged operators in perturbative theories with cubic couplings where the $\mathbb{Z}_2$ symmetry is broken. This correlator contains $\binom{5}{2}=10$ different permutations which contribute to three different sectors in the OPE corresponding to having single- or double-trace operators in the left and right exchanged channels. We can write
\begin{equation}
    \mathcal{G}_{\rm{discon}}(\chi_1,\chi_2) = C_{\phi\phi\phi}\left(\mathcal{G}_{\rm{sing-Id}}(\chi_1,\chi_2)+\mathcal{G}_{\rm{sing-doub}}(\chi_1,\chi_2)+\mathcal{G}_{\rm{doub-doub}}(\chi_1,\chi_2)\right)\,,
\end{equation}
with
\begin{align}
    \mathcal{G}_{\rm{sing-Id}}(\chi_1,\chi_2) &= \chi_1^{\Delta_\phi}+\chi_2^{\Delta_\phi}\,,\\
    \mathcal{G}_{\rm{sing-doub}}(\chi_1,\chi_2)&=\chi_1^{2\Delta_\phi}\chi_2^{\Delta_\phi}\left( \frac{1}{(1-\chi_2)^{\Delta_\phi}} + \frac{1}{\big((1-\chi_1)(1-\chi_1-\chi_2)\big)^{\Delta_\phi}} \right) \,+ (\chi_1\leftrightarrow\chi_2)\,, \\
    \mathcal{G}_{\rm{doub-doub}}(\chi_1,\chi_2) &= \chi_1^{2\Delta_\phi}\chi_2^{2\Delta_\phi}\left(\frac{1}{\big((1-\chi_1)^2(1-\chi_2)\big)^{\Delta_\phi}}+\frac{1}{(1-\chi_1-\chi_2)^{\Delta_\phi}} \right. \nonumber\\ & \left.+\frac{1}{\big((1-\chi_1-\chi_2)^2\big)^{\Delta_\phi}} + \frac{1}{\big((1-\chi_1)(1-\chi_2)^2\big)^{\Delta_\phi}} \right)\,,
\end{align}
where we isolated different power-laws in the cross-ratio to make manifest the scaling dimensions of the exchanged operators. From this we can read off several families of OPE coefficients. It is easy to see directly from the structure of the OPE that
\begin{equation}
    a_{\phi,1}= a_{1,\phi}=C_{\phi\phi\phi}\,, \quad a_{\phi,[\phi\phi]_{2n}}= a_{[\phi\phi]_{2n},\phi} = C_{\phi\phi\phi} (C_{\phi \phi [\phi\phi]_{2n}}^{\rm{GFF}})^2\,.
\end{equation}
Here $C_{\phi \phi [\phi\phi]_{2n}}^{\rm{GFF}}=C_{\Delta_n}$ from \eqref{gffopecoeff}. Furthermore, it is straightforward to expand $\mathcal{G}_{\rm{doub-doub}}$ in blocks and read-off the product of OPE coefficients $a_{[\phi\phi]_{2n},[\phi \phi]_{2m}}$ from which we can infer the non-trivial OPE coefficients $C_{[\phi\phi]_{2n}\phi[\phi\phi]_{2m}}$. The first few are given by
\begin{align}
\label{disc5ope}
   & a_{2\Delta_\phi,2\Delta_\phi}=2\,C_{\phi\phi\phi}\,,\,  a_{2\Delta_\phi+2,2\Delta_\phi}=2 C_{\phi\phi\phi} \frac{\Delta_\phi^2(1+\Delta_\phi)}{1+4\Delta_\phi}\,,\\ &  a_{2\Delta_\phi+2,2\Delta_\phi+2} = C_{\phi\phi\phi}  \frac{\Delta_\phi^2(1+\Delta_\phi)^2(2+\Delta_\phi(12+\Delta_\phi))}{2(1+4\Delta_\phi)^2}\,.
\end{align}

\subsection{Polyakov Blocks}
\label{ssec:Pblocks}

\begin{figure}[h]
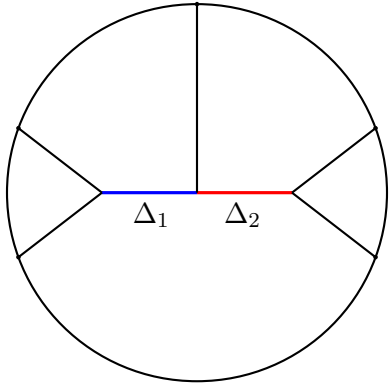

    \centering
    \tikz[baseline=0ex]{\draw[thick] (0,0) circle(2.5);

\foreach \angle in {200, 160, 90, 20, -20} {
    \fill[black] (\angle:2.5) circle(0.03);
}

\coordinate (L) at ($(0,0) + (-1.25,0)$);
\coordinate (R) at ($(0,0) + (1.25,0)$);
\coordinate (C) at ($(L)!0.5!(R)$);

\draw[very thick, blue] (L) -- (C) ; \node[below,xshift=-0.6cm]{$\Delta_1$};
\draw[very thick, red] (C) -- (R); \node[below,xshift=0.6cm] {$\Delta_2$};
;

\draw[thick] (160:2.5) -- (L);
\draw[thick] (200:2.5) -- (L);

\draw[thick] (20:2.5) -- (R);
\draw[thick] (-20:2.5) -- (R);

\draw[thick] (90:2.5) -- (C);
}
\caption{Double-exchange Witten diagram with a $\Delta_1$ exchange in the 12 OPE and a  $\Delta_2$ exchange in the 45 OPE }
\label{fig:doubexch}
\end{figure}

Let us now introduce the Polyakov blocks for 5-point functions. We will do this entirely in Mellin space following our discussions for four points in section  \ref{ssec:mellin}. \\ \\
Consider a correlation function $A(x_{ij})$ of $n$ scalar operators with dimensions and coordinates labelled by $\Delta_i$ and $x_i$ respectively ($i,j=1,\cdots , n$). Its inverse Mellin transform is given by 
\begin{equation}\label{nptMellin-gen}
   A(x_{ij})=\int_{-i\infty}^{i\infty}\, [\text{d}s_{ij}]\prod_{i<j\le n}\Big(|x_{ij}|^{-s_{ij}}\,\Gamma(s_{ij})\,\Big)\, M(s_{ij})\,.
\end{equation}
The Mellin variables $s_{ij}$ are symmetric i.e. $s_{ij}=s_{ji}$. Furthermore they  are constrained as follows  
\begin{equation}\label{Mellinvarconst}
  \sum_{i(\neq j)} s_{ij}=\Delta_j\,.
\end{equation}
The notation $[\text{d}s_{ij}]$ for the differential element indicates that the above condition is being imposed. \\ \\
For $n=4$ and with identical scalars $\Delta_i=\Delta_\phi$ the Mellin integral reduces to \eqref{Mellin-def} where the 4-point conformal prefactor has been factored out and the $x_{ij}$'s have reorganized into 4-point cross ratios. Notice that in that case one has only two independent Mellin variables corresponding to two cross ratios $u,v$ in general $d$.\\ \\
For $n=5$ one has 10  ten variables $s_{ij}$ (with $i<j$) among which five are independent. We may choose the independent ones to be $s_{12},s_{23},s_{34},s_{45},s_{15}$. Any other choice $\{s_{pq}\}$ can be independent if $p,q$ are consecutive pairs in a permutation of $1,2,3,4,5$. 
With our choice, \eqref{Mellinvarconst} decides the form of the dependent variables as:
\begin{align}\label{5ptMellin}
    &s_{13}= \frac{\Delta_\phi }{2}-s_{12}-s_{23}+s_{45}\,, \ \  s_{14}=\frac{\Delta_\phi}{2}  - s_{15}+ s_{23}- s_{45}\,, \ \ 
    s_{24}= \frac{\Delta_\phi}{2}  + s_{15}- s_{23}- s_{34}\,,\\
    &s_{25}= \frac{\Delta_\phi}{2}  - s_{12}- s_{15}+ s_{34}\,, \ \ 
    s_{35}= \frac{\Delta_\phi}{2} + s_{12}- s_{34}- s_{45}\nonumber\,.
\end{align}
The forms of the dependent variables are such that the distances $|x_{ij}|$ rearrange into five cross ratios of the $5$-point function in general $d$ \cite{Buric:2021ttm,Buric:2021kgy}. If we further restrict to 1d by rewriting all position dependence  in terms of the only two independent cross-ratios $\chi_1,\chi_2$ we get
\begin{align}
& \mathcal{G}(\chi_1,\chi_2)=\int [\text{d}s_{ij}]\, \prod_{i<j\le 5}\big(\Gamma(s_{ij})\,\big)\, f(\chi_1,\chi_2|\,s_{ij})\, M(s_{12},s_{23},s_{34},s_{45},s_{15})\nonumber\\
\text{where \ }
&f(\chi_1,\chi_2|\,s_{ij})=\frac{\left(1-\chi_1-\chi_2\right)^{-2 s_{15}}  \chi_1^{2 \Delta _{\phi }-2 s_{12}}  \chi_2^{2 \Delta _{\phi }-2 s_{45}}}{\left(1-\chi_1\right)^{\Delta_{\phi }-2 s_{15}+2 s_{23}-2 s_{45}}\left(1-\chi_2\right)^{\Delta _{\phi }-2 s_{12}-2 s_{15}+2 s_{34}}}\,.
\end{align}
Here we have removed the $5$-point prefactor, and we show $M(s_{ij})$ as a function of only the independent variables. \\ \\ 
Notice that the 12-3-45 OPE channel, which corresponds to the expansion in small $\chi_1$ and $\chi_2$, is controlled by the poles of $s_{12}$ and $s_{45}$ respectively. This is analogous to the $s$-channel expansion being controlled by $s_{12}$ poles in the 4-point case. A general channel $ij$-$k$-$lm$ corresponds to poles in the pair of variables $(s_{ij}, s_{lm})$. A pair with a repeated index (e.g. $(s_{12}, s_{23})$) cannot correspond to an OPE channel in this sense.     \\ \\
We will bootstrap $\mathcal{G}(\chi_1,\chi_2)$ with its  Mellin transform $M(s_{ij})$. This means we will impose the following conditions: \\ \\
\underline{Crossing symmetry:} This corresponds to $M(s_{ij})$ remaining invariant under any permutation of $1,2,3,4,5$: the external operator labels. This can be  stated explicitly as
\begin{equation}
    M(s_{12},s_{23},s_{34},s_{45},s_{15}) \stackrel{2\leftrightarrow 3}{=\joinrel=}  M(s_{13},s_{23},s_{24},s_{45},s_{15})\, = \,\cdots  \text{(all permutations)}\,.
\end{equation}
The first equality indicates that under $2\leftrightarrow 3$ we should replace the functional dependence of $M$ on $s_{12}$ and $s_{34}$ by $s_{13}= \frac{\Delta_\phi }{2}-s_{12}-s_{23}+s_{45}$ and $s_{24}= \frac{\Delta_\phi}{2}  + s_{15}- s_{23}- s_{34}$ respectively. \\ \\
\underline{Regge boundedness:} Just like with four points, we restrict our attention to Mellin amplitudes which are polynomially bounded at infinity.\footnote{See also \cite{Costa:2023wfz} for a detailed discussion on the Regge limit of 5-point functions in general $d$.}  Specifically we require:
\begin{equation}\label{5ptRegge}
    M(s_{12},s_{23},s_{34},s_{45},s_{15}) \lesssim s_{12}\, s_{45}\,, \ \  \text{as } s_{12}, s_{45}\gg 1\,.
\end{equation}
The limits on $s_{12}, s_{45}$ are taken in no specific order and $s_{23},s_{34},s_{15}$ are kept fixed. The pair $(s_{12},s_{45})$  corresponds to the 12-3-45 channel. By crossing symmetry the same bound also exists for other channels i.e. large values of two variables $(s_{ij},s_{lm})$ where the indices are not repeated. \\ \\
\underline{Operator product expansion:} The OPE guarantees that the correlator $\mathcal{G}(\chi_1,\chi_2)$ obeys the conformal block expansion \eqref{OPE5pt}. In Mellin space we implement the OPE by a Polyakov block expansion 
\begin{equation}
    M(s_{12},s_{23},s_{34},s_{45},s_{15}) =\sum_{\Delta_1,\Delta_2}a_{\Delta_1,\Delta_2}\, P_{\Delta_1,\Delta_2}(s_{12},s_{23},s_{34},s_{45},s_{15}) \,.
\end{equation}
The resulting $M(s_{ij})$ should be crossing symmetric, Regge bounded and compatible with the conformal block expansion in position space where we sum over $\{\Delta_1,\Delta_2\}$ . As we show below this is the bootstrap condition on $\{\Delta_1,\Delta_2\}$ and $a_{\Delta_1,\Delta_2}$.
\\ \\
We will now lay down the rules for a 5-point Polyakov block, and  we will elaborate on them further below.
\begin{itemize}
    \item $P_{\Delta_1,\Delta_2}(s_{ij})$ is crossing symmetric i.e. invariant under all permutations of the external operators. 
    \item $P_{\Delta_1,\Delta_2}(s_{ij})$ has joint simple poles in two variables $(s_{12}, s_{45})$ and their crossing symmetric counter-parts. The joint poles should produce only $G_{\Delta_1,\Delta_2}(\chi_1,\chi_2)$, but there can be spurious blocks from the Mellin measure. 
    \item The `regular terms' in $P_{\Delta_1,\Delta_2}(s_{ij})$ correspond to less singular terms i.e. isolated simple poles and polynomials. They should not add any new spurious blocks.
    \item The regular terms should respect the Regge boundedness of $\mathcal{G}(\chi_1,\chi_2)$. This amounts to satisfying \eqref{5ptRegge}.
    \item The number $n$ of independent regular terms has to fixed by eliminating  $n$ sum-rules. 
\end{itemize}
These are simple extensions of the four points conditions from Section \ref{ssec:mellin}. The first bullet point simply reflects the crossing property of $M(s_{ij})$. \\ \\ 
In the second point we demand that the Polyakov block has the simplest set of poles whose residues give a conformal block.
A (non-crossing symmetric) Mellin amplitude that has this property is as follows:
\begin{equation}\label{doublexc-Mellin}
M^{\text{12-3-45}}_{\Delta_1,\Delta_2}(s_{ij})=\sum_{n_1,n_2=0}^{\infty}\frac{\mathcal{Q}_{n_1,n_2}(\Delta_1,\Delta_2|\Delta_\phi)}{(s_{12}-(\Delta_\phi-\frac{\Delta_1}{2}-n_1))(s_{45}-(\Delta_\phi-\frac{\Delta_2}{2}-n_2))}\,.
\end{equation}
where
\begin{align}
\mathcal{Q}_{n_1,n_2}&(\Delta_1,\Delta_2|\Delta_\phi)=\nonumber\\
&\mathcal{N}_{\Delta_1,\Delta_2}\Gamma \left(\mbox{$\frac{\Delta _1+\Delta _2+\Delta _{\phi }-1}{2} $}\right)\left[\frac{\Gamma \big(\frac{\Delta _1-1 +2\Delta _{\phi }}{2}\big) \big(\frac{\Delta _1-2\Delta _{\phi }+2}{2}\big)_{n_1} \big(\frac{\Delta _1-\Delta _2-\Delta _{\phi }+2}{2} \big)_{n_1} }{2 \Gamma \left(n_1+1\right)  \Gamma \left(\frac{2n_1+2\Delta _1+1}{2}\right)}\times (\Delta_1\leftrightarrow\Delta_2)\right]\nonumber\\
&\times \, _3F_2\left(\mbox{$-n_1,-n_2,\frac{\Delta_1+\Delta _2+\Delta _{\phi }-1}{2};\frac{\Delta_2-\Delta_1+\Delta _{\phi }-2n_1}{2},\frac{\Delta_1-\Delta_2+\Delta _{\phi }-2n_2}{2};1$}\right)\,.
\end{align}
The amplitude \eqref{doublexc-Mellin} is nothing but the Mellin transform of a 5-point double-exchange Witten diagram (see e.g. \cite{Paulos:2011ie})\footnote{See also \cite{Albayrak:2018tam,Jepsen:2019svc,Ma:2022ihn,Li:2023azu} for other studies of higher-point Witten diagrams.} of Figure \ref{fig:doubexch}. 

\begin{figure}[h]
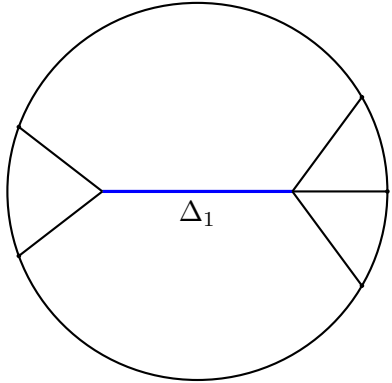

    \centering
    
\tikz[baseline=0ex]{\draw[thick] (0,0) circle(2.5);

\foreach \angle in {200, 160, 30, 0, -30} {
    \fill[black] (\angle:2.5) circle(0.03);
}

\coordinate (L) at ($(0,0) + (-1.25,0)$);
\coordinate (R) at ($(0,0) + (1.25,0)$);

\draw[very thick, blue] (L) -- (R);\node[below] {$\Delta_1$};

\draw[thick] (160:2.5) -- (L);
\draw[thick] (200:2.5) -- (L);

\draw[thick] (30:2.5) -- (R);
\draw[thick] (-30:2.5) -- (R);
\draw[thick] (0:2.5) -- (R);
}
\caption{Single-exchange diagram with a $\Delta_1$ exchange in the 12 OPE channel.}
\label{fig:singexch}
\end{figure}

\begin{figure}
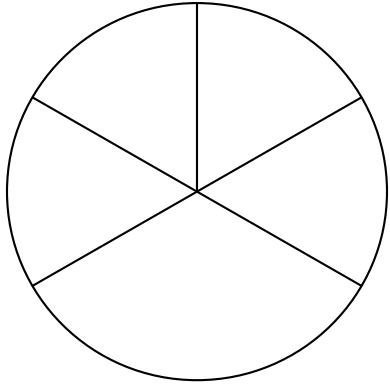

    \centering
\tikz[baseline=0ex]{
  \draw[thick] (0,0) circle (2.5);
  \coordinate (C) at ($(0,0)$);
  \foreach \angle in {30,90,150,-30,210} {
    \fill[black] (\angle:1.);
    circle(0.03); }
  \draw[thick] (30:2.5) -- (C);
  \draw[thick] (90:2.5) -- (C);
  \draw[thick] (150:2.5) -- (C);
  \draw[thick] (-30:2.5) -- (C);
  \draw[thick] (210:2.5) -- (C);
}
      \caption{Five-point contact Witten diagram}
      \label{fig:ctc}
\end{figure}

The residues of the $(s_{12}, s_{45})$ poles correspond to the powers $\chi^{\Delta_1+2n_1}\chi_2^{\Delta_2+2n_2}$\,. The form of $\mathcal{Q}_{n_1,n_2}$ ensures that carrying out the integrals over $s_{23},s_{34},s_{15}$  we recover the exact form of $G_{\Delta_1,\Delta_2}(\chi_1,\chi_2)$. The normalization $\mathcal{N}_{\Delta_1,\Delta_2}$ is chosen such that the block comes with unit coefficient. The superscript 12-3-45 stands for the channel exchanges in the bulk.\\ \\ 
However, there are also spurious blocks, which involve unphysical poles i.e. those from the $\Gamma(s_{ij})$ factors of \eqref{5ptMellin}. The spurious poles can be either in both of the variables $(s_{12},s_{45})$  or in just one of them. In the latter case the other variable is at a `physical' pole of \eqref{doublexc-Mellin}, and one obtains  blocks of the form $G_{\Delta_1,\Delta_{2\phi,n}}(\chi_1,\chi_2)$ or $G_{\Delta_{2\phi,n},\Delta_2}(\chi_1,\chi_2)$, with $\Delta_{2\phi,n}$ being the following shorthand notation for the spurious dimension 
\begin{equation}
    \Delta_{2\phi,n}=2\Delta_\phi+n\,.
\end{equation}
(In this section we will use the above  notation for double twist operator dimensions instead of $\Delta_n$ in order to not confuse with `physical dimensions' $\Delta_1, \Delta_2$.)
Note that these particular spurious blocks may 
be present also for some of the crossed versions of $M_{\Delta_1,\Delta_2}(s_{ij})$ but not all, e.g.  we will get $G_{\Delta_1,\Delta_{2\phi,n}}$ by replacing $(s_{12},s_{45})\leftrightarrow (s_{12},s_{34})$ in \eqref{doublexc-Mellin}  but not with $(s_{12},s_{45})\leftrightarrow (s_{13},s_{25})$. We discuss all the spurious blocks more elaborately in the next subsection.\\ \\
Consider now the `regular' terms in the third bullet point. These could be simple poles in a single Mellin variable or just a polynomial, but we do not want them to add more spurious blocks than the ones from $M^{\text{12-3-45}}_{\Delta_1,\Delta_2}(s_{ij})$. Such an amplitude having simple poles only in $s_{12}$  is given as follows:
\begin{equation}\label{singlexc-Mellin}
M^{\text{12-345}}_{\Delta}(s_{ij})=\sum_{n=0}^{\infty}\frac{f_n(\Delta|\Delta_\phi)}{(s_{12}-(\Delta_\phi-\frac{\Delta}{2}-n))}
\end{equation}
where 
\begin{equation}
    f_n(\Delta|\Delta_\phi)=\frac{ \mathcal{N}_{\Delta}\big(\frac{\Delta -3 \Delta _{\phi }+4}{2} \big)_{n-1} \big(\frac{\Delta -2\Delta_{\phi }+4}{2}\big)_{n-1}}{n!  \big( \Delta+\frac{3}{2}\big)_{n-1}}\,.
\end{equation}
The expression in \eqref{singlexc-Mellin} is actually the Mellin transform of a 5-point Witten diagram that exchanges a single operator in the bulk corresponding to CFT dimension $\Delta$ (see e.g. \cite{Penedones:2010ue}) as represented in Figure \ref{fig:singexch}. 
The residue function $f_{n}(\Delta|\Delta_\phi)$ is such that it leads only to the spurious blocks $G_{\Delta,\Delta_{2\phi,m}}$ and not $G_{\Delta+1,\Delta_{2\phi,m}}$ or $G_{\Delta+2,\Delta_{2\phi,m}}$, etc, and the normalization $\mathcal{N}_{\Delta}$ is chosen such that $G_{\Delta_{2\phi,0}}$ has unit coefficient. In fact we can fix $f_n$ by demanding that there is no spurious block $G_{\Delta+p,\Delta_{2\phi,0}}$ except for $p=0$.\\ \\
It is easy to check that $M_{\Delta}(s_{ij})$ is Regge bounded in the sense of \eqref{5ptRegge}.
There are also crossed versions of $M_{\Delta}(s_{ij})$ where we replace $s_{12}$ with other variables but none of them can give rise to $G_{\Delta,\Delta_{2\phi,m}}$ spurious blocks. On the other hand, the replacement $s_{12}\leftrightarrow s_{45}$ leads to spurious $G_{\Delta_{2\phi,m},\Delta}$ blocks. \\ \\
The other allowed regular terms are polynomials, but the only one that is crossing symmetric and respects the Regge condition \eqref{5ptRegge} is a constant (note that the crossing symmetric linear polynomial $\sum s_{ij}$ is a constant). We choose it to be unity:
\begin{equation}
    M^{(\text{con})}(s_{ij})=1\,.\label{5ptcontactMellin}
\end{equation}
This is the Mellin transform of a contact Witten diagram with a 5-point vertex without derivatives (see \cite{Penedones:2010ue}) as given in Figure \ref{fig:ctc}.

All the Witten diagrams $M_{\Delta_1,\Delta_2}, M_{\Delta}$ and $ M^{(\text{con})}$ can be derived from an AdS$_2$ scalar field theory with the following action:
\begin{align}
&S_{\text{AdS}_2}  \, = \, S_{\text{free}}[\Phi,\Delta_\phi]+S_{\text{free}}[\mathcal{O}_1,\Delta_1]+S_{\text{free}}[\mathcal{O}_2,\Delta_2] \nonumber\\
& \hspace{1cm} + \int d^2x \sqrt{g}\Big[g_1\Phi^2\,\mathcal{O}_1+g_2\Phi^2\,\mathcal{O}_2+g_3\,\Phi\,\mathcal{O}_1\,\mathcal{O}_2+g_4\Phi^3\,\mathcal{O}_1+g_5\Phi^5\Big],
\end{align}
where $S_{\text{free}}[\mathcal{O},\Delta]$ denotes the free scalar action of an operator $\mathcal{O}$ with mass $m^2_{\Delta}=\Delta(\Delta-1)$\,. The double exchange diagram is composed with the vertices $g_1,g_2,g_3$, the single exchange with $g_1, g_4$  (with the relabeling $\Delta_1\to \Delta$) and the contact diagram with $g_5$. \\ \\
Having introduced all the ingredients let us write down the full structure of the 5-point Polyakov block (ignoring $\Delta_\phi$ dependencies in the constants):
\begin{mdframed}
\begin{align}\label{5ptpolyakovblock}
  P_{\Delta_1,\Delta_2}(s_{ij})&=\sum_{n_1,n_2=0}^{\infty}\frac{\mathcal{Q}_{n_1,n_2}(\Delta_1,\Delta_2)+c_1(\Delta_1,\Delta_2)\big(f_n(\Delta_1)s_{45}+f_n(\Delta_2)s_{12}\big)+c_2(\Delta_1,\Delta_2)}{(s_{12}-(\Delta_\phi-\frac{\Delta_1}{2}-n_1))(s_{45}-(\Delta_\phi-\frac{\Delta_2}{2}-n_2))}\nonumber\\
    & \, + \, \text{all permutations} \, + \, (\Delta_1\leftrightarrow\Delta_2)\,.
\end{align}
\end{mdframed}
There are 30 terms in total, one for each pair $(s_{ij},s_{lm})$ with all indices different. Note that we consider $(s_{ij},s_{lm})$ to be different from $(s_{lm},s_{ij})$. \\ \\
The whole expression has only two independent undetermined constants $c_1(\Delta_1,\Delta_2)$ and $c_1(\Delta_1,\Delta_2)$. As indicated in the final bullet point rule for writing this Polyakov block, we will fix $c_1$ and $c_2$ by demanding that the overall coefficients of any two spurious blocks are zero. We will specify our choice for these two spurious blocks below.

\subsection{Sum-rules}
\label{ssec:sumrules}

Let us write down all the spurious blocks that arise from all possible 5-point diagrams. Since we introduced the double-exchange, single-exchange and contact diagrams in Mellin space, let us denote their inverse transforms as $W_{\Delta_1,\Delta_2}(\chi_a)\,, W_{\Delta}(\chi_a)$ and $W^{\text{con}}(\chi_a)$ respectively. Let us also introduce the following shorthand notations for the different spurious exchanges:
\begin{align}
    &\Delta_{3\phi,n}=3\Delta_\phi+n\,,\nonumber\\
    &\Delta_{i,\phi,n}=\Delta_i+\Delta_\phi+n\,, \ i=1,2\,.
\end{align}
With this, the block decomposition of double-exchange diagrams can be written as (denoting $\chi_a:=\chi_1,\chi_2$\,)
\begin{align}
W^{(i)}_{\Delta_1,\Delta_2}(\chi_a)=\,  \delta_{i,\text{dir}} \,  G_{\Delta_1,\Delta_2}&(\chi_a)\nonumber\\ 
-\sum_{n_1,n_2}\Big[a^{2\phi,2\phi,(i)}_{n_1,n_2}\, G_{\Delta_{2\phi,n_1},\Delta_{2\phi,n_2}}&(\chi_a)+ a^{2\phi,3\phi,(i)}_{n_1,n_2}\,G_{\Delta_{2\phi,n_1},\Delta_{3\phi,n_2}}(\chi_a)+a^{3\phi,2\phi,(i)}_{n_1,n_2}\,G_{\Delta_{3\phi,n_1},\Delta_{2\phi,n_2}}(\chi_a)\Big]\nonumber\\
-\sum_{n}\Big[a^{1,2\phi,(i)}_{n}\, G_{\Delta_1,\Delta_{2\phi,n}}&(\chi_a)+a^{1,(1,\phi),(i)}_{n}\,G_{\Delta_1,\Delta_{1,\phi,n}}(\chi_a)\nonumber\\
+a^{2\phi,2,(i)}_{n}\,  G_{\Delta_{2\phi,n},\Delta_2}&(\chi_a)+a^{(2,\phi),2,(i)}_{n}\,G_{\Delta_{2,\phi,n},\Delta_2}(\chi_a)\Big]\,.\label{doubleWitten}
\end{align}
Here the superscript $(i)$ refers to the channel of the exchange and `dir' denotes the direct channel i.e. 12-3-45. Also for the direct channel all $a^{2\phi,2\phi}_{n_1,n_2}, a^{2\phi,3\phi}_{n_1,n_2}, a^{3\phi,2\phi}_{n_1,n_2}, a^{1,2\phi}_{n}, a^{2\phi,2}_{n}, a^{1,(1,\phi)}_{n}$ or  $a^{(2,\phi),2}_{n}$ are zero for $n_1,n_2$ or $n$ odd. The same spurious blocks are also absent when we add all the diagrams. One way to understand this is that, as we will see in the next subsection, the Polyakov block trivializes the disconnected correlator and deformations around it. So the disconnected  correlator plays the role of GFF for the 5-point case, and it does not contain the blocks that correspond to the odd $n$'s, as these do not occur in the OPE of identical scalars. \\ \\
The coefficients  $a^{1,2\phi,(i)}_{n}$ and $a^{1,(1,\phi),(i)}_{n}$ are nonzero only for the channels $i=(12$-$3$-$45)$,\, $(12$-$4$-$35)$,\, $(12$-$5$-$34)$, while $a^{2\phi,(i),2}_{n}$ and $a^{(2,\phi),2,(i)}_{n}$ are nonzero only in the $i=(12$-$3$-$45)$,\, $(13$-$2$-$45)$,\, $(23$-$1$-$45)$ channels.\\ \\
The direct channel block decomposition of single-exchange diagrams is as follows:
\begin{align}
W^{(i)}_{\Delta_1}(\chi_a)=-\sum_{n_1,n_2}&\Big[b^{2\phi,2\phi,(i)}_{n_1,n_2}\,G_{\Delta_{2\phi,n_1},\Delta_{2\phi,n_2}}(\chi_a)+\big(\, b^{2\phi,3\phi,(i)}_{n_1,n_2}\,G_{\Delta_{2\phi,n_1},\Delta_{3\phi,n_2}}(\chi_a)+(2\phi\leftrightarrow3\phi)\,\big)\Big]\nonumber\\
-\sum_{n}\Big[\, \delta_{i,(\text{12-345})}&\big(\, b^{1,2\phi}_{n}\,G_{\Delta_1,\Delta_{2\phi,n}}(\chi_a)+b^{1,(1,\phi)}_{n}\,G_{\Delta_1,\Delta_{1,\phi,n}}(\chi_a)\,\big)\nonumber\\
\delta_{i,(\text{45-123})}&\big(\,b^{2\phi,1}_{n}\,G_{\Delta_{2\phi,n},\Delta_1}(\chi_a)+b^{(1,\phi),1}_{n}\,G_{\Delta_1,\Delta_{1,\phi,n}}(\chi_a)\,\big)\,\Big]\,.\label{singleWitten}
\end{align}
In the above expression $(i)$ refers to all the channels $ij$-$klm$ of a single operator exchange in the bulk. There are ten such channels. Notice that for $i=(12$-$345)$  one can have the spurious blocks involving a $\Delta_1$ in the $\phi(x_1)\phi(x_2)$ OPE whereas  $i=(45$-$123)$ involves spurious blocks with a $\Delta_1$ in the $\phi(x_4)\phi(x_5)$ OPE. \\ \\
The contact diagram decomposes into three types of spurious blocks:
\begin{align}
W^{\text{con}}=-\sum_{n_1,n_2}&\Big[c^{2\phi,2\phi,\text{con}}_{n_1,n_2}\,G_{\Delta_{2\phi,n_1},\Delta_{2\phi,n_2}}(\chi_a)+\big(\, c^{2\phi,3\phi,\text{con}}_{n_1,n_2}\,G_{\Delta_{2\phi,n_1},\Delta_{3\phi,n_2}}(\chi_a)+(2\phi\leftrightarrow3\phi)\,\big)\Big]\,.\label{contact5ptWitten}
\end{align}
These three types of diagrams are all the ingredients one has in the Polyakov block, which we defined in Mellin space in \eqref{5ptpolyakovblock}. Let us denote its inverse Mellin transform by $P_{\Delta_1,\Delta_2}(\chi_1,\chi_2)$. Its block decomposition is given by 
\begin{align}
    P_{\Delta_1,\Delta_2}(\chi_a)&=G_{\Delta_1,\Delta_2}(\chi_a)+G_{\Delta_2,\Delta_1}(\chi_a)\nonumber\\ 
-\sum_{n_1,n_2}&\Big[\,\alpha^{2\phi,2\phi}_{n_1,n_2}\, G_{\Delta_{2\phi,2n_1},\Delta_{2\phi,2n_2}}(\chi_a)+ \big(\alpha^{2\phi,3\phi}_{n_1,n_2}\,G_{\Delta_{2\phi,2n_1},\Delta_{3\phi,2n_2}}(\chi_a)+(2\phi\leftrightarrow3\phi)\,\big)\Big]\nonumber\\
    -\sum_{n}&\Big[\,\alpha^{1,2\phi}_{n}\, G_{\Delta_1,\Delta_{2\phi,2n}}(\chi_a)+\alpha^{1,(1,\phi)}_{n}\,G_{\Delta_1,\Delta_{1,\phi,n}}(\chi_a) \nonumber\\
&+\alpha^{2\phi,1}_{n}\,  G_{\Delta_{2\phi,2n},\Delta_1}(\chi_a)+\alpha^{(1,\phi),1}_{n}\,G_{\Delta_{1,\phi,n},\Delta_1}(\chi_a)\, + \, (\Delta_1\leftrightarrow \Delta_2)\Big]\,.\label{5ptPolyblockdeco}
\end{align}
We will refer to the $\alpha$ coefficients as functionals since they will describe the sum-rules below. Let us write them in terms of the individual Witten diagram coefficients defined before.
For convenience we define the sets  $I'=\{\text{12-3-45, 12-4-35, 12-5-34}\}$, $I=\{\text{all double exchange channels}\}$, and $J=\{\text{all single exchange channels}\}$. Then we have
\begin{align}
    \alpha^{1,2\phi}_{n}& =\sum_{i\in I'}a_{2n}^{1,2\phi,(i)}\, + \, c_1\,b_{2n}^{1,2\phi,\text{(12-345)}}\,.\\
    \alpha^{1,(1,\phi)}_{n}& =\sum_{i\in I'}a_n^{1,(1,\phi),(i)}\, + \, c_1\,b_n^{1,(1,\phi),\text{(12-345)}}\,.\\
    \alpha^{2\phi,2\phi}_{n}& =\sum_{i\in I}a_{2n}^{2\phi,2\phi,(i)}\, + \, c_1\,\sum_{j\in J}b_{2n}^{2\phi,2\phi,(j)}\, + \, c_2\,  c_{2n}^{2\phi,2\phi}\,.\\
    \alpha^{2\phi,3\phi}_{n}& =\sum_{i\in I}a_{2n}^{2\phi,3\phi,(i)}\, + \, c_1\,\sum_{j\in J}b_{2n}^{2\phi,3\phi,(j)}\, + \, c_2\,  c_{2n}^{2\phi,3\phi}\,.
\end{align}
The other coefficients like $a^{2,2\phi}, a^{1,(1,\phi)}_n$, etc that are related by swapping $x_1,x_2\leftrightarrow x_4,x_5$ or $\Delta_1\leftrightarrow\Delta_2$ are defined analogously. Recall that the coefficients $c_1$ and $c_2$ were introduced as undetermined constants coming with the `regular terms' in the definition of the Polyakov block \eqref{5ptpolyakovblock}. We specify how to fix them at the end of this subsection. \\ \\
The 5-point Polyakov bootstrap equation is given by
\begin{equation}
\boxed{\sum_{\Delta_1,\Delta_2}a_{\Delta_1,\Delta_2}G_{\Delta_1,\Delta_2}(\chi_1,\chi_2)=\sum_{\substack{ \Delta_1,\Delta_2\\ \text{ pairs}}}a_{\Delta_1,\Delta_2}P_{\Delta_1,\Delta_2}(\chi_1,\chi_2)}
\end{equation}
This implies that every family of spurious blocks must separately be set to zero, leading to
multiple families of sum-rules. Let us discuss them one by one. 

\paragraph{The $(1,2\phi)$ sum-rule:} This comes from eliminating $G_{\Delta_1,\Delta_{2\phi,2n}}$ blocks. It is given by 
\begin{equation}
    \boxed{\ \sum_{\Delta_2}a_{\Delta_1,\Delta_2}\alpha^{1,2\phi}_{n}(\Delta_1,\Delta_2)=0\,.\, }\label{sumrule1}
\end{equation}
The ($1, 2\phi$) sum-rules  fix an operator $\mathcal{O}_{\Delta_1}(x)$ in the $\phi(x_1) \phi(x_2)$ OPE  and sum over all the operators $\mathcal{O}_{\Delta_2}(y)$ in the $\phi(x_4)\phi(x_5)$ OPE. It is clear that the sum-rules coming from the vanishing of $G_{\Delta_{2\phi,2n},\Delta_1}$, $G_{\Delta_2, \Delta_{2\phi,2n}}$ and $G_{\Delta_{2\phi,2n},\Delta_2}$ are all equivalent to \eqref{sumrule1} since they merely correspond to exchange of external coordinates $(x_1,x_2)\leftrightarrow (x_4,x_5)$ and/or dimensions $\Delta_1\leftrightarrow\Delta_2$. \\ \\
The functional $\alpha_n^{1,2\phi}$ involves coefficients from three double-exchange diagrams and one single exchange diagram. We can think of the `$1, 2\phi$' conditions as a 4-point Polyakov bootstrap sum-rule, for the correlator 
\begin{equation}
    \langle\mathcal{O}_{\Delta_1}(x)\phi(x_3)\phi(x_4)\phi(x_5)\rangle=\sum_{\Delta_2}\lambda_{\phi 12}\lambda_{\phi\phi 2}\,G^{\Delta_1\Delta_\phi\Delta_\phi\Delta_\phi}_{\Delta_2}(x,x_3,x_4,x_5)\,.
\end{equation}
This has three external operator which are identical ($\phi$) and one which is different ($ \mathcal{O}_{\Delta_1}$). Here $\lambda_{\phi 12}$, $\lambda_{\phi\phi 2}$ denote the OPE coefficients in the $\phi\mathcal{O}_{\Delta_1}$ and $\phi\phi$ OPE's respectively. Notice that this is crossing symmetric in all three channels. Therefore, its Polyakov blocks will be defined similarly to \eqref{Polydef} i.e. three exchange and a contact Witten diagram, with the difference that the diagrams will have one non-identical external leg of dimension $\Delta_1$. Unlike \eqref{Polyblockdeco}, the corresponding spurious blocks in this case will be  $G_{\Delta_{2\phi,n}}$ and $G_{\Delta_{1,\phi,n}}$. The resulting sum-rules have been explored in \cite{Ghosh:2023wjn}. Our equation \eqref{sumrule1} is the same as the one there, specifically the sum-rule for eliminating $G_{\Delta_{2\phi,n}}$.
\paragraph{The $(1,(1,\phi))$ sum-rule:} This follows from eliminating the $G_{\Delta_1,\Delta_{1,\phi,n}}$ blocks. It is given by 
\begin{equation}\label{sumrule2}
    \boxed{\ \sum_{\Delta_2}a_{\Delta_1,\Delta_2}\, \alpha^{1,(1,\phi)}_{n}(\Delta_1,\Delta_2)=0\,.\, }
\end{equation}
Just like the previous sum-rule, here we fix one dimension $\Delta_1$ and sum over the other dimension $\Delta_2$ in the 5-point OPE. As before, the functional $\alpha^{1,(1,\phi)}_{n}$ involves three double exchange and a single-exchange diagram. So it is equivalent to the Polyakov bootstrap sum-rule for $\langle\mathcal{O}_{\Delta_1}\phi\phi\phi \rangle$, and in this case corresponds to the vanishing of $G_{\Delta_{1,\phi,n}}$ blocks.  Once again the condition \eqref{sumrule2} automatically eliminates  $G_{\Delta_{1,\phi,n},\Delta_1}$, $G_{\Delta_2,\Delta_{2,\phi,n}}$ and  $G_{\Delta_{2,\phi,n},\Delta_2}$ by appropriate swapping of coordinates and dimensions.

\paragraph{The $(2\phi,2\phi)$ sum-rule:}  This corresponds to the vanishing of $G_{\Delta_{2\phi,2n_1},\Delta_{2\phi,2n_2}}$ blocks, and given by
\begin{equation}\label{sumrule3}
    \boxed{\ \sum_{(\Delta_1,\Delta_2)}a_{\Delta_1,\Delta_2}\, \alpha^{2\phi,2\phi}_{n_1,n_2}(\Delta_1,\Delta_2)=0\,.\, }
\end{equation}
The $(\Delta_1,\Delta_2)$ indicates the sum should be over all such distinct pairs. The functional $\alpha^{2\phi,2\phi}_{n_1,n_2}$ involves all the channels of  double- and single-exchange diagrams as well as the contact diagram. \\ \\
This is a very nontrivial sum-rule since we have to sum over both $\Delta_1$ and $\Delta_2$ in the 5-point OPE. This cannot be derived from any individual 4-point function. Rather it can be thought of as a bootstrap condition on an infinite number of correlators $\langle \mathcal{O}\phi\phi\phi\rangle$ or an infinite number of OPEs $\mathcal{O}\times \phi$ for all $\mathcal{O}\in \phi\times \phi$.
\paragraph{The $(2\phi,3\phi)$ sum-rule:} This sum-rule comes from eliminating $G_{\Delta_{2\phi,2n_1},\Delta_{3\phi,2n_2}}$ blocks. It is given by
\begin{equation}\label{sumrule4}
    \boxed{\ \sum_{(\Delta_1,\Delta_2)}a_{\Delta_1,\Delta_2}\, \alpha^{2\phi,3\phi}_{n_1,n_2}(\Delta_1,\Delta_2)=0\,.\, }
\end{equation}
Like the previous sum-rule the sum is over all pairs $(\Delta_1,\Delta_2)$ and the functional $\alpha^{2\phi,3\phi}_{n_1,n_2}$ involves all the 5-point Witten diagrams. \\ \\
Once again since it involves sum over both $\Delta_1$ and $\Delta_2$ the $(2\phi,3\phi)$ sum-rule cannot be obtained from a single 4-point function, and is a collective condition on an infinite number of OPEs $\mathcal{O}\times \phi$. Also by the exchange of $x_1,x_2\leftrightarrow x_4,x_5$ we can see that  vanishing of the blocks $G_{\Delta_{3\phi,n_1},\Delta_{2\phi,n_2}}$ imposes a sum-rule equivalent to \eqref{sumrule4}.

\subsubsection*{Fixing $c_1$ and $c_2\,$:}

Finally let us discuss how to fix the constants multiplying the `regular terms' in the Polyakov block. We do this by eliminating two of the sum-rules described above, i.e. by setting to zero two functionals which depend on $c_1$ and $c_2$. One possible choice is:
\begin{align}
    &\alpha^{1,2\phi}_0(\Delta_1,\Delta_2)=0\,,\label{fixcond1}\\
    &\alpha^{2\phi,3\phi}_0(\Delta_1,\Delta_2)=0\,.\label{fixcond2}
\end{align}
This is a simple choice since the functional $\alpha^{1,2\phi}_0$ does not depend on $c_2$, so \eqref{fixcond1} directly fixes $c_1$. Note that the $\alpha^{1,2\phi}_n$ family is easy to evaluate since it depends on only four Witten diagrams, and $n=0$ is the simplest one. Since $\alpha^{2\phi,3\phi}_0$ depends on both $c_1,c_2$ we can use \eqref{fixcond2} to fix $c_2$. \\ \\
There are, of course, other choices with functionals that involve both $c_1$ and $c_2$. E.g. having $\alpha^{1,(1,\phi)}_0=0$ and $\alpha^{2\phi,2\phi}_0=0$ would work and is as simple as \eqref{fixcond1} and \eqref{fixcond2}. However, setting $\alpha^{2\phi,2\phi}_0=0$ and $\alpha^{2\phi,3\phi}_0=0$, or $\alpha^{2\phi,3\phi}_0=0$ and $\alpha^{2\phi,3\phi}_2=0$  would be more cumbersome to implement since these functionals individually involve all the Witten diagrams.

\subsection{Structure of functionals}
\label{ssec:funcs}

We will now discuss the orthogonality properties of the 5-point Polyakov bootstrap functionals. 
These are analogous to the ones we showed in \eqref{4ptortho} for four points.
The simple functionals $\alpha^{1,2\phi}$ and $\alpha^{1,(1,\phi)}$ have the following properties
\begin{align}
    &\alpha_n^{1,2\phi}(\Delta_1,\Delta_2=\Delta_{2\phi,m})=\delta_{mn}-d_n^{(1)}\delta_{m0}\,, \ \ \  \alpha_n^{1,2\phi}(\Delta_1,\Delta_2=\Delta_{1,\phi,m})=0\nonumber\\
     &\alpha_n^{1,(1,\phi)}(\Delta_1,\Delta_2=\Delta_{2\phi,m})=-d_n^{(2)}\delta_{m0}\,, \ \ \ \ \ \ \ \ \ \alpha_n^{1,(1,\phi)}(\Delta_1,\Delta_2=\Delta_{1,\phi,m})=\delta_{mn}\label{simpleorthogonal}\,.
\end{align}
Since these functionals are equivalent to those from the correlator $\langle\mathcal{O}_{\Delta_1}\phi\phi\phi\rangle$, \eqref{simpleorthogonal} are the same orthogonality properties found from the 4-point Polyakov block \cite{Ghosh:2023wjn} for that correlator.  Note that $d_n^{(1)}$ and $d_n^{(2)}$ are related to the constant $c_1$ in the Polyakov block and the respective spurious block coefficient of the single-exchange diagram.  E.g. $d^{(1)}_n=\, b_n^{1,2\phi}$ and $d^{(2)}_n=\, b_n^{1,(1,\phi)}$ in our choice of normalization and fixing of $c_1$.
\\ \\
%
The orthogonality properties of $\alpha^{2\phi,2\phi}$ and $\alpha^{2\phi,3\phi}$ are more interesting. They are as follows 
\begin{align}
    &\alpha_{n_1,n_2}^{2\phi,2\phi}(\Delta_{2\phi,m_1},\Delta_{2\phi,m_2})=\delta_{n_1m_1}\delta_{n_2m_2}\,,     \ \  \alpha_{n_1,n_2}^{2\phi,2\phi}(\Delta_{2\phi,m_1},\Delta_{3\phi,m_2})=d^{(3)}\delta_{0m1}\delta_{0m_2}\,,\nonumber\\
     &\alpha_{n_1,n_2}^{2\phi,3\phi}(\Delta_{2\phi,m_1},\Delta_{2\phi,m_2})=0\,, \hspace{2cm} \alpha_{n_1,n_2}^{2\phi,3\phi}(\Delta_{2\phi,m_1},\Delta_{3\phi,m_2})=\delta_{n_1m_1}\delta_{n_2m_2}\nonumber \\  & \hspace{10.5cm}+d^{(4)}\delta_{0m1}\delta_{0m_2}\,.
     \label{moreorthogonal}
\end{align}
The constants $d_n^{(3)}$ and $d_n^{(4)}$ are related to both $c_1$ and $c_2$ in the Polyakov block and the respective coefficients of $G_{\Delta_{2\phi,n_1},\Delta_{2\phi,n_2}}$ and $G_{\Delta_{2\phi,n_1},\Delta_{3\phi,n_2}}$ of  single-exchange and contact diagrams. Seen as functions of two variables ($\Delta_1,\Delta_2$), the functionals have a rather non-trivial additional structure, as they exhibit co-dimension 1 loci where they vanish. Indeed, it is sufficient for one of the exchanged dimensions to be a double- or triple twist value, for the functional to vanish with an arbitrary value of the other dimension. While we observe this experimentally, we leave a more rigorous understanding of this property for future work.  \\

Since they trivialise on the double-twist and triple-twist dimensions, the functionals are ideal to bootstrap certain special correlation functions. We will now discuss some of these examples.

\subsubsection*{Bootstrapping the disconnected correlator}
Consider the disconnected correlator that we introduced in Section \ref{ssec:setup}. It has the following classes of operator pairs in the OPE
\begin{align}
    (\phi,1) \, &: \ \ \Delta_1=\Delta_\phi, \ \Delta_2=0\, \ \ \ \ \ \ \ \ \ \ \ \ \ \ a_{\Delta_1,\Delta_2}: \,  a_{\phi,1}=C_{\phi\phi\phi}\,,\nonumber\\
    (\phi,[\phi\phi]_n) \, &: \ \ \Delta_1=\Delta_\phi, \ \Delta_2=\Delta_{2\phi,n}\ \ \ \ \ \ \ \ \ a_{\Delta_1,\Delta_2}: \,   a_{\phi,[\phi\phi]_n}=C_{\phi\phi\phi}(C^{\text{GFF}}_{\phi\phi[\phi\phi]_n})^2\,,\nonumber\\
    ([\phi\phi]_{n_1},[\phi\phi]_{n_2}) \, &: \ \ \Delta_1=\Delta_{2\phi,n_1}, \ \Delta_2=\Delta_{2\phi,n_2}\, \ \ \ a_{\Delta_1,\Delta_2}: \,   a_{[\phi\phi]_{n_1},[\phi\phi]_{n_2}}\label{disccorrops}\,.
\end{align}
The first few $a_{[\phi\phi]_{n_1},[\phi\phi]_{n_2}}$ coefficients were given in \eqref{disc5ope}. We will show how our sum-rules trivialize on this set of OPE data by using two of them explicitly.\\ \\ 
\underline{Using $(1,2\phi)$ functionals:} First we will use the $n$-th $(1,2\phi)$ sum-rule, given by \eqref{sumrule1}. In that equation one of the exchanged dimensions i.e. $\Delta_1$ is fixed.  Let us take $\Delta_1=\Delta_\phi$. We have 
\begin{equation}\label{12phidemo}
    C_{\phi\phi\phi}\, \alpha^{1,2\phi}_n(\Delta_\phi,0)\,+\, \sum_{m}a_{\phi,[\phi\phi]_m}\,\alpha^{1,2\phi}_n(\Delta_\phi,\Delta_{2\phi,m})=0
\end{equation}
The sum in the second term reduces to $a_{\phi,[\phi\phi]_n}-a_{\phi,[\phi\phi]_0}\, b^{1,2\phi}_n(\Delta_\phi)$ using \eqref{simpleorthogonal}. Now the term $\alpha^{1,2\phi}(\Delta_\phi,0)$ comes from a Polyakov block $P_{\Delta_\phi,0}$, specifically from double-exchange diagrams $W^{(i)}_{\Delta_\phi,0}$ and single-exchange $W^{(j)}_{\Delta_\phi}$. In the direct channel one has
\begin{equation}
    W^{\text{12-3-45}}_{\Delta_\phi,0}(\chi_1,\chi_2)=\mathcal{L}(x_i)^{-1}C_{\phi\phi\phi}^{-1}\,\langle\phi(x_1)\phi(x_2)\phi(x_3)\rangle\langle\phi(x_4)\phi(x_5)\rangle\label{Witten-identity}\,.
\end{equation}
where the exchange of the $\phi$ and the identity correspond to the 3-point function and the 2-point function factors, respectively.  As discussed previously, the disconnected correlator is a sum of products of 2- and 3-point factors. This means that the double-exchange contribution to the $(1,2\phi)$ functional \eqref{12phidemo} is nothing but the OPE coefficient $-a_{\phi,[\phi\phi]_n}$ itself. For a similar reason the constant $c_1=-a_{\phi,[\phi\phi]_0}$. In other words we must have
\begin{equation}
   C_{\phi\phi\phi}\, \alpha^{1,2\phi}_n(\Delta_\phi,0)=-C_{\phi\phi\phi}(C^{\text{GFF}}_{\phi\phi[\phi\phi]_n})^2 + C_{\phi\phi\phi}(C^{\text{GFF}}_{\phi\phi[\phi\phi]_0})^2\,b_n^{1,2\phi}(\Delta_\phi)\,.
\end{equation}
Therefore the equation \eqref{12phidemo} is trivially satisfied. \\ \\
\underline{Using $(2\phi,2\phi)$ functionals:} Now consider the $(2\phi,2\phi)$ sum-rule given by \eqref{sumrule3}, specialized to the case $n_1=n_2=0$ for simplicity.
This sum-rule runs over all the operators in  \eqref{disccorrops} and we hence have the equation
\begin{align}\label{2phi2phidemo}
       C_{\phi\phi\phi}\, \alpha^{2\phi,2\phi}_{0,0}(\Delta_\phi,0)\,&+\, \sum_{n}a_{\phi,[\phi\phi]_n}\,\alpha^{2\phi,2\phi}_{0,0}(\Delta_\phi,\Delta_{2\phi,n})\nonumber\\
       &+\, \sum_{n,m}a_{[\phi\phi]_{n},[\phi\phi]_{m}}\,\alpha^{2\phi,2\phi}_{0,0}(\Delta_{2\phi,n},\Delta_{2\phi,m})=0\,.
\end{align}
In general, the term $\alpha^{2\phi,2\phi}_{0,0}(\Delta_1,\Delta_2)$ is cumbersome to derive. This is because it involves the coefficient of $G_{\Delta_{2\phi,0},\Delta_{2\phi,0}}$ and $G_{\Delta_{2\phi,0},\Delta_{3\phi,0}}$ (to fix $c_2$) from all the Witten diagrams, as well as those corresponding to $G_{\Delta_{2\phi,0},\Delta_2}$ (to fix $c_1$) from four diagrams. The evaluation of some of these coefficients, specifically the ones from crossed channel diagrams e.g. 14-2-35 or 14-3-25 channel double-exchanges or 14-235 single-exchange, are quite involved but we chalk them out in Appendix \ref{app:Mellin2}. \\ \\
However, for the evaluation at dimensions appearing in \eqref{2phi2phidemo} most of the difficult coefficients vanish, \footnote{This is because coefficients from the `difficult' channels do not have poles in $\Delta_1,\Delta_2$ while there are zeroes from the normalization due to which they vanish. See appendix \ref{app:Mellin2}.} so the evaluation of $\alpha^{2\phi,2\phi}_{0,0}(\Delta_1,\Delta_2)$ is easier. One obtains for the various terms in the sum-rule 
\begin{align}\label{a2phi2phivarious}
    &\alpha^{2\phi,2\phi}_{0,0}(\Delta_\phi,0)=-4\,,\nonumber\\
    & \alpha^{2\phi,2\phi}_{0,0}(\Delta_\phi,\Delta_{2\phi,0})=-2\,,\nonumber\\
     & \alpha^{2\phi,2\phi}_{0,0}(\Delta_\phi,\Delta_{2\phi,n})=0\,, \ (n\neq 0)\,,\\
     & \alpha^{2\phi,2\phi}_{0,0}(\Delta_{2\phi,0},\Delta_{2\phi,0})=4\,,\nonumber\\
          & \alpha^{2\phi,2\phi}_{0,0}(\Delta_{2\phi,n},\Delta_{2\phi,m})=0\,, \ (n,m\neq 0)\nonumber\,.
\end{align} 
Let us point out that the term $\alpha^{2\phi,2\phi}_{0,0}(\Delta_\phi,0)$ corresponds to Witten diagrams with an identity exchange. So, as we explained around eqn. \eqref{Witten-identity}, it is obtained from the disconnected correlator itself, specifically from the OPE coefficients of $G_{\Delta_{2\phi,0},\Delta_{2\phi,0}}$ and $G_{\Delta_{2\phi,0},\Delta_{3\phi,0}}$. \footnote{Notice that \eqref{a2phi2phivarious} slightly contradict the orthogonality conditions \eqref{moreorthogonal} because we get, e.g. $\alpha^{2\phi,2\phi}_{0,0}(\Delta_{2\phi,0},\Delta_{2\phi,0})=4$ instead of 1. The value is different from 1 for special cases: if we have $n_1=n_2$ in $\alpha^{2\phi,2\phi}_{n_1,n_2}$, or if one or both of the dimensions $\Delta_1$ or $\Delta_2$ is $\Delta_{2\phi,0}$ (the consequence of the latter being that we get a term $\propto c_1$). The present case satisfies both conditions. }\\ \\ 
If we use \eqref{a2phi2phivarious} along-with the respective OPE coefficients $a_{\phi,[\phi\phi]_0}=2\, C_{\phi\phi\phi}$, $a_{[\phi\phi]_0,[\phi\phi]_0}=2\,C_{\phi\phi\phi}$ we see that \eqref{2phi2phidemo} is satisfied.

\subsubsection*{Deforming the disconnected correlator}
We showed from the orthogonality relations \eqref{simpleorthogonal} and \eqref{moreorthogonal} that our sum-rules trivialize when the 5-point OPE has dimension pairs of the type $(\Delta_{2\phi,n_1},\Delta_{2\phi,n_2})$, $(\Delta_{2\phi,n_1},\Delta_{3\phi,n_2})$, $(\Delta_1,\Delta_{2\phi,n})$ or $(\Delta_1,\Delta_{1,\phi,n})$. From the examples in subsection \ref{ssec:funcs} we see that in such cases, one can determine all OPE coefficients from the knowledge of only a few. For instance, in \eqref{12phidemo} we can determine $a_{\phi,[\phi\phi]_n}$ from $a_{\phi,[\phi\phi]_0}$ and in \eqref{2phi2phidemo} we obtain $a_{[\phi\phi]_0,[\phi\phi]_0}$ in terms of $a_{\phi,[\phi\phi]_0}$. \\ \\ 
This feature of the sum-rules remains valid even if the respective operators acquire anomalous dimensions and OPE coefficient corrections. 
To see this, consider a (somewhat artificial) deformation of the disconnected correlator where the double-twist dimensions are $\Delta=\Delta_{2\phi,n}+\gamma_n$ and the OPE coefficients are modified as follows
\begin{align}
    (\phi,[\phi\phi]_n) \, &:   \ \ \ \ a_{\Delta_1,\Delta_2}\to \, a_{\phi,[\phi\phi]_n}+\delta^{(1)}_n\,,\nonumber\\
    ([\phi\phi]_{n_1},[\phi\phi]_{n_2}) \, &: \ \ \ \  a_{\Delta_1,\Delta_2}\to  \,   a_{[\phi\phi]_{n_1},[\phi\phi]_{n_2}}+\delta^{(2)}_{n_1,n_2}\label{moddisccorrops}\,.
\end{align}
Then the ($2\phi,2\phi$) sum-rule from \eqref{2phi2phidemo} gets corrected and the leading order correction terms look like 
\begin{align}\label{mod2phi2phidemo}
       \, \delta^{(1)}_0\,\alpha^{2\phi,2\phi}_{0,0}(\Delta_\phi,\Delta_{2\phi,0})&+\, \sum_{n}\gamma_n\, a_{\phi,[\phi\phi]_n}\,\partial_{\Delta_2}\alpha^{2\phi,2\phi}_{0,0}(\Delta_\phi,\Delta_2=\Delta_{2\phi,n})+\, \delta^{(2)}_{0,0}\,\alpha^{2\phi,2\phi}_{0,0}(\Delta_{2\phi,0},\Delta_{2\phi,0})\nonumber\\
       &+\, \sum_{(n,m)}2\gamma_m\, a_{[\phi\phi]_{n},[\phi\phi]_{m}}\,\partial_{\Delta_2}\alpha^{2\phi,2\phi}_{0,0}(\Delta_{2\phi,n},\Delta_2=\Delta_{2\phi,m})=0\,.
\end{align}
In the last term we assumed $\partial_{\Delta_1}\alpha^{2\phi,2\phi}_{0,0}(\Delta_1,\Delta_2)=\partial_{\Delta_1}\alpha^{2\phi,2\phi}_{0,0}(\Delta_2,\Delta_1)$\,. This equation determines $\delta^{(2)}_{0,0}$ in terms of $\delta^{(1)}_0$ if we assume that all $\gamma_n$-s are fixed\,. \\ \\ 
It is not hard to see that if the pairs $(\Delta_{2\phi,n_1},\Delta_{2\phi,n_2})$, $(\Delta_{2\phi,n_1},\Delta_{3\phi,n_2})$, $(\Delta_1,\Delta_{2\phi,n})$ or $(\Delta_1,\Delta_{1,\phi,n})$ get modified by some given  anomalous dimensions their OPE coefficient corrections would be obtained by $\alpha^{2\phi,2\phi}_{n_1,n_2}$, $\alpha^{2\phi,3\phi}_{n_1,n_2}$, $\alpha^{1,2\phi}_{n}$ or $\alpha^{1,(1,\phi)}_{n}$ functionals respectively.\footnote{In fact one can achieve the same result in perturbation theory by repeated actions of only $\alpha^{1,2\phi}_{n}$ and $\alpha^{1,(1,\phi)}_{n}$ for suitable $\Delta_1$ and $n$.}  The rationale behind treating modified dimensions as inputs is that we consider them to be fixed at the level of a 4-point function that has previously been determined by bootstrap, or other methods (e.g. the 4-point sum-rules \eqref{Polysumrule} are suitable to fix $\gamma_n$). 
\paragraph{Further checks}
Along with the disconnected correlator and their deformations, we also considered simple correlation functions whose OPE decomposition contains infinite families of operators which are not of double- or triple-twist type, and therefore need an infinite number of contributions to satisfy the sum-rules. Examples of such correlators can be obtained by considering composite operators such as $\phi^2$ and $\phi^4$ in generalized free theory, and we verified the associated sum rules extensively. Instead of presenting these explicit (but cumbersome) checks here, we will instead describe how to bootstrap such a correlator in the following section.
\section{Application: Truncated numerical bootstrap}
\label{sec:Appli}
    In this section we apply our functionals to bootstrap the 5-point crossing equations. 
    Since the OPE decomposition does not have manifest positivity properties, we resort to the truncated bootstrap as done in \cite{Poland:2023vpn,Poland:2023bny,Poland:2025ide}. 
    As a target, we consider an exactly known but non-trivial solution to crossing, introduced in Section \ref{sec:all-to-all}. 
    An important feature of this solution is that its spectrum is not a perturbative deformation of double-twist or triple-twist towers. 
    We treat the spectrum as an input (that in a realistic situation would be provided by the 4-point bootstrap or other methods) and bootstrap the non-trivial OPE coefficients, first using derivative functionals (Section \ref{sec:derv_fun}) and then using the new functionals proposed in this work (Section \ref{sec:our_fun}). 
    Our knowledge of the exact correlator makes this example an ideal setup to compare the two different sets of functionals in a controlled environment. 
    We comment on this comparison at the end of Section \ref{sec:comparison}.

\subsection{Bootstrap target: The `all-to-all' correlator}
\label{sec:all-to-all}
Consider the correlator, 
\begin{equation}
     G_{\rm{all-to-all}}(x_i)=\prod_{i<j}^5 (x_{ij})^{2\Delta_\phi/4}\,,
\end{equation}
which can be obtained by computing the five point function of $:\phi^4:$ operators in a generalized free theory through Wick contractions that connect every operator to every other exactly once.
Here, $\Delta_\phi$ denotes the dimension of the external operator.\footnote{This dimension is four times that of the constituent field $\phi$, but we do not introduce notation for this to avoid confusion.} This function is manifestly crossing symmetric and by pulling out the prefactor, we find the following simple function of the cross-ratios
\begin{equation}
     \mathcal{G}_{\rm{all-to-all}}(\chi_a)=\left( \frac{\chi_1^3 \,\chi_2^3}{(1-\chi_1)(1-\chi_2)(1-\chi_1-\chi_2)}\right)^{\Delta_\phi/2}\,.
\end{equation}
From the above expression it can easily be seen that the OPE is controlled by a tower of operators of scaling dimensions
\begin{equation}
    \Delta_{(n)} = \frac{3 \Delta_\phi}{2} + 2n\,,
\end{equation}
 which as previously advertised are neither double- nor triple-traces of the external operator. They can instead be understood as $:\phi^6:$-type operators in the $:\phi^4:\times :\phi^4:$ OPE. It is straightforward to expand this correlator in conformal blocks and extract OPE coefficients, which will subsequently be compared with those obtained by truncating the crossing equation with derivative and analytic functionals. Specializing to the case $\Delta_\phi=1$ as we will below, and introducing by abuse of notation $a_{n,m}\equiv a_{3/2+2n,3/2+2m}$, the first few coefficients are given by
 \begin{align}
     a_{0,0}=1\,,\, a_{0,1}=1/16\,,\, a_{0,2}=1/256\,,\, a_{0,3}=1/4096\,,\,a_{1,1}=7/128\,,\,a_{1,2}=23/4096\,.
 \end{align}
    
    \subsection{Bootstrap with derivative functionals}\label{sec:derv_fun}
    
        A simple way to extract the constraints on CFT data encoded in the crossing equation \eqref{eq:crossing} is to Taylor expand in $\chi_1$ and $\chi_2$ around the crossing symmetric point 
        \begin{align}
            \chi_1 = \chi_2 = \chi_s \equiv \frac{1}{2} \left( 3 - \sqrt{5}\right),
        \end{align}
        leading to the derivative sum-rules 
        \begin{equation}\label{eq:derv_sr}
           0=\partial_{\chi_1}^n\partial_{\chi_2}^m\left[\sum\limits_{\Delta_1, \Delta_2} a_{\Delta_1,\Delta_2} F(\chi_1,\chi_2) \right]\Bigg|_{\chi_1 = \chi_2 = \chi_s} \,.
        \end{equation}
        labeled by two integers $n$ and $m$. 
        Here,
        \begin{align}
            F_{\Delta_1,\Delta_2}(\chi_1,\chi_2)= \left(\hspace{-0.05 cm}G_{\Delta_1,\Delta_2}(\chi_1, \chi_2) - \left(\frac{\chi_1^2 \chi_2}{(\chi_1+\chi_2-1)^2}\right)^{\Delta_\phi} \hspace{-0.35 cm}G_{\Delta_1,\Delta_2}\left(\chi_2, 1+\frac{\chi_2}{\chi_1-1}\right) \hspace{-0.1 cm}\right)\hspace{-0.05 cm}.
        \end{align}
        We use derivative sum-rules to bootstrap the all-to-all correlator in the following way. 
        Considering six randomly chosen derivative sum-rules, we truncate the spectrum to the lowest seven operators and solve the six resulting equations for the six unknowns $a_{0,1}$, $a_{0,2}$, $a_{0,3}$, $a_{1,1}$, $a_{1,2}$ and $a_{2,2}$ with $a_{1,1}$ normalized to $1$.
        The result of this procedure is shown in figure \ref{fig:comparison} for $10$ iterations selecting out of derivative functionals with $n,m \leq 3$.
        The above procedure is chosen in order to have a simple meaningful comparison with the analytic functionals and not with the goal of formulating an optimal strategy for truncated bootstrap. 
    
    \subsection{Bootstrap with analytic functionals}\label{sec:our_fun}
As before, we will assume the scaling dimensions to be given. We take them to be as follows:
\begin{equation}
    (\Delta_{(p)},\Delta_{(q)}):=\Big(\frac{3}{2}+2p,\frac{3}{2}+2q\Big)\,, \ p,q\in \mathbb{Z}_{\ge 0}\,.
\end{equation}
and $\Delta_\phi=1$. Our goal is to obtain the OPE coefficients $a_{p,q}:=a_{\Delta_{(p)},\Delta_{(q)}}$ with the input of $a_{0,0}=1$\,. \\ \\ 
The strategy is simple: we will choose a set of operator pairs $(\Delta_{(p)},\Delta_{(q)})$ corresponding to $p\leq p_{\text{max}},\, q\leq q_{\text{max}}(p)$. We will choose the same number of sum-rules. 
For this problem each sum-rule will couple multiple unknowns $a_{p,q}$, so we cannot solve them individually. Our hope is to solve them as a set of simultaneous linear equations  and obtain $a_{p,q}$ to a good numerical accuracy. \\ \\ 
Let us choose $p_{\text{max}}=2$, and $q_{\text{max}}(0)=3, q_{\text{max}}(1)=2, q_{\text{max}}(0)=2$. This means we have the following pairs: 
\begin{align}
    (\Delta_{(p)},\Delta_{(q)})=& \, \mbox{$(\frac 32,\frac 32), (\frac 32,\frac 72), (\frac 32,\frac {11}{2}), (\frac 32,\frac {15}{2}), (\frac 72,\frac {7}{2}),  (\frac 72,\frac {11}{2}),  (\frac {11}{2},\frac {11}{2})$}\,,\\
    a_{p,q} \ \ = &\ a_{0,0}\ , \ \ \, a_{0,1}\ ,  \ \ \, a_{0,2}\ , \  \ \, a_{0,3} \ , \  \ \, a_{1,1}\ , \  \ \, a_{1,2}\ , \  \ \, a_{2,2}\,.
\end{align}
There are six variables, so we need six sum-rules. Let us first choose the functional $\alpha^{1,2\phi}_0$. For this we choose the fixed dimension to be $\Delta_1=\frac 32, \frac 72, \frac{11}{2}$ which gives, from \eqref{sumrule1}, three distinct equations. Numerically they read:\footnote{The numerical values reported below are obtained from the techniques discussed in Appendices \ref{app:Mellin} and \ref{app:integratedvertex}. All functional expressions involve infinite sums. We chose suitable truncations so that the results from both techniques match up to a certain accuracy.}\\ \\
\underline{$(1,2\phi)$, $n=0 $ sum-rules:}
\begin{align}
&\Delta_1=\mbox{$\frac{3}{2}$} : \hspace{1cm}    a_{0,0}\, 0.06696 - a_{0,1}\, 1.13115+  a_{0,2}\, 0.84200 + a_{0,3}\, 1.40585 =0\,,\label{numsumrule1}\\
&\Delta_1=\mbox{$\frac{7}{2}$} : \hspace{1cm}    a_{0,1}\,  1.52816-  a_{1,1}\, 1.78614+ a_{1,2}\, 0.34937 =0\,,\label{numsumrule2}\\
&\Delta_1=\mbox{$\frac{11}{2}$} : \hspace{0.8cm} a_{0,2}\,1.30190- a_{1,2}\,0.80423-a_{2,2}\,0.18168=0\,. \label{numsumrule3}
\end{align}
We will not use $(1,2\phi)$ sum-rules with $n>0 $ because for a given $\Delta_1$ they would sum over the same series of operators but would require more of them for numerical convergence. E.g. the $(1,2\phi)$ sum-rule for $n=3, 4,$ etc with $\Delta_1=\frac 32$ we get a sum over $a_{0,q}$ that is different from $n=0$ i.e. \eqref{numsumrule1}, but we would require to truncate $q$ at higher values than what we should have for $n=0,1$. This is evident from their orthogonality property - we cannot satisfy the $n$-th $(1,2\phi)$ sum-rule with operator pairs $(\Delta_1,\Delta_{2\phi,m})$ with all $m<n$. \\ \\ 
We need three more equations. So, we will use the sum-rules $(2\phi,2\phi)$ and $(2\phi,3\phi)$. The former will be for $n_1=n_2=0$ and the latter for $n_1=0, n_2=1$ and $n_1=0,n_2=2$, hence three equations.
The advantage is that each of them sum over all the OPE coefficients, and for low enough $n_1,n_2$ converge quickly.  \\ \\
\underline{$(2\phi,2\phi)$, $n=0 $ sum-rule:}
\begin{align}
-a_{0,0}\,0.20280\,+\,a_{0,1}\,2.00474 & \,+a_{0,2}\,0.38663 \,+ \, a_{0,3}\,0.74856 \,+\,a_{1,1}\,1.40910\nonumber\\
&- \,a_{1,2}\,0.76550\, + \, a_{2,2}\,1.13883\,= \,0\,. \label{numsumrule4}
\end{align} \\
\underline{$(2\phi,3\phi)$ sum-rules:}
\begin{align}
(n_1,n_2)=(0,1): \hspace{0.7cm}  &a_{0,0}\,0.00928 \,+\,a_{0,1}\,0.62012 \,-\,a_{0,2}\, 0.1216 \,-\,a_{0,3}\, 0.23437 \nonumber\\
& \ \ \ \ \ \  -\,a_{1,1}\, 0.99493 \,+\,a_{1,2}\, 0.81915 \,+\,a_{2,2}\, 0.80916=0\,. \label{numsumrule5}\\
(n_1,n_2)=(0,2): \hspace{0.7cm} & \,a_{0,0}\, 0.00099 \,+\,a_{0,1}\, 0.06188 \,+\,a_{0,2}\, 0.37021 \,-\,a_{0,3}\, 0.32135  \nonumber\\
& \ \ \ \ \ \  -\, a_{1,1}\, 0.08503\,-\, a_{1,2}\, 0.05202 \, -\, a_{2,2}\, 0.62975\,=\,0\,. \label{numsumrule6}
\end{align} 
Solving the equations \eqref{numsumrule1}-\eqref{numsumrule6} we obtain
\begin{align}\label{eq:results_analytic_func}
    &a_{0,1}= 0.0623,\,a_{0,2}= 0.0043,\,a_{0,3}= 0.00030,\nonumber\\
    & a_{1,1}= 0.055,\, a_{1,2}= 0.0065,\,a_{2,2}= 0.0022\,.
\end{align}
    \subsubsection*{Comparison} \label{sec:comparison}
        \begin{figure}
            \centering
            \includegraphics[width=\linewidth]{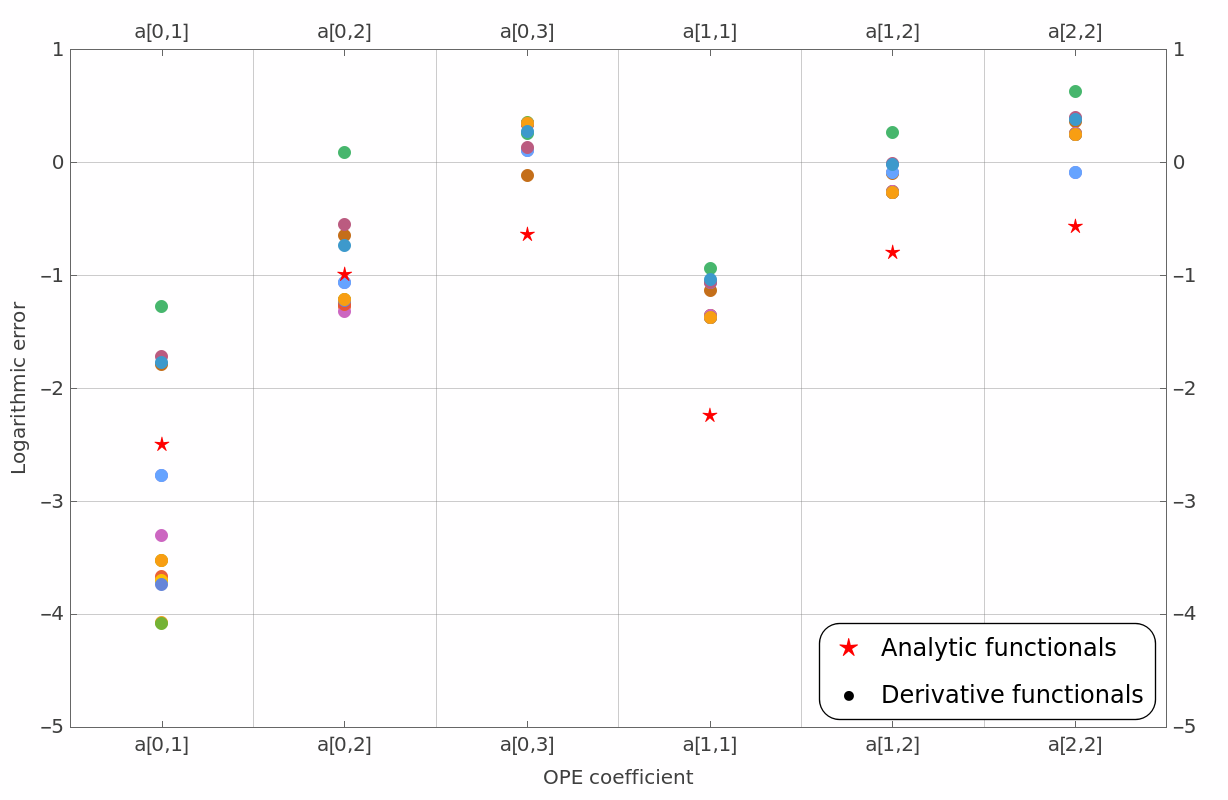}
            \caption{The absolute value of the relative error of numerical estimates for the OPE coefficients $a_{p,q}$ for derivative and analytic functionals plotted on a logarithmic scale.}
            \label{fig:comparison}
        \end{figure}
        Finally, let us compare the results \eqref{eq:results_analytic_func} obtained from the analytic functionals with the results obtained by using derivative functionals as described in Section \ref{sec:derv_fun}.
        For the comparison, we have plotted the logarithmic error
        \begin{align}
            \log_{10}\left(\frac{|a_{p,q}^{\text{exact}}-a_{p,q}|}{a_{p,q}^{\text{exact}}}\right)
        \end{align}
        of the numerically determined OPE coefficients in Figure \ref{fig:comparison}.

        We observe that the accuracy of the numerical estimate for the OPE coefficients of light operators varies considerably between different choices of derivative sum-rules with a mean that is comparable to the error of the solution obtained from analytic functionals. 
        For heavier operators, however, the analytic functionals considerably outperform derivative functionals, producing significantly smaller errors than the best choice of derivative sum-rules.
\section{Conclusions}
\label{sec:Conclusions}

In this paper we took the first steps to extend the Polyakov bootstrap approach to higher-point correlation functions, focusing on the 5-point function in one dimension. We constructed the crossing symmetric Polyakov block, and derived the associated sum-rules which we extensively tested. Finally, we applied the sum-rules to the truncated bootstrap, from which we were able to obtain good results for the OPE coefficients of a non-trivial correlator. There are several unexplored extensions and applications that we discuss here in increasing order of complexity.

\paragraph{More on the 5-point 1d bootstrap} Our sum-rules are a natural way to study $\mathbb{Z}_2$ breaking models, such as $\phi^3$ theory in AdS$_2$ or certain symmetry breaking line defects. It would be nice to study the sum-rule equations in perturbation theory where many functional actions vanish.
It is also straightforward to generalize our analysis to the case of non-identical scalars which is useful to study the lightest correlator in $\mathbb{Z}_2$ preserving theories: $\langle \phi \phi \phi^2\phi\phi\rangle$. Finally, it would be interesting to  understand the formal properties of our functionals more systematically, namely their asymptotics and analytic structure.

\paragraph{Generalization to six-points and fermions} From our analysis, it becomes clear that constructing higher-point Polyakov blocks should be straightforward but technically very involved, due to the necessity of subtraction terms corresponding to higher-point vertices $\Phi^n$ in AdS$_2$ which are relevant deformations preserving the UV features of the solution. A simple way to bypass this is to consider the fermionic version of the Polyakov blocks, which will not involve vertices with more than three legs, since these correspond to irrelevant interactions. Of course, additional technical subtleties due to the fermionic nature of the Witten diagrams is to be expected as well.

\paragraph{Higher-dimensions} Observing the success of our functionals in the one-dimensional truncated Bootstrap makes the three-dimensional analysis of \cite{Poland:2023vpn,Poland:2023bny,Poland:2025ide} a particularly appealing target for the Polyakov Bootstrap. With modest resources, the 3-dimensional truncated bootstrap produced reasonable estimates for OPE coefficients involving two stress-tensors in the 3d Ising model. A higher-dimensional generalization of our approach has clear potential to significantly improve these predictions. The additional technical difficulties of including spin and multiple tensor structures in the OPE are, however, not to be underestimated.

\paragraph{S-matrix Bootstrap}
Finally, we should mention that our results are intimately tied to S-matrices in two-dimensions, with $2\to3$ scattering amplitudes being the key observable. On the one hand, the flat-space limit of conformal 4-point functions is known to map to S-matrices in one higher-dimension \cite{Komatsu:2020sag}. The greater control over the analytic structure of conformal correlators can be used as an alternative to standard axiomatic approaches to scattering amplitudes in deriving fundamental properties of these objects \cite{vanRees:2022itk,vanRees:2023fcf}. The QFT$_2$/CFT$_1$ setup \cite{Cordova:2022pbl,Ghosh:2025sic} is particularly powerful in this regard, and our work opens up an avenue to investigate the analytic properties of 5-point amplitudes which, even in two space-time dimensions, are far from well understood. On the other hand, the Polyakov Bootstrap makes crossing symmetry manifest and therefore has a closely related cousin at the level of scattering amplitudes: the so-called crossing-symmetric dispersion relations (CSDR), which have proved to be a very useful tool in many recent bootstrap applications \cite{Gopakumar:2021dvg, Bhat:2025zex}. We expect that a synergy between these approaches might help make higher-point amplitudes more amenable to a non-perturbative understanding.

\acknowledgments
We are grateful to Volker Schomerus for collaboration at the early stages of this project and for many useful discussions. We further thank Kaushik Ghosh, Vasco Gon\c{c}alves, Marco Meineri, Miguel Paulos and Aninda Sinha for useful conversations.  AA additionally thanks Andreia Gon\c{c}alves for continued support. AK thanks the participants of ``Aspects of CFTs" workshop, held at IIT Kanpur in January 2024, for discussions. AA is funded by the European Union (ERC, FUNBOOTS, project number 101043588).
Views and opinions expressed are however those of the author(s) only and do not necessarily reflect those of the European Union or the European Research Council Executive Agency. Neither the European Union nor the granting authority can be held responsible for them. 
SH received funding from the German Research Foundation DFG under Germany's Excellence Strategy - EXC 2121 Quantum Universe - 390833306, the Collaborative Research Center - SFB 1624 “Higher structures, moduli spaces and integrability” - 506632645 and the Studienstiftung des Deutschen Volkes.

\appendix
\section{Details of Mellin space computations}
\label{app:Mellin} 
\subsection{4-point Witten diagrams}
\label{app:Mellin1} 
 In this appendix we will provide some details of the definitions of Witten diagrams in Mellin space  and the computations related to them. Recall the 4-point $s$-channel diagram that is given by
\begin{equation}
    M^{(s)}_{\Delta}(s_{12},s_{14})=\sum_{m=0}^{\infty}\frac{\mathcal{Q}_m^{(\Delta_\phi)}}{s_{12}+\frac{\Delta}{2}-\Delta_\phi+m}\,.\label{Msapp}
\end{equation}
Here the numerator is given by
\begin{equation}
\mathcal{Q}_m^{(\Delta_\phi)}=\frac{\Gamma (\Delta ) \Gamma \big(\Delta +\frac{1}{2}\big) \big(\left(\frac{\Delta }{2}-\Delta _{\phi }+1\right)_m\big)^2}{\Gamma \big(\frac{\Delta }{2}\big)^4\, \Gamma (m+1) \Gamma \big(\Delta _{\phi }-\frac{\Delta }{2}\big)^2 \,\Gamma \big(m+\Delta +\frac{1}{2}\big)}\,.
\end{equation}
The $t$-channel exchange diagram is given by replacing $s_{12}\leftrightarrow s_{14}$. Let us demonstrate computation of a spurious term $\chi^{2\Delta_\phi}\log \chi$ which we initiated for a general amplitude in \eqref{Mellindemolog}. 
For the  $s$-channel diagram the $s_{14}$ integral is the same as the contact diagram (see \eqref{contactlog}) since $M^{(s)}$ is independent of $s_{14}$, and we carry it out using Barnes first lemma. So we only have the sum over $m$ in \eqref{Msapp} which is straightforward to carry out. We obtain 
\begin{align}
     &\chi^{2 \Delta \phi} \log(\chi)\int_{-i\infty}^{i\infty} \frac{ds_{14}}{2\pi i}\,    \, 2\, \Gamma \left(s_{14}\right){}^2 \Gamma \left(\Delta_\phi -s_{14}\right){}^2\sum_{m=0}^{\infty}\frac{\mathcal{Q}_m^{(\Delta_\phi)}}{m+\frac{\Delta}{2}-\Delta_\phi}\,\nonumber\\
     &=\frac{\chi^{2 \Delta \phi} \log(\chi)\, 4 \Gamma (\Delta ) \Gamma \left(\Delta +\frac{1}{2}\right) \Gamma \big(\Delta _{\phi }\big){}^4 \Gamma \left(2 \Delta _{\phi }-\frac{1}{2}\right)}{\big(\Delta -2 \Delta _{\phi }\big) \Gamma \left(\frac{\Delta }{2}\right)^4 \Gamma \big(2 \Delta _{\phi }\big) \Gamma \big(\frac{\Delta -1+2\Delta _{\phi }}{2}\big) \Gamma \big(\frac{2\Delta _{\phi }-\Delta }{2}\big){}^2 \Gamma \big(\frac{\Delta +1+2\Delta _{\phi }}{2}\big)}\,.\label{schannellog}
\end{align} \\
Let us now show the evaluation of the analogous coefficient in the $t$-channel diagram. This is harder  since the $M^{(t)}$ has an $s_{14}$ in the denominator and so the $s_{14}$ integral cannot be evaluated using Barnes lemma. 
To carry out the integral we write $(s_{14}+\frac{\Delta}{2}-\Delta_\phi+m)^{-1}=\int_0^1 \frac{dy}{y}\, y^{s_{14}+\frac{\Delta}{2}-\Delta_\phi+m}$\,. So for $t$-channel diagram we obtain
\begin{align}
 &\chi^{2 \Delta \phi} \log(\chi)\sum_{m=0}^{\infty}\int_{-i\infty}^{i\infty} \frac{ds_{14}}{2\pi i}   \, 2\,\mathcal{Q}_m^{(\Delta_\phi)} \,\Gamma \left(s_{14}\right){}^2 \Gamma \left(\Delta_\phi -s_{14}\right){}^2 \,\int_0^1 \frac{dy}{y}  y^{s_{14}+\frac{\Delta}{2}-\Delta_\phi+m}\nonumber\\
    &=\chi^{2 \Delta \phi} \log(\chi)\sum_{m=0}^{\infty}\, \int_0^1 \frac{dy}{y}\, y^{\frac{\Delta}{2}-\Delta_\phi+m} \, 2\,\mathcal{Q}_m^{(\Delta_\phi)} \,\Gamma(\Delta_\phi)^4 {}_2\widetilde{F}_1(\Delta_\phi,\Delta_\phi,2\Delta_\phi,\mbox{$1-\frac{1}{y}$})\nonumber\\
    &=\chi^{2 \Delta \phi} \log(\chi)\sum_{m=0}^{\infty}\,2\,\mathcal{Q}_m^{(\Delta_\phi)}\Gamma \left(\Delta _{\phi }\right)^4\, \Gamma \left(\mbox{$\frac{\Delta }{2}+m$}\right)\, _3\widetilde{F}_2\left[\mbox{$1,\Delta _{\phi },\Delta _{\phi };2 \Delta _{\phi },\frac{2m+\Delta +2}{2};1$}\right]\label{tchannellog}
\end{align}
Here $\widetilde{F}$ denotes the regularized hypergeometric functions. To go from the second to third step we used the Pfaff transformation of ${}_2F_1$ to convert the argument $(1-\frac{1}{y})$ to $1-y$, then expanded in $(1-y)$ powers - then we integrated the $k$-th expansion term over $y$ and summed over $k$. So the final expression involves an infinite sum. It converges fast and we can evaluate it numerically by truncating.\\ \\
The $u$-channel coefficient of $\chi^{2\Delta_\phi}\log(\chi)$ is carried out similarly to $t$-channel. It also comes out to be identical to that of the $t$-channel. This shows the property $b_n^{(t)}=b_n^{(u)}$ pointed out below \eqref{Wcondeco} in the main text.

\subsection{5-point Witten diagrams}
\label{app:Mellin2} 
In this appendix we will show how to obtain the coefficients of the various conformal blocks from 5-point Witten diagrams. Let us start with the definition of inverse Mellin transform \eqref{5ptMellin} which we repeat for convenience of the reader
\begin{align}
& \mathcal{G}(\chi_1,\chi_2)=\int [\text{d}s_{ij}]\, \prod_{i<j\le 5}\big(\Gamma(s_{ij})\,\big)\, f(\chi_1,\chi_2|\,\{s_{ij}\})\, M(s_{12},s_{23},s_{34},s_{45},s_{15})\label{InvMellin5point}\\
\text{where \ }
&f(\chi_1,\chi_2|\,\{s_{ij}\})=\frac{\left(1-\chi_1-\chi_2\right)^{-2 s_{15}}  \chi_1^{2 \Delta _{\phi }-2 s_{12}}  \chi_2^{2 \Delta _{\phi }-2 s_{45}}}{\left(1-\chi_1\right)^{\Delta_{\phi }-2 s_{15}+2 s_{23}-2 s_{45}}\left(1-\chi_2\right)^{\Delta _{\phi }-2 s_{12}-2 s_{15}+2 s_{34}}}\label{kinematic}\,.
\end{align}
Let us first consider the blocks $G_{\Delta_{2\phi,n_1},\Delta_{2\phi,n_2}}$ and $G_{\Delta_{2\phi,n_1},\Delta_{3\phi,n_2}}$, $G_{\Delta_{3\phi,n_1},\Delta_{2\phi,n_2}}$ which feature in the decomposition of all the Witten diagrams (see \eqref{doubleWitten}, \eqref{singleWitten}, \eqref{contact5ptWitten}). 
\subsubsection*{Coefficients of $G_{\Delta_{2\phi,n_1},\Delta_{3\phi,n_2}}$} Let us first focus on the blocks  $G_{\Delta_{2\phi,n_1},\Delta_{3\phi,n_2}}$. This involves the power laws $\chi_1^{2\Delta_\phi+m'}\chi_2^{3\Delta_\phi+n'}$ for $m',n'\in \mathbb{Z}_{\ge 0}$\,. They are obtained from the following residues:
\begin{equation}
    s_{12}=-m\,,\, s_{45}=-n-m+s_{23}-\frac{\Delta_\phi}{2}\,,\, s_{23}=-p\,.\label{2phi3phipoles}
\end{equation}
In choosing the contour we stay careful about the following two things: 1. if we are considering
negative integer poles of a $\Gamma(s_{ij})$  then all those poles are enclosed in the contour, and 2. we take all the terms $\chi_1,\chi_2, (1-\chi_1)$, $(1-\chi_2)$, $(1-\chi_1-\chi_2)$ are less than 1 individually,  which would require all $s_{ij}$ contours to be closed on the left.  These make it clear that \eqref{2phi3phipoles} is the only combination of poles how the required power laws can emerge. \\ \\
With these poles the kinematic part \eqref{kinematic} becomes 
\begin{align}
    &\frac{\left(1-\chi_1-\chi_2\right)^{-2 s_{15}}  \chi_1^{2 \Delta _{\phi }+2 m}  \chi_2^{3 \Delta _{\phi }+2 m+2n+2p}M(-m,-p,s_{34},-n-m-p-\frac{\Delta_\phi}{2},s_{15})}{\left(1-\chi_1\right)^{2\Delta_{\phi }-2 s_{15}+2m+2n}\left(1-\chi_2\right)^{\Delta _{\phi }+2 m-2 s_{15}+2 s_{34}}}\nonumber\\
&=\chi_1^{2 \Delta _{\phi }+2 m}  \chi_2^{3 \Delta _{\phi }+2 m+2n+2p}\,M(\mbox{$-m,-p,s_{34},-n-m-p-\frac{\Delta_\phi}{2},s_{15}$})\sum_{q,r} \,\chi_1^{q}\,\chi_2^{r}\,N_{q,r}(s_{34},s_{15})\,.\label{kinematicalexpand1}
\end{align}
Here $N_{q,r}$ are the coefficients appearing in the expansion of the powers of $(1-\chi_1-\chi_2)$, $(1-\chi_1)$ and $(1-\chi_2)$. Collecting the powers of $\chi_1$ and $\chi_2$ we obtain
\begin{equation}
\chi_1^{2\Delta_\phi+m'}\chi_2^{3\Delta_\phi+n'}\int \,  \text{d}s_{34}\, \text{d}s_{15}\,\prod_{\stackrel{(ij)\neq}{12,\,45,\,23}}\big(\Gamma(s_{ij})\,\big)\, \widetilde{M}(s_{34},s_{15})\label{2phi3phicoeffgen}
\end{equation}
where
\begin{equation}
    \widetilde{M}(s_{34},s_{15})=\sum_{(q,r)=(0,0)}^{(m',n')}\mbox{$M\big(\,\frac{q-m'}{2},m+n+\frac{r-n'}{2},s_{34},\frac{r-n'-\Delta_\phi}{2},s_{15}\,\big)\,N_{q,r}(s_{34},s_{15})$}\,.
\end{equation}
There are only two integrals to do in in \eqref{2phi3phicoeffgen} , so it is simpler to evaluate than the coefficients for $G_{\Delta_{2\phi,n_1},\Delta_{2\phi,n_2}}$ (see \eqref{2phi2phicoeffgen} below). \\ \\ 
Eqn. \eqref{2phi3phicoeffgen} is easiest to evaluate when $M(\{s_{ij}\})$ independent of $s_{34}, s_{15}$. This  happens for the following cases:
\begin{itemize}
    \item Contact diagram with $M^{\text{(con)}}=1$ (see \eqref{5ptcontactMellin}).
    \item Single-exchange diagrams $M^{\text{12-345}}_{\Delta}$, $M^{\text{13-245}}_{\Delta}$, $M^{\text{23-145}}_{\Delta}$,  $M^{\text{45-123}}_{\Delta}$ (see \eqref{singlexc-Mellin}).
    \item Double-exchange diagrams $M^{\text{12-3-45}}_{\Delta_1,\Delta_2}$, $M^{\text{13-2-45}}_{\Delta_1,\Delta_2}$, $M^{\text{23-1-45}}_{\Delta_1,\Delta_2}$ (see \eqref{doublexc-Mellin}).
\end{itemize}
For all these diagrams the integral can be evaluated using first Barnes lemma twice. E.g. for the case of $m'=n'=0$ one has
\begin{align}
\chi_1^{2\Delta_\phi}\chi_2^{3\Delta_\phi}\, & M\big(\,s_{12}=0,s_{23}=0,s_{45}=-\mbox{$\frac{\Delta_\phi}{2}$}\,\big)\,\Gamma \big(\mbox{$-\frac{\Delta _{\phi }}{2}$}\big)\int \,  \text{d}s_{34}\, \text{d}s_{15}\,\Gamma \left(s_{15}\right) \Gamma \left(s_{34}\right)\nonumber\\
& \times \Gamma \big(\Delta _{\phi }-s_{15}\big) \Gamma \big(\mbox{$ s_{15}- s_{34}+\frac{\Delta _{\phi }}{2} $}\big) \Gamma \big(\Delta _{\phi }-s_{34}\big) \Gamma \big(\mbox{$s_{34}- s_{15}+\frac{\Delta_\phi}{2}$} \big)\nonumber\\
=\chi_1^{2\Delta_\phi}\chi_2^{3\Delta_\phi}\, & M\big(\,s_{12}=0,s_{23}=0,s_{45}=-\mbox{$\frac{\Delta_\phi}{2}$}\,\big)\,\frac{\Gamma \big(\mbox{$-\frac{\Delta _{\phi }}{2}$}\big)\Gamma \left(\Delta _{\phi }\right){}^3 \Gamma \big(\frac{3 \Delta _{\phi }}{2}\big){}^2}{\Gamma \left(3 \Delta _{\phi }\right)}\,. \label{easyintegral}
\end{align}
For higher $m',n'$ which involves higher order polynomials, it remains an easy exercise to evaluate the integrals by Barnes lemma, simply by shifting the arguments of the Gamma functions, e.g. writing $s_{15}\, \Gamma(s_{15})=\Gamma(s_{15}+1)$, etc\,. \\ \\ 
A slightly harder case is when $M(\{s_{ij}\})$ has a pole in either $s_{34}$ or $s_{15}$. This happens for the following caeses:
\begin{itemize}
    \item Single-exchange diagrams $M^{\text{14-235}}_{\Delta}$, $M^{\text{15-234}}_{\Delta}$, $M^{\text{24-135}}_{\Delta}$,  $M^{\text{25-134}}_{\Delta}$, $M^{\text{34-125}}_{\Delta}$,  $M^{\text{35-124}}_{\Delta}$\,.
    \item Double-exchange diagrams $M^{\text{12-5-34}}_{\Delta_1,\Delta_2}$, $M^{\text{12-4-35}}_{\Delta_1,\Delta_2}$, $M^{\text{13-4-25}}_{\Delta_1,\Delta_2}$, $M^{\text{13-5-24}}_{\Delta_1,\Delta_2}$, $M^{\text{23-4-15}}_{\Delta_1,\Delta_2}$, $M^{\text{23-5-14}}_{\Delta_1,\Delta_2}$.
\end{itemize}
Consider that there is a pole in $(s_{34}+\frac{\Delta}{2}-\Delta_\phi+p)^{-1}$ with residue $A_p$ (the form of $A_p$ depends on the diagram we have), followed by a sum over $p$. We can carry out the $s_{15}$ integral using Barnes lemma, and the $s_{34}$ integral with the same technique as in $t$-channel Witten diagram for four points (see \eqref{tchannellog}). So, for the $m'=n'=0$ case, for instance, one has for the coefficient of $\chi_1^{2\Delta_\phi}\chi_2^{3\Delta_\phi}$
\begin{align}
&  \,\Gamma \big(\mbox{$-\frac{\Delta _{\phi }}{2}$}\big)\int \,  \text{d}s_{34}\, \text{d}s_{15}\,\left[\sum_{p=0}^{\infty}\frac{A_p}{s_{34}+\frac{\Delta}{2}-\Delta_\phi+p}\right]\,\big(\Gamma \left(s_{15}\right) \Gamma \left(s_{34}\right)\cdots\big)\nonumber\\
    &=\int \,  \text{d}s_{34}\frac{\Gamma \left(s_{34}\right) \Gamma \big(-\frac{\Delta _{\phi }}{2}\big) \Gamma \big(\Delta _{\phi }\big)^2 \Gamma \big(s_{34}+\frac{\Delta _{\phi }}{2}\big) \Gamma \left(\Delta _{\phi }-s_{34}\right) \Gamma \left(\frac{3 \Delta _{\phi }}{2}-s_{34}\right)}{\Gamma \left(2 \Delta _{\phi }\right)}\nonumber\\
&\hspace{8cm}\times\left[\sum_{p=0}^{\infty}A_p\,\int_0^1\,\frac{dy}{y}  y^{s_{34}+\frac{\Delta}{2}-\Delta_\phi+p}\right]\nonumber\\
&=\sum_{p=0}^\infty A_p\,\Gamma \big(\mbox{$-\frac{\Delta _{\phi }}{2}$}\big) \Gamma \big(\Delta _{\phi }\big)^3 \Gamma \big(\mbox{$\frac{3 \Delta _{\phi }}{2}$}\big)^2 \Gamma \big(\mbox{$p+\frac{\Delta }{2}$}\big) \, _3\widetilde{F}_2\big(\mbox{$1,\Delta _{\phi },\frac{3 \Delta _{\phi }}{2};p+\frac{\Delta }{2}+1,3 \Delta _{\phi };1$}\big)\label{harderintegral}\,.
\end{align}
In the last line we evaluated the $s_{34}$ integral which gives a ${}_2F_1$ function with $(1-y)$ argument - and expanded the ${}_2F_1$ in the powers $(1-y)^q$ to evaluate the $y$ integral, and finally summed over $q$ to obtain the ${}_3\widetilde{F}_2$. 
Once again the higher polynomials are obtained by a simple generalization of the above treatment. \\ \\
The hardest case in evaluating \eqref{2phi3phicoeffgen} is that where the $M(\{s_{ij}\})$ has poles in both $s_{34}$ and $s_{15}$. This happens for 
\begin{itemize}
    \item Double-exchange diagrams $M^{\text{14-3-25}}_{\Delta_1,\Delta_2}$, $M^{\text{15-3-24}}_{\Delta_1,\Delta_2}$, $M^{\text{14-2-35}}_{\Delta_1,\Delta_2}$, $M^{\text{15-2-34}}_{\Delta_1,\Delta_2}$, $M^{\text{24-1-35}}_{\Delta_1,\Delta_2}$,  $M^{\text{25-1-34}}_{\Delta_1,\Delta_2}$.
\end{itemize}
Consider the following pole structure: $\mathcal{Q}_{kl}(s_{34}+\frac{\Delta_1}{2}-\Delta_\phi+k)^{-1}(s_{15}+\frac{\Delta_2}{2}-\Delta_\phi+l)^{-1}$ followed by sums over $k,l$. It is easy to see that this is the $M^{\text{15-2-34}}$ diagram. The other diagrams can be brought to the same integral form by shifting the Mellin variables.
Let us carry out the $s_{15}$ integral first. We may convert the $s_{15}$ pole to a $y_2$-integral. Then for the $m'=n'=0$ term we get the coefficient
\begin{align}
    &\int \, \text{d}s_{15}\,\Gamma \left(s_{15}\right) \Gamma \big(\Delta _{\phi }-s_{15}\big) \Gamma \big(\mbox{$ s_{15}- s_{34}+\frac{\Delta _{\phi }}{2} $}\big) \Gamma \big(\mbox{$s_{34}- s_{15}+\frac{\Delta_\phi}{2}$} \big)\int_0^1 \,\frac{dy_2}{y_2}\,y_2^{s_{15}+\frac{\Delta_2}{2}-\Delta_\phi+l}\nonumber\\
    &=\int_0^1\frac{dy_2}{y_2}\mbox{$\Gamma \big(\Delta _{\phi }\big){}^2 \Gamma \big(s_{34}+\frac{\Delta _{\phi }}{2}\big) \Gamma \big(\frac{3 \Delta _{\phi }}{2}-s_{34}\big)$} \, _2\widetilde{F}_1\big(\mbox{$s_{34}+\frac{\Delta _{\phi }}{2},\Delta _{\phi };2 \Delta _{\phi };1-\frac{1}{y_2}$}\big) y_2^{\frac{\Delta _2}{2}-\Delta _{\phi }+l}\nonumber\\
    &=\sum_{p=0}^{\infty}\frac{\mbox{$\Gamma \big(\Delta _{\phi }\big)^2 \Gamma \big(s_{34}+\frac{\Delta _{\phi }}{2}\big) \Gamma \big(\frac{3 \Delta _{\phi }}{2}-s_{34}+p\big)$}\, (\Delta_\phi)_p\, \Gamma \big(l+\frac{\Delta_2}{2}\big)}{\Gamma(2\Delta_\phi+p)\Gamma \left(l+p+\frac{\Delta_2}{2}+1\right)}\label{hardestintegral1}
\end{align}
In the last line we expanded the $\, _2\widetilde{F}_1$ and then integrated the $p$-th expansion term over $y_2$. The $s_{34}$ integral can then be carried out. If we write the $s_{34}$ pole as a $y_1$ integral then we get (isolating only the $s_{34}$ dependent terms)
\begin{align}
    &\int \, \text{d}s_{34}\,\Gamma \left(s_{34}\right)\Gamma \mbox{$\big(s_{34}+\frac{\Delta _{\phi }}{2}\big) \Gamma \big(\frac{3 \Delta _{\phi }}{2}-s_{34}+p\big)$}\Gamma(\Delta_\phi-s_{34})\,\int \frac{dy_1}{y_1}\,y_1^{s_{34}+\frac{\Delta_1}{2}-\Delta_\phi+k}\nonumber\\
    &=\mbox{$\Gamma \big(\Delta _{\phi }\big) \Gamma \big(\frac{3 \Delta _{\phi }}{2}\big) \Gamma \big(k+\frac{\Delta _1}{2}\big) \Gamma \big(p+\frac{3 \Delta _{\phi }}{2}\big) \,\Gamma \big(p+2 \Delta _{\phi }\big) $} \,\nonumber\\
    & \hspace{2cm} \times{}_3\widetilde{F}_2\big(\mbox{$1,\Delta _{\phi },\frac{3 \Delta _{\phi }}{2};\frac{\Delta _1}{2}+1+k,p+3 \Delta _{\phi };1$}\big)\label{hardestintegral2}\,.
\end{align}
The steps leading to \eqref{hardestintegral2} is similar to what we used for \eqref{harderintegral}. The final answer  hence combines \eqref{hardestintegral1} and \eqref{hardestintegral2} and gives
\begin{align}
    &\sum_{p,k,l=0}^{\infty}\frac{\mbox{$\Gamma \big(\Delta _{\phi }\big)^3 $}\, (\Delta_\phi)_p\, \Gamma \big(l+\frac{\Delta_2}{2}\big)\mbox{$ \Gamma \big(\frac{3 \Delta _{\phi }}{2}\big) \Gamma \big(k+\frac{\Delta _1}{2}\big) \Gamma \big(p+\frac{3 \Delta _{\phi }}{2}\big) \,\Gamma \big(p+2 \Delta _{\phi }\big) $}}{\Gamma(2\Delta_\phi+p)\Gamma \left(l+p+\frac{\Delta_2}{2}+1\right)}\nonumber\\
    &\hspace{1cm}\times \mathcal{Q}_{kl}(\Delta_1,\Delta_2)\,\Gamma(-\mbox{$\frac{\Delta_\phi}{2}$})\,{}_3\widetilde{F}_2\big(\mbox{$1,\Delta _{\phi },\frac{3 \Delta _{\phi }}{2};\frac{\Delta _1}{2}+1+k,p+3 \Delta _{\phi };1$}\big)\,.\label{hardestintegral}
\end{align}
The coefficient of $\chi_1^{2\Delta_\phi}\chi_2^{3\Delta_\phi}$, from $M^{\text{15-2-34}}$ and related diagrams, is hence three nested infinite sums. It can be evaluated numerically by truncation. These nested sums also feature in the higher powers of the same family. 
We observed that for higher powers the truncation needs to be at higher values for same numerical accuracy.

\subsubsection*{Coefficients of $G_{\Delta_{2\phi,n_1},\Delta_{2\phi,n_2}}$ } The $G_{\Delta_{2\phi,n_1},\Delta_{2\phi,n_2}}$ block expands into the power laws $\chi_1^{2\Delta_\phi+m'}\chi_2^{2\Delta_\phi+n'}$ for $m',n'\in \mathbb{Z}_{\ge 0}$. Looking at \eqref{InvMellin5point} it is clear that these powers follow from the residues of the poles
\begin{equation}
    s_{12}=-m, \ s_{45}=-n\,, \ \forall m,n\in \mathbb{Z}_{\ge 0}\,,\label{2phi2phipoles}
\end{equation}
from the $\Gamma(s_{12})\Gamma(s_{45})$ in the measure. \\ \\
The kinematic part \eqref{kinematic} then becomes 
\begin{align}
    &\frac{\left(1-\chi_1-\chi_2\right)^{-2 s_{15}}  \chi_1^{2 \Delta _{\phi }+2 m}  \chi_2^{2 \Delta _{\phi }+2 n}M(-m,s_{23},s_{34},-n,s_{15})}{\left(1-\chi_1\right)^{\Delta_{\phi }-2 s_{15}+2 s_{23}+2 n}\left(1-\chi_2\right)^{\Delta _{\phi }+2 m-2 s_{15}+2 s_{34}}}\nonumber\\
&=\chi_1^{2 \Delta _{\phi }+2 m}  \chi_2^{2 \Delta _{\phi }+2 n}\,M(-m,s_{23},s_{34},-n,s_{15})\sum_{p,q} \,\chi_1^{p}\,\chi_2^{q}\,N_{p,q}(s_{23},s_{34},s_{15})\label{kinematicalexpand}\,,
\end{align}
where we further expanded $(1-\chi_1-\chi_2)$, $(1-\chi_1)$ and $(1-\chi_2)$ powers in $\chi_1$ and $\chi_2$. Each coefficient $N_{p,q}$ appearing here is a polynomial in $s_{23}, s_{34}, s_{15}$ and we have to carry out the respective integrals. We can then collect the powers of $\chi_1$ and $\chi_2$ to write
\begin{equation}
\chi_1^{2\Delta_\phi+m'}\chi_2^{2\Delta_\phi+n'}\int \, \text{d}s_{23}\, \text{d}s_{34}\, \text{d}s_{15}\,\prod_{\stackrel{(ij)\neq}{12,\,45}}\big(\Gamma(s_{ij})\,\big)\, \widetilde{M}(s_{23},s_{34},s_{15})\label{2phi2phicoeffgen}
\end{equation}
where
\begin{equation}
    \widetilde{M}(s_{23},s_{34},s_{15})=\sum_{(p,q)=(0,0)}^{(m',n')}M\big(\,\frac{p-m'}{2},s_{23},s_{34},\frac{q-n'}{2},s_{15}\,\big)\,N_{p,q}(s_{23},s_{34},s_{15})\,.
\end{equation}
The eqn. \eqref{2phi2phicoeffgen} is more involved than \eqref{2phi3phicoeffgen} since there are three integrals instead of two.
But the difficulty in evaluating \eqref{2phi2phicoeffgen} also depends on the form of Mellin amplitude $M(\{s_{ij}\})$ i.e. what diagram we have at hand.  \\ \\
Once again it is easiest when $M\big(\,\frac{p-m'}{2},s_{23},s_{34},\frac{q-n'}{2},s_{15}\,\big)$ has no poles in $s_{23}, s_{34}, s_{15}$. This is the case when we have 
\begin{itemize}
    \item Contact diagram, $M^{\text{(con)}}=1$.
    \item Single-exchange diagrams $M^{\text{12-345}}_{\Delta}$,  $M^{\text{45-123}}_{\Delta}$.
    \item Double-exchange diagram $M^{\text{12-3-45}}_{\Delta_1,\Delta_2}$.
\end{itemize}
For hese  diagrams the integrals can be evaluated using Barnes lemma three times. The procedure mimics \eqref{easyintegral} and we will not repeat it here. \\ \\
The integral is harder when $M\big(\,\frac{p-m'}{2},s_{23},s_{34},\frac{q-n'}{2},s_{15}\,\big)$ has one pole in $s_{23}, s_{34}$ or $s_{15}$. This is the case when we have
\begin{itemize}
    \item Single-exchange diagrams $M^{\text{13-245}}_{\Delta}$,  $M^{\text{23-145}}_{\Delta}$,  $M^{\text{14-235}}_{\Delta}$,  $M^{\text{15-234}}_{\Delta}$,   $M^{\text{24-135}}_{\Delta}$,  $M^{\text{25-134}}_{\Delta}$, $M^{\text{34-125}}_{\Delta}$,  $M^{\text{35-124}}_{\Delta}$.
    \item Double-exchange diagram $M^{\text{13-2-45}}_{\Delta_1,\Delta_2}$, $M^{\text{23-1-45}}_{\Delta_1,\Delta_2}$, $M^{\text{12-4-35}}_{\Delta_1,\Delta_2}$, $M^{\text{12-5-34}}_{\Delta_1,\Delta_2}$.
\end{itemize}
In the above cases two of the Mellin integrals can be evaulated using Barnes lemma, and for the one with the pole has to be done using the treatment similar to \eqref{harderintegral}. \\ \\
Our integral \eqref{2phi2phicoeffgen} is even harder when $M\big(\,\frac{p-m'}{2},s_{23},s_{34},\frac{q-n'}{2},s_{15}\,\big)$ has two simple poles. But now there are two cases to distinguish
\begin{itemize}
    \item Double-exchange diagrams $M^{\text{14-2-35}}_{\Delta_1,\Delta_2}$, $M^{\text{15-2-34}}_{\Delta_1,\Delta_2}$, $M^{\text{24-1-35}}_{\Delta_1,\Delta_2}$, $M^{\text{25-1-34}}_{\Delta_1,\Delta_2}$.
    \item Double-exchange diagram $M^{\text{14-3-25}}_{\Delta_1,\Delta_2}$, $M^{\text{15-3-24}}_{\Delta_1,\Delta_2}$.
\end{itemize}
For the first set of four diagrams, it is possible carry out one of the three Mellin integrals using Barnes lemma, as the respective variable will have no pole, e.g. with $M^{\text{15-2-34}}_{\Delta_1,\Delta_2}$ this will be the $s_{23}$ variable integral.  The other two integrals will have a pair of simple poles, and we have to follow the steps shown in \eqref{hardestintegral1} and \eqref{hardestintegral2}.\\ \\
For the remaining two double-exchange diagrams, it is not possible to carry out any of the Mellin integrals using Barnes lemma, as all the variables will feature in one or the other pole. We can still try to evaluate them with steps similar to \eqref{hardestintegral1} and \eqref{hardestintegral2}. However the resulting expression is either a set of very poorly convergent nested infinite sums, or infinite sums where with a $y$-integral (from 0 to 1) at every summand. In both cases it is not feasible to use the expression efficiently for numerical evaluation. So, for these two diagrams, we have to use the method of App. \ref{app:integratedvertex} .

\subsubsection*{Coefficients of $G_{\Delta_1,\Delta_{2\phi,n}}$ and $G_{\Delta_1,\Delta_{1,\phi,n}}$}
We will discuss the coefficients of these blocks together as they both feature in the same four diagrams: $M^{\text{12-3-45}}_{\Delta_1,\Delta_2}$, $M^{\text{12-4-35}}_{\Delta_1,\Delta_2}$, $M^{\text{12-5-34}}_{\Delta_1,\Delta_2}$ and $M^{\text{12-345}}_{\Delta_1}$. \\ \\
The block $G_{\Delta_1,\Delta_{2\phi,n}}$ involves the terms $\chi_1^{\Delta_1+m'}\chi_2^{2\Delta_\phi+n'}$\,. They are obtained from the following poles 
\begin{equation}
    s_{12}=-\frac{\Delta_1}{2}+\Delta_\phi-m, \ s_{45}=-n\,, \ \forall m, n\in \mathbb{Z}_{\ge 0}\,.\label{Del12phipoles}
\end{equation}
We will expand the kinematical part in Mellin integral just like  \eqref{kinematicalexpand} and collect the necessary powers of $\chi_1$ and $\chi_2$ to write
\begin{equation}
\chi_1^{\Delta_1+m'}\chi_2^{2\Delta_\phi+n'}\int \, \text{d}s_{23}\, \text{d}s_{34}\, \text{d}s_{15}\,\prod_{\stackrel{(ij)\neq}{12,\,45}}\big(\Gamma(s_{ij})\,\big)\, \widetilde{M}(s_{23},s_{34},s_{15})\label{Del12phicoeffgen}
\end{equation}
where
\begin{equation}
    \widetilde{M}(s_{23},s_{34},s_{15})=\sum_{(p,q)=(0,0)}^{(m',n')}M\mbox{$\big(-\frac{\Delta}{2}+\Delta_\phi+\frac{p-m'}{2},s_{23},s_{34},\frac{q-n'}{2},s_{15}\big)$}\,N_{p,q}(s_{23},s_{34},s_{15})\,.
\end{equation}
The coefficient $N_{p,q}$ is defined similar to \eqref{kinematicalexpand}.
It is easy to evaluate for $M^{\text{12-3-45}}_{\Delta_1,\Delta_2}$ and $M^{\text{12-345}}_{\Delta_1}$ as there is no pole in $s_{23}, s_{34}$ or $s_{45}$ from $M(\{s_{ij}\})$.  We can use Barnes lemma three times to evaluate it. For $M^{\text{12-4-35}}_{\Delta_1,\Delta_2}$ and $M^{\text{12-5-34}}_{\Delta_1,\Delta_2}$ there is a pole one of the three variables. We have already seen how to handle such integrals in \eqref{harderintegral} and we proceed as shown there to evaluate it. \\ \\ 
The block $G_{\Delta_1,\Delta_{1,\phi,n}}$ will follow from the terms $\chi_1^{\Delta_1+m'}\chi_2^{\Delta_1+\Delta_\phi+n'}$\,. They are obtained from the following poles 
\begin{equation}
    s_{12}=-\frac{\Delta_1}{2}+\Delta_\phi-m, \ s_{45}=\frac{\Delta_\phi-\Delta_1}{2}-m-n-p\,, s_{23}=-p\,, \ \forall m, n\in \mathbb{Z}_{\ge 0}\,.\label{Del1Del1phipoles}
\end{equation}
This results in the powers $\chi_1^{\Delta_1+2m}\chi_2^{\Delta_1+\Delta_\phi+2m+2n+2p}$.
The kinematical part in Mellin integral should be expanded as in \eqref{kinematicalexpand1}. Collecting the appropriate powers of $\chi_1,\chi_2$ one gets
\begin{equation}
\chi_1^{\Delta_1+m'}\chi_2^{\Delta_1+\Delta_\phi+n'}\int \, \text{d}s_{34}\, \text{d}s_{15}\,\prod_{\stackrel{(ij)\neq}{12,\,45}}\big(\Gamma(s_{ij})\,\big)\, \widetilde{M}(s_{34},s_{15})\label{Del1Del1phicoeffgen}
\end{equation}
where
\begin{equation}
    \widetilde{M}(s_{34},s_{15})=\sum_{(q,r)=(0,0)}^{(m',n')}M\mbox{$\big(\Delta_\phi-\frac{\Delta}{2}+\frac{q-m'}{2},m+n+\frac{r-n'}{2},s_{34},\frac{r-n'+\Delta_\phi-\Delta_1}{2},s_{15}\big)$}\,N_{q,r}(s_{34},s_{15})\,.
\end{equation}
Here the definition of $N_{q,r}$ is analogous to \eqref{kinematicalexpand1}. Once again the integral is simple for $M^{\text{12-3-45}}_{\Delta_1,\Delta_2}$ and $M^{\text{12-345}}_{\Delta_1}$ (no pole in $s_{34}$ or  $s_{15}$) and a bit harder for  $M^{\text{12-4-35}}_{\Delta_1,\Delta_2}$ and $M^{\text{12-5-34}}_{\Delta_1,\Delta_2}$ (one pole). As appropriate, we follow the steps of \eqref{easyintegral} and \eqref{harderintegral} respectively.

\section{Spectral decomposition of five-point Witten diagrams}
\label{app:spectral}
In this appendix, we review the spectral approach to Witten diagrams which is very useful to derive the conformal block expansion of these objects, particularly in certain OPE channels which do not require complicated crossing moves \cite{Zhou:2018sfz,Jepsen:2019svc}. This will serve as a useful crosscheck to the Mellin techniques, which, while more versatile, often lead to more complicated expressions. 

This formalism attempts to use the conformal structure of three-point integrals to trade complicated AdS position space integrals by spectral integrals that can be evaluated in the end of the computation by picking residues. There are two key identities in this process: the spectral representation of the scalar bulk-to-bulk propagator
\begin{equation}
    G_{BB}(\zeta)=\int d \nu\, \rho_\Delta(\nu)\,\Omega_\nu(\zeta)\,,
\end{equation}
which is key for exchange diagrams, and the spectral representation of the Dirac delta function
\begin{equation}
   \delta(\zeta)=\int d \nu\, \rho_\delta(\nu)\,\Omega_\nu(\zeta)\,,
\end{equation}
which is needed both for single-exchange and contact diagrams. Here, we introduced the AdS$_{d+1}$ harmonic function $\Omega_\nu(\zeta)$, which is an eigenfunction of the AdS laplacian, satisfying
\begin{equation}
    \square_{\rm{AdS} } \, \Omega_\nu(\zeta)= -\left(\nu^2+\frac{d^2}{4}\right) \Omega_\nu(\zeta)\,,
\end{equation}
as well as the spectral functions
\begin{equation}
    \rho_\delta(\nu)= \frac{\Gamma(\frac{d}{2}+i\nu)\Gamma(\frac{d}{2}-i\nu)}{2 \pi^d \Gamma(-i\nu)\Gamma(i\nu)}\,, \quad \rho_\Delta= \frac{1}{\nu^2+ (\Delta-d/2)^2}\,.
\end{equation}
These relations are supplemented by the so-called  `split representation' of the Harmonic function
\begin{equation}
    \Omega_\nu(\zeta(x_1,y_1,x_2,y_2))= \int d^dx K_{d/2+i\nu}(x_1,y_1;x) K_{d/2-i\nu}(x_2,y_2;x)\,,
\end{equation}
which allows one to trade integrals with bulk point dependence by integrals with only three-boundary points, which can always be performed with the help of the mythical AdS 3-point integrals
\begin{equation}
\int \sqrt{g}\, d^{d+1}y\, K_{\Delta_1}(y;x_1)K_{\Delta_2}(y;x_2)K_{\Delta_3}(y;x_3) = \frac{\lambda_{\Delta_1,\Delta_2,\Delta_3}}{x_{12}^{\Delta_1+\Delta_2-\Delta_3}x_{13}^{\Delta_1+\Delta_3-\Delta_2}x_{23}^{\Delta_2+\Delta_3-\Delta_1}}\,,
\end{equation}
where here and henceforth we simply use the shorthand notation $y_i$ and $x_i$ to denote a bulk and boundary point, respectively. We have also defined the bulk-to-boundary propagator in Poincar\'e coordinates
\begin{equation}
     K_{\Delta_i}(y;x_i)= \left(\frac{y}{y^2+(x-x_i)^2}\right)^{\Delta_i}\,,
\end{equation}
as well as the AdS 3-point factor
\begin{equation}
   \lambda_{\Delta_1,\Delta_2,\Delta_3}= \frac{\pi^{d/2} \Gamma(\frac{\Delta_1+\Delta_2-\Delta_3}{2})\Gamma(\frac{\Delta_2+\Delta_3-\Delta_1}{2})\Gamma(\frac{\Delta_1+\Delta_3-\Delta_2}{2})\Gamma(\frac{\Delta_1+\Delta_2+\Delta_3-d}{2})}{2 \Gamma(\Delta_1)\Gamma(\Delta_2)\Gamma(\Delta_3)} \,.
\end{equation}
We will also use the more compact notation
\begin{equation}
    \langle \mathcal{O}_1(x_1)\mathcal{O}_2(x_2)\mathcal{O}_3(x_3)\rangle \equiv \frac{1}{x_{12}^{\Delta_1+\Delta_2-\Delta_3}x_{13}^{\Delta_1+\Delta_3-\Delta_2}x_{23}^{\Delta_2+\Delta_3-\Delta_1}}\,,
\end{equation}
for readability in intermediate steps of the calculations.
Let us see now see this machinery in action.
\paragraph{Contact diagram}
The simplest contact diagrams, which have non-derivative couplings of $\phi^5$ type, are given by
\begin{equation}
    G_{\rm{ctc}}(x_i)= \int d^{d+1}y \prod_{i=1}^5 K_{\Delta_i}(y;x_i)= \tikz[baseline=0ex]{
  \draw[thick] (0,0) circle (1);
  \coordinate (C) at ($(0,0)$);
  \foreach \angle in {30,90,150,-30,210} {
    \fill[black] (\angle:1.);
    circle(0.03); }
  \draw[thick] (30:1) -- (C);
  \draw[thick] (90:1) -- (C);
  \draw[thick] (150:1) -- (C);
  \draw[thick] (-30:1) -- (C);
  \draw[thick] (210:1) -- (C);
}\,,
\end{equation}
we then introduce two auxiliary integration points which we connect to the original integration point with Dirac delta functions (denoted diagramatically by a dashed line)
\begin{align}
    &\int d^{d+1}y \,d^{d+1}y'\,d^{d+1}y'' \left(K_{\Delta_1}(y',x_1) K_{\Delta_2}(y',x_2)\delta^{d+1}(y-y')K_{\Delta_3}(y,x_3)\right.  \nonumber\\ &\left.\delta^{d+1}(y-y'') K_{\Delta_4}(y'',x_4) K_{\Delta_5}(y'',x_5)\right)
    =
\tikz[baseline=0ex]{\draw[thick] (0,0) circle(1.);
\foreach \angle in {200, 160, 90, 20, -20} {
    \fill[black] (\angle:1.) circle(0.03);
}
\coordinate (L) at ($(0,0) + (-0.5,0)$);
\coordinate (R) at ($(0,0) + (0.5,0)$);
\coordinate (C) at ($(L)!0.5!(R)$);
\draw[dashed] (L) -- (C) ;
\draw[dashed] (C) -- (R);
\draw[thick] (160:1.) -- (L);
\draw[thick] (200:1.) -- (L);
\draw[thick] (20:1.) -- (R);
\draw[thick] (-20:1.) -- (R);
\draw[thick] (90:1.) -- (C);
}\,,
\end{align}
in a way that reflects our chosen OPE channel 12-3-45. This also makes it clear that certain diagrams will be harder to express in a given channel as they will require crossing moves.
We subsequently use the spectral-split representation for the Dirac deltas, yielding
\begin{align}
        &\int d^{d+1}y \,d^{d+1}y'\,d^{d+1}y''  d^dw d^dw' d\nu_1 d\nu_2 \, \rho_\delta(\nu_1) \rho_\delta(\nu_2)\left(K_{\Delta_1}(y',x_1) K_{\Delta_2}(y',x_2)K_{d/2+i \nu_1}(y',w)\right.  \nonumber\\
      & \left. K_{d/2-i \nu_1}(y,w)K_{\Delta_3}(y,x_3)  K_{d/2+i \nu_2}(y,w')K_{d/2-i \nu_2}(y'',w')K_{\Delta_4}(y'',x_4) K_{\Delta_5}(y'',x_5)\right) = \nonumber\\
      & \qquad\qquad\qquad\qquad\qquad\qquad\qquad\qquad \tikz[baseline=0ex]{\draw[thick] (0,0) circle(1.);
\foreach \angle in {200, 160, 90, 20, -20} {
    \fill[black] (\angle:1.) circle(0.03);
}
\coordinate (L) at ($(0,0) + (-0.5,0)$);
\coordinate (R) at ($(0,0) + (0.5,0)$);
\coordinate (C) at ($(L)!0.5!(R)$);
\draw[dashed] (L) -- (125:1.) ;
\draw[dashed] (125:1.) -- (C);
\draw[dashed] (R) -- (55:1.) ;
\draw[dashed] (55:1.) -- (C);
\draw[thick] (160:1.) -- (L);
\draw[thick] (200:1.) -- (L);
\draw[thick] (20:1.) -- (R);
\draw[thick] (-20:1.) -- (R);
\draw[thick] (90:1.) -- (C);
}\,.
\end{align}
At this stage, all the AdS integrals are standard three-point integrals which can readily be performed yielding
\begin{align}
    &G_{\rm{ctc}}=\int d\nu_1 d\nu_2 \, \rho_\delta(\nu_1)\rho_\delta(\nu_2) a_{\Delta_1,\Delta_2,d/2+i\nu_1}a_{d/2-i\nu_1,\Delta_3,d/2+i\nu_2}a_{d/2-i\nu_2,\Delta_4,\Delta_5} \Big[ \int dw dw'  \\
    &\langle \mathcal{O}_1(x_1)\mathcal{O}_2(x_2)\mathcal{O}_{d/2+i \nu_1}(w) \rangle \langle \mathcal{O}_{d/2-i \nu_1}(w)\mathcal{O}_3(x_3)\mathcal{O}_{d/2+i \nu_2}(w') \rangle \langle \mathcal{O}_{d/2-i \nu_2}(w') \mathcal{O}_4(x_4)\mathcal{O}_5(x_5) \rangle \Big]\,, \nonumber
\end{align}
where in the second line we identify the 5-point conformal partial wave (CPW)
\begin{equation}
    \Psi_{\nu_1,\nu_2}^{12,3,45}(x_i)=\int dw dw' \langle \mathcal{O}_1\mathcal{O}_2\mathcal{O}_{d/2+i \nu_1}(w) \rangle \langle \mathcal{O}_{d/2-i \nu_1}(w)\mathcal{O}_3\mathcal{O}_{d/2+i \nu_2}(w') \rangle \langle \mathcal{O}_{d/2-i \nu_2}(w') \mathcal{O}_4\mathcal{O}_5
    \rangle\,,
\end{equation}
and can therefore compactly write the spectral representation of the 5-point contact diagram
\begin{equation}
    G_{\rm{ctc}}(x_i)= \int d\nu_1 d\nu_2 \, \rho_{\rm{ctc}}(\nu_1,\nu_2)     \Psi_{\nu_1,\nu_2}^{12,3,45}(x_i)\,,
\end{equation}
with
\begin{equation}
    \rho_{\rm{ctc}}(\nu_1,\nu_2) =  \rho_\delta(\nu_1)\rho_\delta(\nu_2) a_{\Delta_1,\Delta_2,d/2+i\nu_1}a_{d/2-i\nu_1,\Delta_3,d/2+i\nu_2}a_{d/2-i\nu_2,\Delta_4,\Delta_5} \,.
\end{equation}
The spectral representation is related to the conformal block decomposition by deformation of the spectral contour on the principal series. Indeed, physical operators have real scaling dimensions which are captured by the poles of the spectral function $\rho(\nu_1,\nu_2)$. However, the partial wave is not a pure conformal block, it is instead a linear combination of conformal blocks all four possible combination of their shadows ($\nu_i\to-\nu_i$) weighted by the so-called shadow coefficients. See for example \cite{Meltzer:2019nbs,Antunes:2021kmm} for a detailed discussion. 

Once the dust settles, the OPE coefficients are simply obtained as residues of the spectral function
\begin{equation}
\label{Residues}
    a_{\Delta_L,\Delta_R}= 4\,\textrm{Res}_{\nu_2\to \Delta_R}\textrm{Res}_{\nu_1\to \Delta_L} K_{d/2-i \nu_1}^{\Delta_3,d/2-i\nu_2} K_{d/2-i \nu_2}^{\Delta_3,d/2+i\nu_1}  \rho(\nu_1,\nu_2)\,,
\end{equation}
where $K$ are the shadow factors. With this expression in hand, we can analyze the pole structure and detect what families of operators are exchanged in this correlation function. There are three families of exchanges
\begin{itemize}
    \item $\Delta_L=\Delta_1+\Delta_2+2n_L\,,\Delta_R=\Delta_4+\Delta_5+2n_R\,,$ `double-trace, double-trace'
     \item $\Delta_L=\Delta_1+\Delta_2+2n_L\,,\Delta_R=\Delta_1+\Delta_2+\Delta_3+2n_L+2m\,,$ `double-trace, triple-trace'
    \item $\Delta_L=\Delta_3+\Delta_4+\Delta_5+2n_R+2m\,,\Delta_R=\Delta_4+\Delta_5+2n_R\,,$ `triple-trace, double-trace'
\end{itemize}
corresponding exactly to the two most interesting families of functionals discussed in the main text. The respective OPE coefficients are obtained precisely by computing the residues in \eqref{Residues} at these values of scaling dimensions.

\paragraph{Single- and double-exchange diagrams (direct channel)}
It is very straightforward to adapt the previous derivation to single- and double-exchange diagrams where the channel matches our choice of OPE 12-3-45. The only difference is that either one or two of the Dirac delta functions are replaced by bulk-to-bulk propagators, and therefore at the level of the spectral decomposition we simply need to perform the replacement $\rho_\delta\to\rho_\Delta$. This simple rule can also be seen in the formalism used in the main text, since the bulk-to-bulk propagator is an eigenfunction of the laplacian operator. Concretely, we have
\begin{align}
    \tikz[baseline=0ex]{\draw[thick] (0,0) circle(1.);
\foreach \angle in {200, 160, 30, 0, -30} {
    \fill[black] (\angle:1.) circle(0.03);
}
\coordinate (L) at ($(0,0) + (-0.5,0)$);
\coordinate (R) at ($(0,0) + (0.5,0)$);
\draw[very thick, blue] (L) -- (R);\node[below] {$\Delta$};
\draw[thick] (160:1.) -- (L);
\draw[thick] (200:1.) -- (L);
\draw[thick] (30:1.) -- (R);
\draw[thick] (-30:1.) -- (R);
\draw[thick] (0:1.) -- (R);
} = G_{\rm{sing, \Delta}} = \int d\nu_1 d\nu_2 \, \rho_{\rm{sing, \Delta}}(\nu_1,\nu_2)     \Psi_{\nu_1,\nu_2}^{12,3,45}(x_i)\,,
\end{align}
with the spectral function
\begin{equation}
    \rho_{\rm{sing, \Delta}}(\nu_1,\nu_2) = \rho_{\Delta}(\nu_1)\rho_\delta(\nu_2) a_{\Delta_1,\Delta_2,d/2+i\nu_1}a_{d/2-i\nu_1,\Delta_3,d/2+i\nu_2}a_{d/2-i\nu_2,\Delta_4,\Delta_5}\,.
\end{equation}
while the change to the spectral function is superficially small, it introduces new families of exchanged operators because of the new pole present in $\rho_{\Delta}$. Apart from the families enumerated for the contact diagram, we also have
\begin{itemize}
    \item $\Delta_L=\Delta\,, \Delta_R=\Delta_4+\Delta_5+2n_R$ `single-trace, double-trace '
    \item $\Delta_L=\Delta\,, \Delta_R=\Delta_3+\Delta+2n_R$ `single-trace, exchange-double-trace '
\end{itemize}
which are related to 4-point functional with $\Delta$ as an external operator, as discussed in the main text. 

Similarly, for the double exchange we have
\begin{align}
    \tikz[baseline=0ex]{\draw[thick] (0,0) circle(1.);
\foreach \angle in {200, 160, 90, 20, -20} {
    \fill[black] (\angle:1.) circle(0.03);
}
\coordinate (L) at ($(0,0) + (-0.5,0)$);
\coordinate (R) at ($(0,0) + (0.5,0)$);
\coordinate (C) at ($(L)!0.5!(R)$);
\draw[very thick, blue] (L) -- (C) ;\node[below] {$\Delta,\Delta'$};
\draw[very thick, red] (C) -- (R);
\draw[thick] (160:1.) -- (L);
\draw[thick] (200:1.) -- (L);
\draw[thick] (20:1.) -- (R);
\draw[thick] (-20:1.) -- (R);
\draw[thick] (90:1.) -- (C);}
 = G_{\rm{doub},\Delta,\Delta'}= \int d\nu_1 d\nu_2 \, \rho_{\rm{doub, \Delta,\Delta'}}(\nu_1,\nu_2)     \Psi_{\nu_1,\nu_2}^{12,3,45}(x_i)\,,
\end{align}
with the spectral function
\begin{equation}
    \rho_{\rm{doub},\Delta,\Delta'}(\nu_1,\nu_2) = \rho_{\Delta}(\nu_1)\rho_{\Delta'}(\nu_2) a_{\Delta_1,\Delta_2,d/2+i\nu_1}a_{d/2-i\nu_1,\Delta_3,d/2+i\nu_2}a_{d/2-i\nu_2,\Delta_4,\Delta_5}\,.
\end{equation}
which again introduces new families of exchanged operators
\begin{itemize}
    \item $\Delta_L=\Delta\,, \Delta_R=\Delta'\,\,$`single-trace, single-trace '
    \item {$\Delta_L=\Delta_1+\Delta_2+2n_L\,,\Delta_R=\Delta'\,\,$ `double-trace, single-trace} 
     \item {$\Delta_L=\Delta_3+\Delta'+2n_L\,,\Delta_R=\Delta'\,\,$ `exchange-double-trace, single-trace'}
\end{itemize}
We conclude by noting that single- and double- exchange diagrams in other channels can also be obtained in this formalism at the cost of introducing crossing-kernels/6j-symbols in the spectral functions. In the main text, we instead use the Mellin formalism, integrated vertex identities, and the equation of motion/Laplacian to derive recursion relations for the OPE coefficients of these diagrams.

\section{Integrated vertex identities}
\label{app:integratedvertex}

    In this appendix, we discuss a standard tool for the evaluation of Witten diagrams: Integrated vertex identities.
    Concretely, we first give the scalar integrated vertex identity in arbitrary $d$ (App.~\ref{app:scalar_vertex}) and then showcase its usage for the evaluation of $d=1$ exchange Witten diagrams, that are relevant for the construction of 5-point Polyakov blocks discussed in this paper (App.~\ref{app:SE_5}).
    Finally, we describe a notebook shared in the ancillary files provided on the arxiv page of this publication, which contains further integrated vertex identity results on 1d 5-point Witten diagrams (App.~\ref{app:notebook}).
    
    \subsection{Scalar integrated vertex identity}\label{app:scalar_vertex}
        Integrated vertex identities allow us to express AdS$_2$ vertices involving bulk-to-bulk and bulk-to-boundary propagators integrated over a common point in AdS as a sum of bulk-to-boundary propagators. 
        The concrete example discussed in this appendix is the integrated vertex identity expressing the integral
        \begin{equation}
            I(y';x_1,x_2)=\int d^{d+1}y K_{\Delta_1}(y;x_1)K_{\Delta_2}(y;x_2)G_{BB}^\Delta(y,y')
        \end{equation}
        as the sum
        \begin{align}\label{ividentity}
              I(y';x_1,x_2)=&\sum_{i=0}^\infty (x_{12})^{2i} T_i(\Delta,\Delta_1,\Delta_2)K_{\Delta_1+1}(y';x_1)K_{\Delta_2+1}(y';x_2) \\
             +&\sum_{i=0}^\infty (x_{12})^{\Delta-\Delta_1-\Delta_2+2i} Q_i(\Delta,\Delta_1,\Delta_2) K_{\frac{\Delta+\Delta_1-\Delta_2}{2}+i}(y',x_1)K_{\frac{\Delta-\Delta_1+\Delta_2}{2}+i}(y',x_2) \nonumber \;,
          \end{align}
          where
          \begin{equation}\label{eq:def_T}
            T_i(\Delta,\Delta_1,\Delta_2)=\frac{(\Delta_1)_i (\Delta_2)_i}{(\Delta -\Delta_{12}) (\Delta -\widetilde\Delta_{12}) \left(\frac{\Delta_{12}-\Delta +2}{2}\right)_i \left(\frac{\Delta_{12}-\widetilde
            \Delta +2}{2}\right)_i}
          \end{equation}
          and
          \begin{equation}\label{eq:def_Q}
            \begin{split}
              Q_i(\Delta,\Delta_1,\Delta_2)={}&\frac{(-1)^i \Gamma(\frac{d-2 i-2\Delta}{2})\sin(\frac{\pi  (d-2 \Delta )}{2})\Gamma(\frac{-d+\Delta +\Delta_1+\Delta_2}{2})\Gamma(\frac{-\Delta -\Delta_1+\Delta_2+2}{2}) }{4 \pi  \Gamma (i+1)\Gamma (\Delta_1) \Gamma (\Delta_2)}\\
              {}&\times \frac{\Gamma (\frac{\Delta -\Delta_1+\Delta_2}{2}) \Gamma (\frac{\Delta +\Delta_1-\Delta_2}{2}) \Gamma(\frac{-\Delta +\Delta_1+\Delta_2}{2})\Gamma(\frac{-\Delta +\Delta_1-\Delta_2+2}{2})  }{\Gamma(\frac{-\Delta +\Delta_1-\Delta_2-2 i+2}{2})\Gamma(\frac{-\Delta -\Delta_1+\Delta_2-2 i+2}{2})}\;.
            \end{split}
          \end{equation}
        For a derivation of this result and similar identities see \cite{DHoker:1999mqo}.
        \subsection{Exchange diagrams}\label{app:SE_5}
        Using Identity \eqref{ividentity}, tree level scalar exchange Witten diagrams can be expressed as sums of contact diagrams, which can then directly be decomposed in to conformal blocks. 
        Concretely, this leads to a decomposition into two infinite sums of contact diagrams
        \begin{align}
            \begin{tikzpicture}[scale=0.5,baseline={([yshift=-.5ex]current bounding box.center)}]
    		\node  (1) at (3, 0) {};
    		\node  (3) at (-3, 0) {};
    		\node  (6) at (0, -3) {};
    		\node  (8) at (0, 3) {};
    		\node  (11) at (2, 2.25) {};
    		\node  (12) at (2, -2.25) {};
    		\node  (13) at (-2.75, 1.25) {};
    		\node  (16) at (-2.75, -1.25) {};
    		\node  (17) at (-1.75, 0) {};
    		\node  (18) at (0.75, 0) {};
    		\node  (19) at (-0.5, 0.5) {$\Delta$};
    		\node  (20) at (-3.5, -1.75) {};
    		\node  (21) at (-3.5, -1.75) {$\Delta_1$};
    		\node  (22) at (-3.5, 1.75) {$\Delta_2$};
    		\node  (23) at (2.75, -2.75) {$\Delta_5$};
    		\node  (24) at (2.75, 2.75) {$\Delta_3$};
    		\node  (25) at (4, 0) {$\Delta_4$};
    		\draw [bend left=45] (3.center) to (8.center);
    		\draw [bend left=45] (8.center) to (1.center);
    		\draw [bend left=45] (1.center) to (6.center);
    		\draw [bend left=45] (6.center) to (3.center);
    		\draw (16.center) to (17.center);
    		\draw (17.center) to (13.center);
    		\draw (17.center) to (18.center);
    		\draw (18.center) to (11.center);
    		\draw (18.center) to (1.center);
    		\draw (18.center) to (12.center);
            \end{tikzpicture}
            =
            \sum_{i=0}^\infty (x_{12})^{2i} &T_i(\Delta,\Delta_1,\Delta_2)
            \begin{tikzpicture}[scale=0.5,baseline={([yshift=-.5ex]current bounding box.center)}]
    		\node  (1) at (3, 0) {};
    		\node  (3) at (-3, 0) {};
    		\node  (6) at (0, -3) {};
    		\node  (8) at (0, 3) {};
    		\node  (11) at (2.75, 1.25) {};
    		\node  (12) at (2, -2.25) {};
    		\node  (17) at (-2.75, 1.25) {};
    		\node  (18) at (0, 0) {};
    		\node  (21) at (-3.25, -2.75) {$\Delta_1+i$};
    		\node  (22) at (-4, 1.75) {$\Delta_2+i$};
    		\node  (23) at (2.75, -2.75) {$\Delta_5$};
    		\node  (24) at (0, 3.75) {$\Delta_3$};
    		\node  (25) at (3.75, 1.75) {$\Delta_4$};
    		\node  (26) at (-2, -2.25) {};
    		\draw [bend left=45] (3.center) to (8.center);
    		\draw [bend left=45] (8.center) to (1.center);
    		\draw [bend left=45] (1.center) to (6.center);
    		\draw [bend left=45] (6.center) to (3.center);
    		\draw (17.center) to (18.center);
    		\draw (18.center) to (11.center);
    		\draw (18.center) to (12.center);
    		\draw (18.center) to (26.center);
    		\draw (18.center) to (8.center);
            \end{tikzpicture} \\
            +
            \sum_{i=0}^\infty (x_{12})^{\Delta-\Delta_1-\Delta_2+2i} &Q_i(\Delta,\Delta_1,\Delta_2)
            \begin{tikzpicture}[scale=0.5,baseline={([yshift=-.5ex]current bounding box.center)}]
    		\node  (1) at (3, 0) {};
    		\node  (3) at (-3, 0) {};
    		\node  (6) at (0, -3) {};
    		\node  (8) at (0, 3) {};
    		\node  (11) at (2.75, 1.25) {};
    		\node  (12) at (2, -2.25) {};
    		\node  (17) at (-2.75, 1.25) {};
    		\node  (18) at (0, 0) {};
    		\node  (21) at (-4, -2.75) {$\frac{\Delta+\Delta_1-\Delta_2}{2}+i$};
    		\node  (22) at (-5, 1.75) {$\frac{\Delta+\Delta_2-\Delta_1}{2}+i$};
    		\node  (23) at (2.75, -2.75) {$\Delta_5$};
    		\node  (24) at (0, 3.75) {$\Delta_3$};
    		\node  (25) at (3.75, 1.75) {$\Delta_4$};
    		\node  (26) at (-2, -2.25) {};
    		\draw [bend left=45] (3.center) to (8.center);
    		\draw [bend left=45] (8.center) to (1.center);
    		\draw [bend left=45] (1.center) to (6.center);
    		\draw [bend left=45] (6.center) to (3.center);
    		\draw (17.center) to (18.center);
    		\draw (18.center) to (11.center);
    		\draw (18.center) to (12.center);
    		\draw (18.center) to (26.center);
    		\draw (18.center) to (8.center);
            \end{tikzpicture} \nonumber
        \end{align}
        for single exchange diagrams and analogously four double infinite sums of contact diagrams for double exchange diagrams.
        
    \subsection{Comments on \texttt{five\_point\_iv\_data.nb}}\label{app:notebook}

        The purpose of this appendix is to guide the reader through the ancillary mathematica note book \texttt{five\_point\_iv\_data.nb}, which contains expressions for coefficients occuring in the decomposition of 5-point single and double exchange diagrams into conformal blocks that were obtained using the decomposition into contact diagrams discussed in the previous two subsections.

       The notebook is structured into the following sections.
       \begin{enumerate}
           \item \textbf{Preliminary Definitions.} This section defines useful functions such as the coefficients $T_i$ and $Q_i$ of eq.~\eqref{eq:def_Q} and \eqref{eq:def_T}.
           \item \textbf{Double Exchange: Raw Coefficients.} The coefficient of the leading blocks in the decomposition of exchange diagrams into 12-3-45 conformal blocks can directly be read off from the decomposition of exchange diagrams into contact diagrams as the normalization of three-point functions obtained by taking OPE limits of the contact diagrams. 
           The second section of the notebook contains these coefficients for the 12-3-45, 12-4-35, 14-2-35 and 14-3-25 double exchange diagrams. All other double exchange diagrams are related to one of these four by relabeling of the external legs.
           \item \textbf{Double Exchange: Sums.} The data collected in the second section of the notebook is then arranged into series expressions in the third section of the notebook, which can be evaluated by truncating the sum at some finite order.
           \item \textbf{Double Exchange: Integrals.}
           The series encountered can be viewed as two variable hypergeometric functions evaluated at 1. 
           The convergence of these series is quite poor which makes simply performing truncated sums unsuitable for a high precision numerical evaluation. 
           Instead, the fourth section of the notebook converts the sums into standard contour integral representations of the relevant hypergeometric functions, which can be evaluated more efficiently. 
           \item \textbf{Double Exchange: Discontinuity.}
           The integral contours can be deformed to further simplify the integrand to its discontinuity before numerical evaluations. 
           This strategy is implemented in the fifth section of the notebook.
           \item \textbf{Double Exchange: The Special Case $\Delta_\phi=1$.}
           The integrals obtained in the previous step simplify considerably in the case $\Delta_\phi = 1$. 
           To exploit this simplification, this case is implemented separately. 
           \item \textbf{Single Exchange: Raw Coefficients.} Finally, the above strategy is repeated for the single exchange diagrams.
           \item \textbf{Single Exchange: Simplified Coefficients.}  
           In the case of single exchange diagrams only single variable hypergeometric functions, which are by default already efficiently implemented mathematica, need to be evaluated. 
           We can rely on those and do not need to resort on rewriting the sums as integrals in this case.
    \end{enumerate}

\section{Recurrence relations for cross-channel Witten diagrams}
\label{app:recrel}
In this appendix, we develop recurrence relations which make use of Casimir operators to determine the conformal block expansions of cross-channel Witten diagrams in terms of certain (families of) seed blocks. These relations were used as cross-checks of the Mellin and integrated vertex methods used above, and can also be used to speed up the determination of functionals with higher values of the labels $n_1,n_2$. A summary of the results (as well as many relations omitted below) is given in the ancillary notebook \texttt{casimir5ptrecrel.nb}.
\subsection{Four-point recap}
We start by reviewing the logic in the case of 4-point exchange Witten diagrams. Using the fact that the bulk-to-bulk propagator is a Green's function of the AdS Laplacian
\begin{equation}
   (\square_{\rm{AdS}_2}-\Delta(\Delta-1)) G_{BB}(y_1,y_2)= \frac{1}{\sqrt{g}}\delta^{(2)}(y_1-y_2) \,,
\end{equation}
as well as AdS isometry/boundary conformal invariance at the boundary-boundary-bulk vertex, we can rewrite the action of the bulk Laplacian in terms of the boundary quadratic Casimir associated to the pair of points attached to the integration vertex. Hence, when acting with the $s$-channel Casimir $\mathcal{C}_{12}$ on an $s$-channel exchange diagram, we have
\begin{equation}
    (\mathcal{C}_{12}-\Delta(\Delta-1)) A^{s-\Delta}(x_i)= A^{\rm{ctc}}(x_i)  \,.
\end{equation}
Now, we can use the $s$-channel OPE and the fact that the $s$-channel conformal blocks are eigenfunctions of the Casimir of the respective channel, to linearly relate the OPE coefficients of the exchange diagram to those of the contact
\begin{equation}
   (C_{n}^{(s-\Delta)})^2 = \kappa  \frac{(C_{n}^{(\rm{ctc})})^2}{\Delta_n(\Delta_n-1)-\Delta(\Delta-1)} \,,
\end{equation}
where $\Delta_n$ denotes the dimension of double-twist operators and we allowed for a constant $\kappa$ to soak up normalizations of the propagators. If instead we consider a cross-channel exchange diagram, say in the $t$-channel, we can use the $t$-channel Casimir $\mathcal{C}_{13}$ to obtain a contact diagram. However, the $s$-channel conformal blocks $G_\Delta$ are not eigen-functions of the $t$-channel Casimir. Instead, when acted by this cross-channel differential operator, they become a linear combination of blocks with integer shifts in the scaling dimension $\Delta-1,\Delta,\Delta+1$, given by
\begin{equation}
   \mathcal{C}_{13} G_\Delta = \mu_-(\Delta) G_{\Delta-1} + \mu_0(\Delta)G_\Delta +\mu_+(\Delta) G_{\Delta-1} \,,
\end{equation}
with the coefficients
\begin{equation}
    \mu_-(\Delta)=-(\Delta-2\Delta_\phi)^2\,, \, \mu_0(\Delta)=(\frac{c_\Delta}{2}-2c_{\Delta_\phi})\,,\, \mu_+(\Delta)=- \frac{\Delta^2(2\Delta_\phi+\Delta-1)^2}{4(\Delta^2-1)}\,,
\end{equation}
where $\Delta_\phi$ is the external dimension and we introduced the shorthand notation $c_\Delta \equiv \Delta(\Delta-1)$. 
Now, matching the conformal block expansions, we can derive a recurrence relation for the OPE coefficients
\begin{equation}
   \mu_-(\Delta_n+1) (C^{t,\Delta}_{\Delta_n+1})^2+\mu_0(\Delta_n) (C^{t,\Delta}_{\Delta_n})^2 +\mu_+(\Delta_n-1) (C^{t,\Delta}_{\Delta_n-1})^2 -c_\Delta(C^{t,\Delta}_{\Delta_n})^2= \kappa (C^{ctc}_{\Delta_n})^2 \,,
\end{equation}
which can be solved to a desired order (or in this case, exactly), up to a seed coefficient, that of the leading double twist.
\subsection{Shift relations for five-point blocks}
To derive recurrence relations for the 5-point double- and single-exchange diagrams we need to understand the action of the different Casimir operators on the 5-point blocks in the 12-3-45 OPE channel. By definition, the block is an eigenfunction of the Casimirs $\mathcal{C}_{12}$ and $\mathcal{C}_{45}$, satisfying the eigenvalue equations
\begin{align}
    \mathcal{C}_{12} G_{\Delta_1,\Delta_2}(\chi_1,\chi_2)&=c_{\Delta_1} G_{\Delta_1,\Delta_2}(\chi_1,\chi_2)\,, \\
    \mathcal{C}_{45} G_{\Delta_1,\Delta_2}(\chi_1,\chi_2)&=c_{\Delta_2} G_{\Delta_1,\Delta_2}(\chi_1,\chi_2) \,.
\end{align}
Instead, when acted by the Casimir $\mathcal{C}_{13}$, it acquires shifts by $-1,0$ and $1$ units in the scaling dimension $\Delta_1$, (which will eventually lead to three term recurrence relations)
\begin{equation}
   \mathcal{C}_{13} G_{\Delta_1,\Delta_2} = \nu_-(\Delta_1,\Delta_2) G_{\Delta_1-1,\Delta_2}+ \nu_0(\Delta_1,\Delta_2) G_{\Delta_1,\Delta_2} + \nu_+(\Delta_1,\Delta_2) G_{\Delta_1+1,\Delta_2}  \,,
\end{equation}
where the coefficients are given by
\begin{align}
   \nu_-(\Delta_1,\Delta_2)&= -(\Delta_1-\Delta_\phi)(\Delta_1-\Delta_2-\Delta_\phi)\,,\, \nu_0(\Delta_1,\Delta_2)=\frac{1}{2}(c_{\Delta_2}-c_{\Delta_1}+3c_{\Delta_\phi})\,, \\ 
   \nu_+(\Delta_1,\Delta_2)&=-\frac{(\Delta_1+\Delta_2-\Delta_\phi)(\Delta_1-\Delta_2-\Delta_\phi)(\Delta_1+\Delta_2+\Delta_\phi-1)(\Delta_1+2\Delta_\phi-1)}{4(\Delta_1^2-1)}\,,\nonumber
\end{align}
with a similar relation for $\mathcal{C}_{23}$, and equivalent ones with $\Delta_1 \leftrightarrow\Delta_2$ for $\mathcal{C}_{34}$ and $\mathcal{C}_{45}$. 
Finally, we can consider Casimir operators with points attached to both of the endpoints of the OPE diagram, i.e. $\mathcal{C}_{14},\mathcal{C}_{15},\mathcal{C}_{24}$ and $\mathcal{C}_{25}$. In this case, there are shifts by $-1,0,1$ units in $\Delta_1,\Delta_2$ in all possible combinations, leading to nine terms, for example
\begin{align}
  & \mathcal{C}_{15} G_{\Delta_1,\Delta_2}= \nu_{--}(\Delta_1,\Delta_2) G_{\Delta_1-1,\Delta_2-1}+ \nu_{0-}(\Delta_1,\Delta_2) G_{\Delta_1,\Delta_2-1} + \nu_{+-}(\Delta_1,\Delta_2) G_{\Delta_1+1,\Delta_2-1}  \nonumber\\
   &+ \nu_{-0}(\Delta_1,\Delta_2) G_{\Delta_1-1,\Delta_2}+ \nu_{00}(\Delta_1,\Delta_2) G_{\Delta_1,\Delta_2} + \nu_{+0}(\Delta_1,\Delta_2) G_{\Delta_1+1,\Delta_2}  \\&+\nu_{-+}(\Delta_1,\Delta_2) G_{\Delta_1-1,\Delta_2+1}+ \nu_{0+}(\Delta_1,\Delta_2+1) G_{\Delta_1,\Delta_2-1} + \nu_{++}(\Delta_1,\Delta_2) G_{\Delta_1+1,\Delta_2+1} \,, \nonumber
\end{align}
where the coefficients are given by
\begin{align}
  &  \nu_{--}(\Delta_1,\Delta_2)=(\Delta_1-2\Delta_\phi)(\Delta_2-2\Delta_\phi)\,,\, \nu_{-0}(\Delta_1,\Delta_2)=\frac{1}{2}(\Delta_1-2\Delta_\phi)(\Delta_1-\Delta_2-\Delta_\phi)\,, \nonumber\\
  &  \nu_{-+}(\Delta_1,\Delta_2)= \frac{\left(\Delta _1-2 \Delta _{\phi }\right) \left(\Delta _{\phi }-\Delta _1+\Delta _2\right)
   \left(\Delta _{\phi }-\Delta _1+\Delta _2+1\right) \left(2 \Delta _{\phi }+\Delta _2-1\right)}{4(\Delta_2^2-1)}\,,\nonumber\\
   &\nu_{0-}(\Delta_1,\Delta_2)= -\frac{1}{2}(\Delta_2-2\Delta_\phi)(\Delta_1-\Delta_2+\Delta_\phi)\,,\, \nu_{00}(\Delta_1,\Delta_2)= \frac{1}{4}(9c_{\Delta_\phi}-c_{\Delta_1}-c_{\Delta_2})\,, \nonumber\\
    &\nu_{0+}(\Delta_1,\Delta_2)= \frac{\left(\Delta _1+\Delta _2-\Delta _{\phi }\right) \left(\Delta _{\phi }-\Delta _1+\Delta _2\right)
   \left(\Delta _{\phi }+\Delta _1+\Delta _2-1\right) \left(2 \Delta _{\phi }+\Delta _2-1\right)}{8(4\Delta_2^2-1)}\,, \nonumber\\
   &\nu_{+-}(\Delta_1,\Delta_2)= \frac{\left(\Delta _2-2 \Delta _{\phi }\right) \left(\Delta _{\phi }+\Delta _1-\Delta _2\right)
   \left(\Delta _{\phi }+\Delta _1-\Delta _2+1\right) \left(2 \Delta _{\phi }+\Delta _1-1\right)}{4(4\Delta_1^2-1)}\,, \nonumber\\
  & \nu_{+0}(\Delta_1,\Delta_2)= \frac{\left(\Delta _1+\Delta _2-\Delta _{\phi }\right) \left(\Delta _{\phi }+\Delta _1-\Delta _2\right)
   \left(\Delta _{\phi }+\Delta _1+\Delta _2-1\right) \left(2 \Delta _{\phi }+\Delta _1-1\right)}{8(4\Delta_1^2-1)}\,,\nonumber\\
   &  \nu_{++}= \frac{((\Delta_1+\Delta_2)^2-\Delta_\phi^2)\left((\Delta _1+\Delta _2)^2-(\Delta _{\phi }-1)^2\right) \left(2 \Delta _{\phi }+\Delta _1-1\right) \left(2 \Delta _{\phi }+\Delta _2-1\right)}{16(4\Delta_1^2-1)(4\Delta_2^2-1)} 
\end{align}
with similar identities for the remaining Casimirs. 
\subsection{Recurrence relations for five-point Witten diagrams}
Having derived the shift relations for the conformal blocks we can now apply them to reduce the number of exchanges in Witten diagrams from two to one and from one to zero (i.e. to a contact diagram). 

\paragraph{Double-exchange $\to$ Single-exchange}
We begin by analyzing double-exchange diagrams in terms of single-exchanges. Clearly, we can derive two independent reduction formulae by acting with the Casimirs in the two-cubic vertices. Treating the single-exchange OPE coefficients as known (we will determine them later), we can then work out how many seed double-exchange coefficients are needed to reconstruct the full diagram.
We will discuss the diagrams in increasing order of complexity from the point of view of the recursion relations.

\medskip
The easiest diagrams are in the channels $12-3-45,\, 12-5-34, \, 12-4-35,\,13-2-45,\,23-1-45$ since they are completely determined by the single-exchange diagrams and no additional seed coefficients are needed. For example, we have the relation
\begin{equation}
   a_{\Delta_n,\Delta_m}^{12-5-34}= \kappa_{\Delta_1}  \frac{a_{\Delta_n,\Delta_m}^{125-34,\Delta_2}}{c_{\Delta_n}-c_{\Delta_1}}\,,
\end{equation}
and similarly for the remaining channels.
The second hardest diagrams are  in the channels $13-5-24,\,13-4-25,\,14-5-23,\,15-4-23,\,15-2-34,\,25-1-34,\,14-2-35,\,24-1-35$, and can be reduced to single-exchange coefficients, up to a single seed coefficient with $n=m=0$. These diagrams satisfy one three-term recurrence relation and one nine-term recurrence relation. An example of a three-term recurrence relation is
\begin{align}
  \nu_-(\Delta_n+1,\Delta_m)\, a^{13-4-25}_{\Delta_n+1,\Delta_m}+\nu_0(\Delta_n,\Delta_m)\, a^{13-4-25}_{\Delta_n,\Delta_m}+\nu_+(\Delta_n-1,\Delta_m)\, a^{13-4-25}_{\Delta_n-1,\Delta_m} \,,\nonumber\\
  =c_{\Delta_1}a^{13-4-25}_{\Delta_n,\Delta_m}+ \kappa_{\Delta_1}a^{134-25,\Delta_2}_{\Delta_n,\Delta_m}
\end{align}
and similarly in the other channels. The nine term recurrence relations are instead similar to the ones given below for the hardest diagrams.
These are the diagrams in the channels $15-3-24,\,14-3-25,\,$ which satisfy two nine-term recurrence relations such that two seed blocks are needed, for example those with $n=m=0$ and $n=1, m=0$. An example of these relations is
\begin{align}
 & \nu_{--}(\Delta_n+1,\Delta_m+1) a^{15-3-24}_{\Delta_n+1,\Delta_m+1}+ \nu_{-0}(\Delta_n+1,\Delta_m) a^{15-3-24}_{\Delta_n+1,\Delta_m} \nonumber\\
 & +  \nu_{0-}(\Delta_n,\Delta_m+1) a^{15-3-24}_{\Delta_n,\Delta_m+1}+ \nu_{00}(\Delta_n,\Delta_m) a^{15-3-24}_{\Delta_n,\Delta_m}+ \nu_{0+}(\Delta_n,\Delta_m-1) a^{15-3-24}_{\Delta_n,\Delta_m-1} \nonumber\\
 & +  \nu_{+-}(\Delta_n-1,\Delta_m+1) a^{15-3-24}_{\Delta_n-1,\Delta_m+1}+ \nu_{+0}(\Delta_n-1,\Delta_m) a^{15-3-24}_{\Delta_n-1,\Delta_m} \nonumber\\
 &+ \nu_{-+}(\Delta_n+1,\Delta_m-1) a^{15-3-24}_{\Delta_n+1,\Delta_m-1}+ \nu_{++}(\Delta_n-1,\Delta_m-1) a^{15-3-24}_{\Delta_n-1,\Delta_m-1}  \nonumber\\
 & = c_{\Delta_1} a^{15-3-24}_{\Delta_n,\Delta_m}+ \kappa_{\Delta_1} a^{153-24,\Delta_2}_{\Delta_n,\Delta_m}\,,
\end{align}
and similarly for the other coefficients and in the other channel.
\paragraph{Single-exchange $\to$ Contact}
Reducing the double- to single-exchanges is already a drastic reduction in complexity, as both the Mellin (only single pole) and integrated vertex methods (only single integral) become much more manageable in this case. Nonetheless, nothing stops us from further simplifying the problem by using recurrence relations to reduce OPE coefficients to those of the contact diagram. 

The nature of this reduction is now very different, since we have only one instead of two Casimir operators to act with, the number of independent recurrence relations is lower and in some channels, we will be left with an infinite number of seed blocks. Fortunately, even in the worst case scenario we reduce the `doubly infinite' number of coefficients (labeled by $n,m$) to a `single infinity' labeled by either $n=0,m\geq0$ or $n\geq0,m=0$ (or both). We again proceed by increasing level of complexity. 

\medskip

The easiest channels are the $12-345$ and $123-45$ exchanges which are completely determined in terms of the contact diagram. For example
\begin{equation}
   a^{12-345,\Delta_1}_{\Delta_n,\Delta_m}=\kappa_{\Delta_1} \frac{ a^{\rm{ctc}}_{\Delta_n,\Delta_m}}{c_{\Delta_n}-c_{\Delta_1}}  \,,
\end{equation}
and similarly for the other diagram.
The channels $125-34$ and $124-35$ require as boundary conditions the full $n\geq0,m=0$ set of coefficients and satisfy recursion relations of the form
\begin{align}
  \nu_-(\Delta_n,\Delta_m+1)\, a^{125-34}_{\Delta_n,\Delta_m+1}+\nu_0(\Delta_n,\Delta_m)\, a^{125-34}_{\Delta_n,\Delta_m}+\nu_+(\Delta_n,\Delta_m-1)\, a^{125-34}_{\Delta_n,\Delta_m-1} \,,\nonumber\\
  =c_{\Delta_1}a^{125-34}_{\Delta_n,\Delta_m}+ \kappa_{\Delta_1} a^{\rm{ctc}}_{\Delta_n,\Delta_m}
\end{align}
and similarly the channels $13-245$ and $23-145$ require as boundary conditions the full $n=0,m\geq0$ set of coefficients and satisfy the same recursion relations with the role of $n$ and $m$ swapped.
Finally, the most complicated diagrams are in the channels $14-235,\, 15-234,\, 24-135$ and $25-134$, and require as seeds the full set of coefficients satisfying $n m=0$. They satisfy recursion relations of the form
\begin{align}
 & \nu_{--}(\Delta_n+1,\Delta_m+1) a^{14-235}_{\Delta_n+1,\Delta_m+1}+ \nu_{-0}(\Delta_n+1,\Delta_m) a^{14-235}_{\Delta_n+1,\Delta_m} \nonumber\\
 & +  \nu_{0-}(\Delta_n,\Delta_m+1) a^{14-235}_{\Delta_n,\Delta_m+1}+ \nu_{00}(\Delta_n,\Delta_m) a^{14-235}_{\Delta_n,\Delta_m}+ \nu_{0+}(\Delta_n,\Delta_m-1) a^{14-235}_{\Delta_n,\Delta_m-1} \nonumber\\
 & +  \nu_{+-}(\Delta_n-1,\Delta_m+1) a^{14-235}_{\Delta_n-1,\Delta_m+1}+ \nu_{+0}(\Delta_n-1,\Delta_m) a^{14-235}_{\Delta_n-1,\Delta_m} \nonumber\\
 &+ \nu_{-+}(\Delta_n+1,\Delta_m-1) a^{14-235}_{\Delta_n+1,\Delta_m-1}+ \nu_{++}(\Delta_n-1,\Delta_m-1) a^{14-235}_{\Delta_n-1,\Delta_m-1}  \nonumber\\
 & = c_{\Delta_1} a^{14-235}_{\Delta_n,\Delta_m}+ \kappa_{\Delta_1} a^{\rm{ctc}}_{\Delta_n,\Delta_m}\,,
\end{align}
and similarly for the other channels.
A complete list of all the recursion relations that can be used to reduced the computations to the minimal amount of seed coefficients is given in \texttt{casimir5ptrecrel.nb}.
\bibliographystyle{JHEP}
\bibliography{bibALM.bib}

\end{document}